\documentclass{article}

\usepackage[top=1.in, bottom=1.in, left=0.9in, right=0.9in]{geometry}

\usepackage[T1]{fontenc} 
\usepackage{graphicx} 
\usepackage{color} 
\usepackage{amsmath,amssymb,theorem} 
\usepackage{mathtools}
\usepackage{caption}
\usepackage{sectsty}
\usepackage[round]{natbib}
\usepackage{changepage}
\usepackage{bm}
\usepackage{verbatim}
\usepackage{graphicx}

\definecolor{deepcarmine}{rgb}{0.66, 0.13, 0.24}

\usepackage{hyperref}
\hypersetup{
	colorlinks=true,       
	linkcolor=black,          
	citecolor=deepcarmine,        
}

\newcommand*\diff{\mathop{}\!\mathrm{d}}

\DeclareCaptionLabelFormat{adja-page}{\hrulefill\\#1 #2 \emph{(previous page)}}
\DeclareCaptionLabelFormat{adja-page2}{\hrulefill\\#1 #2}
\allsectionsfont{\sffamily}

\begin{document}

\vspace{2cm}

\begin{flushleft}
\huge{ {Linking connectivity, dynamics and computations in low-rank recurrent neural networks}}

\vspace{0.5cm}
\large{\textsf{Francesca Mastrogiuseppe \textsuperscript{\bf 1,2}, Srdjan Ostojic \textsuperscript{\bf 1 *}}}
\normalsize{
	\\
	\bigskip
	\textsuperscript{\sffamily{\bf 1}} Laboratoire de Neurosciences Cognitives, INSERM U960 and
	\\
	\textsuperscript{\sffamily{\bf 2}} Laboratoire de Physique Statistique, CNRS UMR 8550 \\
	\'Ecole Normale Sup\'erieure - PSL Research University, 75005 Paris, France
	\\
	\bigskip
         \textsuperscript{\bf*} Lead contact. Correspondence: \textsf{srdjan.ostojic@ens.fr}}

\end{flushleft}

\par\noindent\rule{\textwidth}{0.3pt}

\section*{ Summary }
Large  scale neural recordings have established  that the  transformation  of sensory  stimuli into  motor outputs relies  on low-dimensional  dynamics at the  population level, while individual neurons exhibit complex selectivity. Understanding how  low-dimensional computations on  mixed, distributed representations   emerge   from  the   structure   of  the   recurrent connectivity  and inputs to  cortical networks  is a  major challenge. Here, we  study a class of recurrent network models in which the connectivity is a sum of a random part and  a minimal, low-dimensional  structure. We show  that, in such  networks, the  dynamics are low dimensional and can be directly  inferred from connectivity using  a geometrical approach. We  exploit  this  understanding  to  determine  minimal  connectivity required to implement  specific computations, and find that the dynamical range and computational  capacity  quickly increase   with  the   dimensionality   of  the   connectivity  structure. This  framework produces  testable experimental predictions for the relationship between connectivity, low-dimensional dynamics and computational features of recorded neurons.


\newpage

\section*{ Introduction }

Understanding the  relationship between synaptic  connectivity, neural
activity  and  behavior  is  a central  endeavor  of  neuroscience.
Networks  of neurons encode  incoming stimuli  in terms  of electrical
activity  and transform  this  information into  decisions and  motor
actions through synaptic  interactions, thus implementing computations
that underly  behavior. Reaching a  simple, mechanistic grasp  of the
relation between connectivity, activity and behavior is, however, highly
challenging.   Cortical  networks, which  are  believed to  constitute
the fundamental  computational units  in the  mammalian brain,  consist of
thousands   of  neurons  that   are  highly   inter-connected  through
recurrent synapses.   Even if one  were able to  experimentally record
the activity of every neuron  and the strength of each synapse in a
behaving  animal,  understanding the causal relationships between these
quantities  would  remain  a  daunting  challenge  because  an  appropriate
conceptual        framework         is        currently        lacking
\citep{GaoGanguli}.       Simplified,
computational  models  of  neural   networks  provide  a  testbed  for
developing such a framework.   In computational models and trained artificial neural networks, the strengths
of all  synapses and  the activity  of all neurons  are known,  yet an
understanding  of  the  relation  between connectivity,  dynamics  and input-output
computations   has  been   achieved  only   in  very   specific  cases
(e.g.~\citet{Hopfield, BenYishai1995, Wang2002}).

One of the most popular  and best-studied classes of network models is
based  on fully  random recurrent  connectivity \citep{Sompolinsky1988,
  Brunel2000,   VanVreeswijkSompolinsky96}.   Such  networks   display
internally generated irregular  activity that closely  resembles spontaneous
cortical    patterns   recorded   {\em    in-vivo}   \citep{ShadlenNewsome98}.    However, randomly  connected recurrent networks display
only very  stereotyped responses to external inputs \citep{Rajan2010}, can implement
only  a limited  range of  input-output  computations  and  their spontaneous dynamics  are  typically high dimensional
\citep{Williamson2016}.  To  implement more elaborate  computations and
low-dimensional  dynamics, classical  network models  rely  instead on
highly  structured connectivity, in  which every  neuron belongs  to a
distinct cluster,  and is  selective to only  one feature of  the task
(e.g.~\citet{Wang2002, AmitBrunel, LitwinKumarDoiron}). Actual  cortical connectivity appears  to be
neither  fully random  nor  fully structured  \citep{HarrisMrsicFlogel}, and the activity of individual neurons displays a similar
mixture   of  stereotypy  and   disorder  \citep{Rigotti2013,Mante2013,
  ChurchlandShenoy2007}.  To take these observations into account and
implement  general-purpose   computations,  a  large variety   of  functional
approaches  have been  developed for  training recurrent  networks and
designing  appropriate connectivity  matrices  \citep{Hopfield, JaegerHaas,
  Maass2007,       SussilloAbbott,      EliasmithAnderson,Boerlin2013, Pascanu2013,MartensSutskever}.   A unified  conceptual  picture of  how
connectivity determines dynamics and computations is, however, currently
missing \citep{Barak2017, Sussillo2014}.

Remarkably, albeit developed  independently and motivated by different
goals, several of the functional approaches for designing connectivity
appear       to       have       reached       similar       solutions
\citep{Hopfield,JaegerHaas,SussilloAbbott,EliasmithAnderson,Boerlin2013},
in which the implemented computations  do not determine every single entry in
the  connectivity  matrix but  instead  rely  on  a specific  type  of
minimal, low-dimensional  structure, so that in mathematical terms the
obtained connectivity matrices are \emph{low rank}.  In classical Hopfield networks
\citep{Hopfield, Amit1985}, a rank-one term is added to the
connectivity matrix for every item  to be memorized, and each of these
terms fixes  a single dimension, i.e.~row/column  combination, of the
connectivity matrix.  In echo-state \citep{JaegerHaas, Maass2007} and FORCE
learning  \citep{SussilloAbbott},  and   similarly  within  the  Neural
Engineering   Framework  \citep{EliasmithAnderson},   computations  are
implemented through feedback  loops from readout units to  the bulk of
the network. Each feedback loop is mathematically equivalent to adding
a rank-one component and fixing a single row/column combination of the
otherwise  random  connectivity  matrix.   In the  predictive  spiking
theory   \citep{Boerlin2013}  the   requirement  that   information  is
represented  efficiently leads  again  to a  connectivity matrix  with
similar  low-rank form.   Taken  together, the  results of  these
studies  suggest that a  minimal,  low-rank structure  added  on top  of
random  recurrent  connectivity may  provide  a  general and  unifying
framework for implementing computations in recurrent networks.

Based  on this observation,  here we  study a  class of recurrent networks in
which the  connectivity is  a sum of a structured, low-rank part  and a
random part.  We  show that in such  networks, both
spontaneous and  stimulus-evoked activity are  low-dimensional and can
be predicted from the  geometrical relationship between a small number
of high-dimensional vectors  that represent the connectivity structure
and the  feed-forward inputs.  This understanding  of the relationship
between connectivity and network dynamics allows us to directly design
minimal,  low-rank  connectivity  structures that  implement  specific
computations.   We  focus  on  four tasks  of  increasing  complexity,
starting   with   basic   binary   discrimination  and   ending   with
context-dependent evidence integration \citep{Mante2013}.  We find that
the  dynamical repertoire of  the network  increases quickly  with the
dimensionality  of  the   connectivity  structure,  so  that  rank-two
connectivity structures  are already sufficient  to implement complex,
context-dependent tasks  \citep{Mante2013,Saez2015}. For each  task, we
illustrate  the  relationship  between  connectivity,  low-dimensional
dynamics and  the performed computation. In  particular, our framework
naturally   captures   the    ubiquitous   observation   that single-neuron responses are highly heterogeneous  and  mixed  \citep{Rigotti2013,
  Mante2013,    ChurchlandShenoy2007,    Machens2010},    while    the
dimensionality  of the  dynamics  underlying computations  is low  and
increases with task  complexity \citep{GaoGanguli}. Crucially, for each
task, our framework  produces experimentally testable predictions that directly
relate connectivity, the dominant  dimensions of the dynamics, and the
computational features of individual neurons.

\section*{Results}

We  studied  a  class  of  models which  we  call  low-rank  recurrent
networks. In these networks, the connectivity matrix was given by  a sum
of an  uncontrolled, random matrix and a  structured, controlled matrix
$P$. The structured matrix $P$ was low rank, i.e. it consisted only
of a  small number of independent  rows and columns, and its entries were assumed to be weak (of order $1/N$, where $N$ is the number of units in the network).  We considered $P$
moreover to be fixed and known, and uncorrelated with the random part $g\chi$,
which  was  considered unknown  except  for  its  statistics (mean  0,
variance ${g^2}/{N}$). As in  classical models, the networks consisted
of $N$ firing rate units with a sigmoid input-output transfer function
\citep{Sompolinsky1988, SussilloAbbott}:

\begin{equation}
	\dot{x}_i(t)=-x_i(t)+\sum_{j=1}^N J_{ij} \phi(x_j(t))+I_i,
\end{equation}
where $x_i(t)$ is the total input current to unit $i$, $J_{ij}=g\chi_{ij}+P_{ij}$ is the connectivity matrix, $\phi(x)=\tanh(x)$ is the current-to-rate transfer function, and $I_i$ is the external, feed-forward input to unit $i$.

To connect with the previous literature and introduce the methods that
underlie our results, we  start by describing the spontaneous dynamics ($I_i=0$)
in a  network with  a unit-rank  structure $P$.  We  then turn  to the
response to external  inputs, the core of our  results that we exploit
to  demonstrate how  low-rank networks  can implement  four  tasks of
increasing complexity.

\subsection*{One-dimensional spontaneous activity in networks with unit-rank structure}

\begin{figure}[t]

	\begin{adjustwidth}{-0.15in}{0in} 
	\centering
	\includegraphics{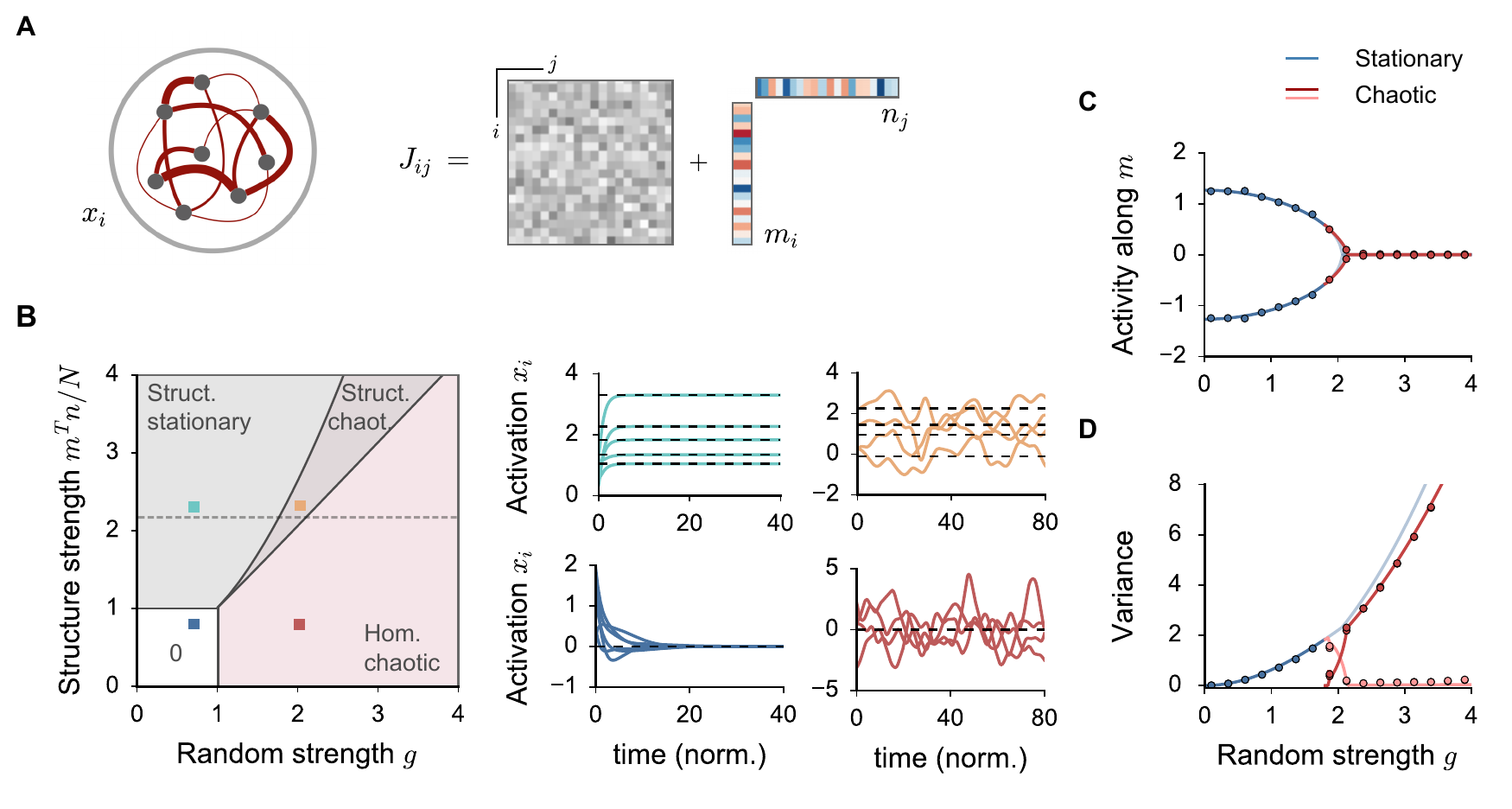}
	\end{adjustwidth}
	
	\caption{ 
		{\bf \textsf{Spontaneous activity in random networks with unit-rank connectivity structure.}}
		\textsf{\bfseries A.} The recurrent network model, whose connectivity matrix consists of the sum of a random (gray) and of a structured unit-rank (colored) component.
		\textsf{\bfseries B.} Left: dynamical regimes of the network activity as function of the structure connectivity strength $m^Tn/N$  and the random strength $g$. Gray areas: bistable activity; red: chaotic activity. Side panels: samples of dynamics from finite networks simulations (parameters indicated by colored dots in the phase diagram).
		\textsf{\bfseries C-D}. Activity statistics as the random strength $g$ is increased and the structure strength is fixed to $2.2$ (dashed line in \textsf{B}).
		\textsf{\bfseries C}: Activity along the vector $m$, as quantified by $\kappa = \langle n_i [\phi_i] \rangle$. Blue (resp. red) lines: theoretical prediction for stationary (resp. chaotic) dynamics. 
		\textsf{\bfseries D}: Activity variance due to random connectivity. Blue and pink lines: static heterogeneity, red: temporal variance that quantifies chaotic activity. Dots: simulations of finite-size networks. 
		See \emph{Methods} for details. }	
	
	\label{fig:p1}
	
\end{figure}

We  started  with  the   simplest  possible  type  of  low-dimensional
connectivity,  a matrix  $P$ with  unit-rank (Fig.~\ref{fig:p1} \textsf{A}). Such  a matrix  is
specified  by two $N$-dimensional vectors  $m=\{m_i\}$ and  $n=\{n_j\}$, which  fully
determine all its  entries. Every column in this  matrix is a multiple
of the vector $m$,  and every row is a multiple of  the vector $n$, so
that the individual entries are given by
\begin{equation}
P_{ij}=\frac{m_in_j}{N}.  \label{eq:rank1_def}
\end{equation}
We will call $m$ and $n$  respectively the right- and left-connectivity vectors (as they correspond to the right and left eigenvectors of the matrix $P$, see \emph{Methods}), and we
consider them  arbitrary, but fixed  and uncorrelated with  the random
part  of the connectivity.  As we  will show,   the spontaneous
network dynamics can  be directly understood from the  geometrical arrangement of the
vectors $m$ and $n$.

 In absence of  structured connectivity, the
dynamics   are  determined  by   the  strength   $g$  of   the  random
connectivity: for $g<1$,  the activity in absence of  inputs decays to
zero,  while  for  $g>1$  it  displays  strong,  chaotic  fluctuations
\citep{Sompolinsky1988}.  Our first aim was to understand  how the interplay
between the  fixed, low-rank  part and the  random part  of the
connectivity shapes the spontaneous activity in the network. 

Our analysis  of network dynamics relies on  an effective, statistical
description that can be mathematically derived if the network is large
and the  low-dimensional part  of the connectivity  is weak  (i.e.  if
$P_{ij}$ scales inversely with the  number of units $N$ in the network
as in Eq.~\ref{eq:rank1_def}).   Under those assumptions, the activity
of  each unit can be described  in terms of the mean and variance of
the total input it  receives. Dynamical equations for these quantities
can be derived by extending  the classical dynamical mean-field theory
\citep{Sompolinsky1988}.   This   theory   effectively   leads   to   a
low-dimensional description of network  dynamics in terms of equations
for a couple of macroscopic  quantities.  Full details of the analysis
are provided  in the  \emph{Methods}; here, we  focus only on  the main
results.

The central ingredient of the theory is an equation for the average equilibrium input $\mu_i$ to unit $i$: 
\begin{equation}\label{eq:mu_main}
\mu_i= \kappa m_i ,  \quad  \mathrm{    where } \quad \kappa=\frac{1}{N} \sum_{j=1}^N n_j  \big[\phi_j \big].  
\end{equation}

The scalar  quantity $\kappa$ represents the overlap  between the left-connectivity    vector    $n$     and    the    $N$-dimensional    vector
$[\phi]=\{\big[\phi_j  \big]  \}$   that  describes  the  mean  firing
activity of  the network  ($[\phi_j]$ is the  firing rate of  unit $j$
averaged over  different realizations of  the random component  of the
connectivity, and depends implicitly on $\kappa$). The overlap $\kappa$ therefore quantifies the degree of
structure  along the vector  $n$ in  the activity  of the  network. If
$\kappa>0$, the equilibrium activity of each neuron is correlated with
the  corresponding  component  of  the vector  $n$,  while  $\kappa=0$
implies no such structure is present. The overlap $\kappa$ is the key macroscopic quantity describing the network dynamics, and our theory provides equations specifying its dependence on network parameters.

If  one   represents  the   network  activity  as   a  point   in  the
$N-$dimensional state-space where every dimension  corresponds to the
activity  of  a  single  unit,  Eq.~\ref{eq:mu_main}  shows  that  the
structured  part   of  the  connectivity   induces  a  one-dimensional
organization of  the spontaneous activity along the  vector $m$.  This
one-dimensional  organization,  however,  emerges  only if  the  overlap
$\kappa$ does not vanish.  As the activity of the network is organized
along the  vector $m$, and  $\kappa$ quantifies the projection  of the
activity onto the vector $n$, non-vanishing values of $\kappa$ require
a  non-vanishing overlap between  vectors $m$  and $n$.  This overlap,
given by $m^Tn/N=\sum_jm_jn_j/N$,  directly  quantifies the strength
of  the  structure in  the  connectivity.  The connectivity  structure
strength  $m^Tn/N$ and  the activity  structure strength  $\kappa$ are
therefore directly related, but in a highly non-linear manner.  If the
connectivity structure is weak, the network only exhibits homogeneous,
unstructured activity corresponding to $\kappa=0$ (Fig.~\ref{fig:p1}  \textsf{B}
blue).    If  the   connectivity  structure   is   strong,  structured
heterogeneous activity  emerges ($\kappa>0$), and the  activity of the
network at equilibrium is organized  in one dimension along the vector
$m$     (Fig.~\ref{fig:p1}    \textsf{B}     green    and
\textsf{C}),  while   the   random  connectivity   induces
additional  heterogeneity along  the remaining  $N-1$  directions. Note
that, because of the symmetry  in the specific input-output function we
use, when a heterogeneous  equilibrium state exists, the configuration
with  the opposite  sign  is an  equilibrium  state too,  so that  the
network  activity is  bistable (for  more general  asymmetric transfer
functions, this bistability is still present, although the symmetry is
lost, see Fig.~S7).

The random part  of the connectivity disrupts the  organization of the
activity induced  by the connectivity structure  through two different
effects. The first effect is that as the random strength $g$ is increased, 
for any  given realization  of the  random part  of the  connectivity, the
total input to  unit $i$ will deviate more  strongly from the expected
mean  $\mu_i$ (Fig.~\ref{fig:p1} \textsf{D}). As a consequence, the activity along the $N-1$ directions that are orthogonal to $m$ increases, resulting in a noisy input to individual neurons that smoothens the gain of the non-linearity. This
  effectively leads  to  a
reduction of  the overall  structure in the  activity as  quantified by
$\kappa$ (Fig.~\ref{fig:p1} \textsf{C}).
 A  second, distinct effect  is that  increasing the  random strength
eventually  leads to  chaotic activity  as in  purely  random networks.   Depending on
the  strength of the  structured connectivity,  two different  types of
chaotic dynamics can  emerge.  If the disorder in  the connectivity is
much   stronger  than   structure,  the   overlap  $\kappa$   is  zero
(Fig.~\ref{fig:p1}  \textsf{C}).   As  a result,  the  mean
activity  of   all  units  vanishes   and  the  dynamics   consist  of
unstructured, $N-$dimensional temporal fluctuations (Fig.~\ref{fig:p1}
\textsf{D}), as  in the  classical chaotic  state  of fully
random  networks  (Fig.~\ref{fig:p1}  \textsf{B}  red).   In
contrast, if  the strengths of the random  and structured connectivity
are  comparable,   a structured
type of chaotic activity emerges, in which $\kappa>0$ so that the mean
activity of  different units is  organized in one dimension  along the
direction $m$  as shown by  Eq.~\ref{eq:mu_main}, but the  activity of
different   units   now    fluctuates   in   time   (Fig.~\ref{fig:p1}
\textsf{B}  orange). As  for structured static  activity, in
this situation the system is bistable as states with opposite signs of
$\kappa$  always exist.

The phase diagram in Fig.~\ref{fig:p1} \textsf{B} summarizes
the  different  types  of  spontaneous  dynamics that  can  emerge  as
function of  the strength of  structured and random components  of the
connectivity   matrix.   Altogether,   the  structured   component  of
connectivity   favors  a   one-dimensional  organization   of  network
activity, while the  random component favors high-dimensional, chaotic
fluctuations.   Particularly  interesting  activity emerges  when  the
structure and disorder are comparable, in which case the dynamics show
one-dimensional  structure  combined  with  high-dimensional  temporal
fluctuations that can  give  rise  to
dynamics   with   very   slow  timescales   (see  Fig.~S6).

\subsection*{Two-dimensional activity in response to an external input}

\begin{figure}[h!]
	
	\centering
	\includegraphics{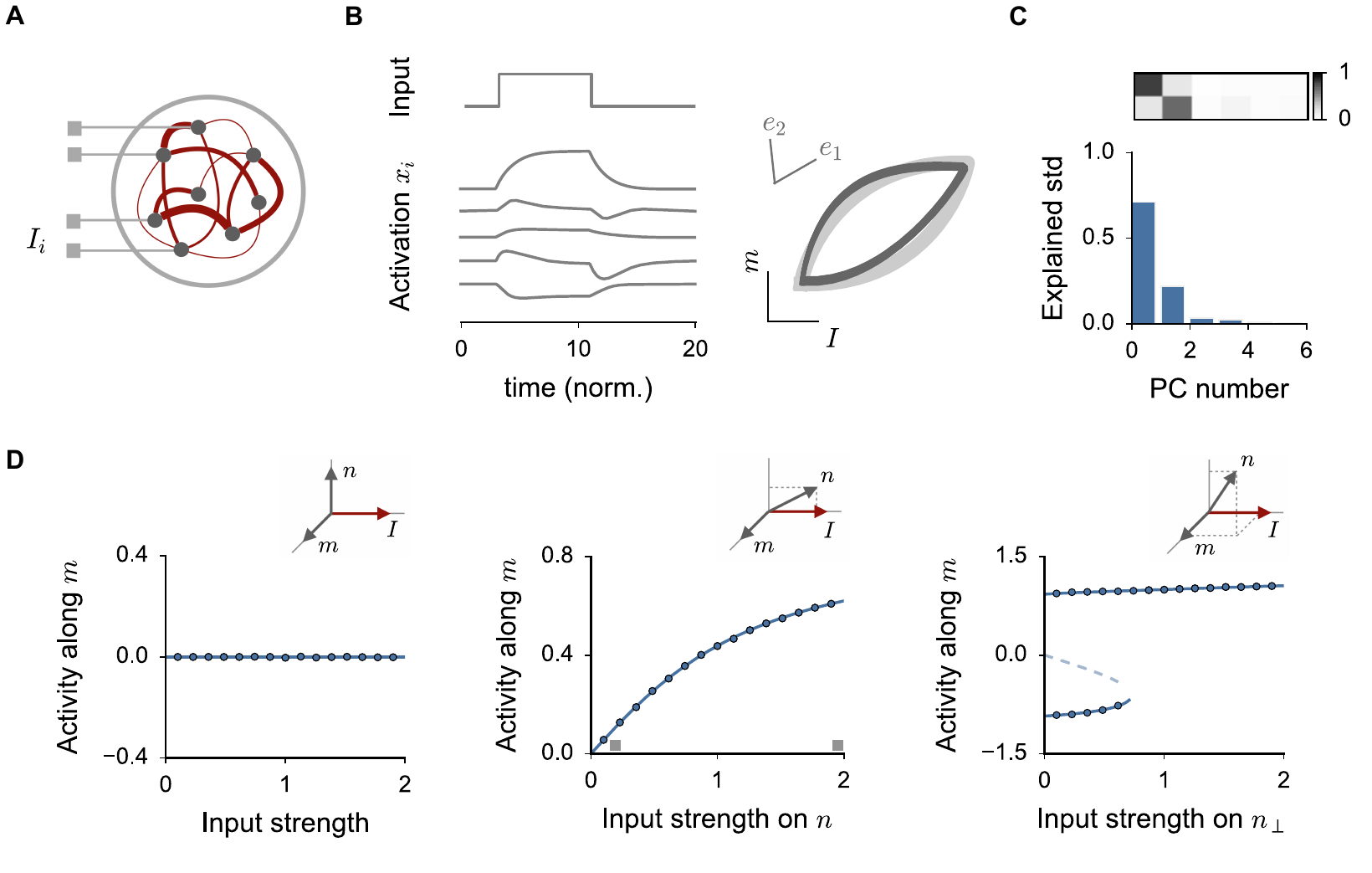}
	
	\caption{ {\bf \textsf{External inputs generate two-dimensional activity in random networks with unit-rank structure.} }
		\textsf{\bfseries A.} The pattern of external inputs can be represented by an $N$-dimensional vector $I=\{I_i\}$, where $I_i$ is the input to unit $i$.
		\textsf{\bfseries B.} Transient dynamics in response to a step input along $I$ in a sample network of $N=3500$ units. Left: activity traces for five units. Right: projections of the population trajectory  onto the plane defined by the right-connectivity vector $m$ and the input vector $I$. Light trace: theoretical prediction. Dark traces: simulations.
		\textsf{\bfseries C.} Principal components (PC) analysis of the average activity trajectory. Bottom: fraction of standard deviation explained by successive PCs. Top: correlation between PCs and the vectors $m$ and $I$. The direction of the projections onto the $m - I$ plane of the two top PCs $e_1$ and $e_2$ are represented in \textsf{B}. See also Fig.~S3.
		\textsf{\bfseries D.} The activity $\kappa$ along $m$ is determined by the geometrical arrangement of the vector $I$ and the connectivity vectors $m$ and $n$. Three different cases are illustrated: (left) $I$, $m$ and $n$ mutually orthogonal; (center) $m$ and $n$ mutually orthogonal, but $I$ has a non-zero overlap with $n$; (right) $m$ and $n$ have non-zero overlap, leading to bistable activity in absence of inputs. Increasing the external input along $n$ suppresses one of the two stable states. Continuous lines: theoretical predictions. Dots: simulations. See \emph{Methods} for details. }
	
	\label{fig:p2}
	
\end{figure}

We  now  turn  to  the  response  to an  external, feed-forward input
 (Fig.~\ref{fig:p2} \textsf{A}). At equilibrium, the total average input to unit $i$ is the sum of a recurrent input $\kappa m_i$ and the feed-forward input $I_i$:
\begin{equation}\label{eq:mu_inputs}
\mu_i=  \kappa m_i +I_i,  \quad  \mathrm{    where } \quad \kappa=\frac{1}{N} \sum_{j=1}^N n_j  \big[\phi_j \big].
\end{equation}
Transient, temporal dynamics close to this equilibrium are obtained by including temporal dependencies in $\kappa$ and $I_i$
(see \emph{Methods}, Eq.~\ref{eq:mu_temp}).

Fig.~\ref{fig:p2} \textsf{B} illustrates the response of the
network to a  step input.  The response of individual units is highly heterogeneous, different units showing increasing, decreasing or multi-phasic responses. While every unit  responds differently, the
theory predicts that,  at the level of the  $N$-dimensional state space
representing the  activity of the whole population,  the trajectory of
the activity  lies on average on the two-dimensional  plane spanned by
the  right-connectivity  vector $m$  and  the  vector  $I =\{I_i\}$  that
corresponds  to the  pattern  of external inputs (Fig.~\ref{fig:p2}
\textsf{B}).   Applying  to  the simulated  activity  a
dimensionality  reduction  technique (see \citet{CunninghamYu} for a recent review)   such  as  Principal  Components
Analysis    confirms  that  the   two
dominant dimensions of the activity indeed lie in the $m-I$ plane (Fig.~\ref{fig:p2}
\textsf{C}), while the
random part  of connectivity leads  to additional activity  in the
remaining  $N-2$ directions  that grows  quickly with  the  strength of
random  connectivity  $g$ (see  Fig.~S3).  This approach therefore  directly links the connectivity in  the network to
the  emerging low-dimensional  dynamics, and  shows that  the dominant
dimensions of activity are determined by a combination of feed-forward
inputs and connectivity \citep{Wang2018}.


The contribution of the connectivity vector $m$ to the two-dimensional
trajectory of  activity is quantified by the  overlap $\kappa$ between
the  network  activity  $[\phi]$  and the  left-connectivity  vector  $n$
(Eq.~\ref{eq:mu_inputs}).  If  $\kappa=0$, the activity  trajectory is
one-dimensional,  and simply  propagates the  pattern  of feed-forward
inputs. This is in particular  the case for fully random networks.  If
$\kappa\neq  0$,  the  network   response  is  instead  a  non-trivial
two-dimensional  combination of the  input and  connectivity structure
patterns.   In  general, the  value  of  $\kappa$,  and therefore  the
organization of network activity, depends on the geometric arrangement
of the  input vector  $I$ with respect  to the  connectivity
vectors  $m$  and $n$,  as  well  as on  the  strength  of the  random
component of the connectivity $g$.

As  the neural  activity  lies  predominantly in  the  $m-I$ plane,  a
non-vanishing  $\kappa$,  together  with  non-trivial  two-dimensional
activity is obtained  when the vector $n$ has  a non-zero component in
the $m-I$ plane. Two  qualitatively different input-output regimes can
be distinguished. The first one is obtained when the connectivity vectors
$m$  and   $n$  are   orthogonal  to  each   other  (Fig.~\ref{fig:p2}
\textsf{D} left and center). In that case, the overlap between them is zero,
and  the spontaneous  activity in  the network  bears no  sign  of the
underlying  connectivity  structure.   Adding  an external  input  can,
however, reveal this connectivity structure and generate non-trivial two-dimensional
activity  if the  input vector  $I$ has  a non-zero  overlap  with the
left-connectivity vector $n$. In such a situation, the vector $n$ picks up  the component of the activity 
along the feed-forward input direction $I$. This leads   to    a    non-zero    overlap   $\kappa$,    which  in turn  implies  that the network activity  will have a
component  along  the   right-connectivity  vector  $m$.  Increasing  the
external input along the direction of $n$ will therefore progressively
increase the  response along $m$  (Fig.~\ref{fig:p2} \textsf{D} center), leading to a two-dimensional output.

A  second, qualitatively  different input-output  regime  is obtained
when the connectivity vectors $m$ and $n$ have a strong enough overlap along a
common direction (Fig.~\ref{fig:p2} \textsf{D} right).  As already
shown in  Fig.~\ref{fig:p1}, an overlap larger than  unity between $m$
and  $n$  induces    bistable,
structured spontaneous  activity along the dimension $m$.  Adding an external  input along the vector $n$  increases the activity along $m$, but
also eventually  suppresses one of the bistable  states.  Large external
inputs along the $n$ direction therefore reliably set the network
into a state in which the activity is a two-dimensional combination of
the input direction and the  connectivity direction $m$. This can lead to
a strongly  non-linear input-output transformation if  the network was
initially set in the state that lies on the opposite branch (Fig.~\ref{fig:p2} \textsf{D} right).

An additional effect of an external input is that it generally tends to
suppress chaotic activity present when the random part of connectivity
is strong  (Figs.~S3 and S4).  This suppression  occurs irrespectively of
the  specific  geometrical configuration  between  the  input $I$  and
connectivity vectors $m$ and $n$,  and therefore independently of the two
input-output  regimes  described  above. 
Altogether, external inputs suppress both
chaotic  and bistable  dynamics (Fig.~S4), and  therefore always
decrease    the    amount    of    variability   in    the    dynamics
\citep{ChurchlandM2010,Rajan2010}.

In summary, external, feed-forward inputs  to a network with unit-rank
connectivity structure in general lead to two-dimensional trajectories
of  activity.   The elicited  trajectory  depends  on the  geometrical
arrangement of the pattern of  inputs with respect to the
connectivity  vectors  $m$ and  $n$,  which  play  different roles.   The
right-connectivity vector $m$ determines the output pattern of network
activity,  while  the   left-connectivity  vector  $n$  instead
selects the inputs that give rise to outputs along $m$.
An output structured along $m$ can be obtained  when $n$ selects
recurrent inputs  (non-zero overlap  between $n$ and  $m$) or  when it
selects external inputs (non-zero overlap between $n$ and $I$).

\subsection*{Higher-rank structure leads to a rich dynamical repertoire}

This  far we  focused  on unit-rank  connectivity  structure, but  our
framework can  be directly  extended to higher  rank structure.  A more
general structured component of rank $r\ll N$ can be written as a superposition of $r$ independent unit-rank terms

\begin{equation}
P_{ij}=\frac{m_i^{(1)}n_j^{(1)}}{N}+\ldots+\frac{m_i^{(r)}n_j^{(r)}}{N},
\end{equation}
and  is  in principle  characterized  by  $2r$  vectors $m^{(k)}$  and
$n^{(k)}$.   In  such a  network,  the  average  dynamics lie  in  the
$(r+1)$-dimensional  subspace  spanned  by  the $r$  right-connectivity
vectors $m^{(k)}, k=1, \ldots, r$ and the input vector $I$, while the left
connectivity vectors  $n^{(k)}$ select  the inputs amplified  along the
corresponding  dimension $m^{(k)}$.  The  details of  the dynamics  will in general
depend  on the  geometrical arrangement  of these  $2r$  vectors among
themselves  and with  respect  to  the input  pattern.  The number  of
possible  configurations  increases   quickly  with the  structure
rank, leading to a wide repertoire of dynamical states that includes continuous
attractors  (Fig. S5)  and sustained oscillatory activity  (Fig.  S8).  In  the remainder of this  manuscript, we will  explore only the
rank-two  case.

\subsection*{Implementing a simple discrimination task}

\begin{figure}[h!]

	\begin{adjustwidth}{-0.05in}{0in} 
	\centering
	\includegraphics{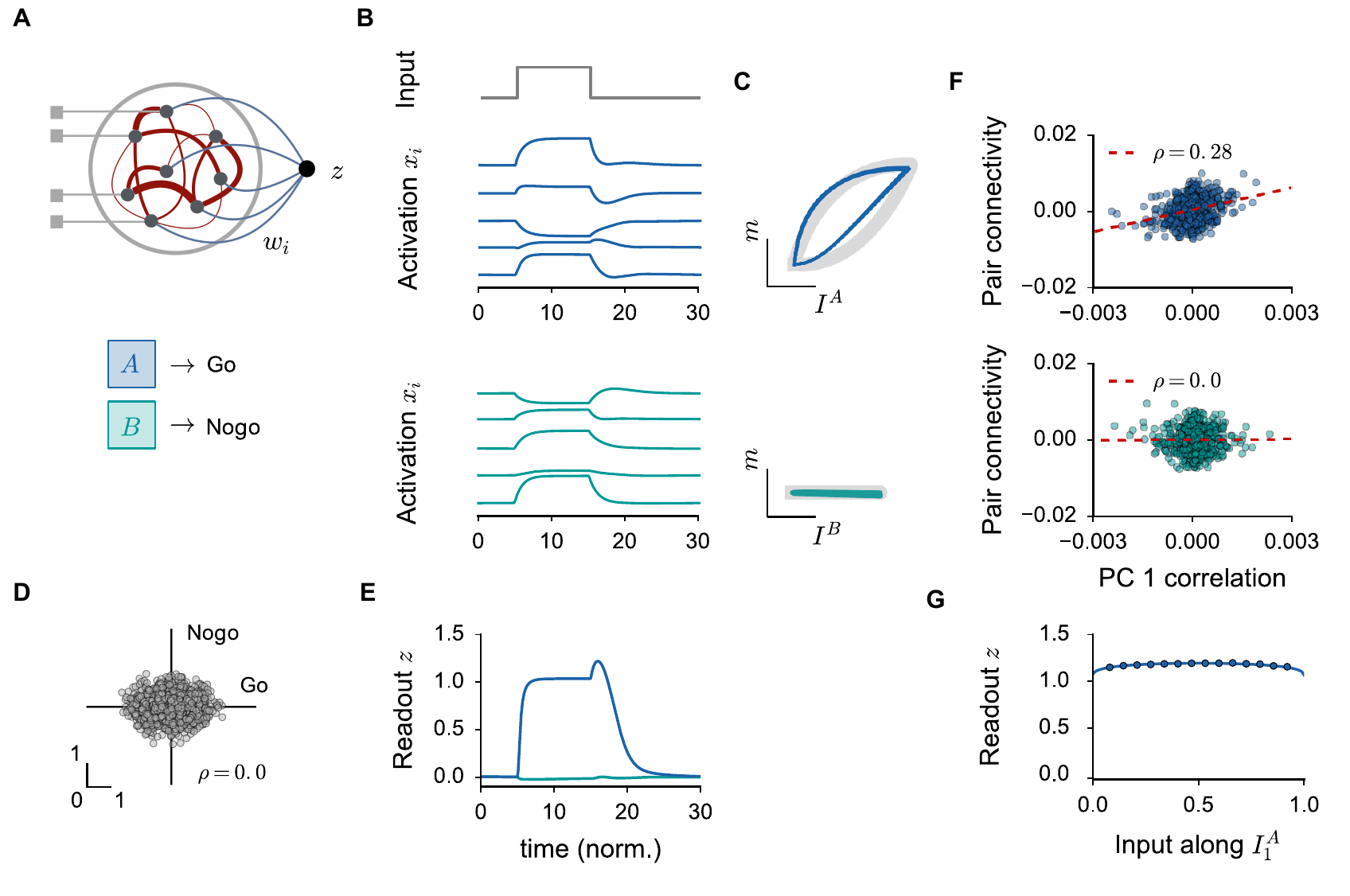}
	\end{adjustwidth}
	
	\caption{{\bf \textsf{Implementing a simple Go-Nogo discrimination task  with a unit-rank  connectivity structure.} }
		\textsf{\bfseries A.}  A linear readout is added to the network,  with randomly chosen weights  $w_i$. The stimuli are represented by random input patterns $I^A$ and $I^B$. The task consists in producing an output in response to stimulus $A$, but not $B$. The simplest unit-rank structure that implements the task is given by $m=w$ and $n=I^{A}$. 
		\textsf{\bfseries B.} Response of a sample network to the Go (blue) and Nogo (green) inputs. Activity traces for five units. 
		\textsf{\bfseries C.} Projections of the population trajectories onto the planes predicted to contain the dominant part of the dynamics. Gray: predicted trajectory. Colored traces: simulations. 
		\textsf{\bfseries D.} Linear regression coefficients for the Go and the Nogo stimuli. Every dot corresponds to a network unit. 
		\textsf{\bfseries E.} Readout dynamics for the Go (blue) and the Nogo (green) stimulus. 
		\textsf{\bfseries F.} Average connectivity strength as a function of the product between the coefficients of the first PC. Every dot corresponds to a pair of units. 
		\textsf{\bfseries G.} Generalization properties of the network. We select two Go stimuli $I^{A}_1$ and $I^{A}_2$, and we set $n = I^{A}_1+I^{A}_2$. We build the input pattern as a normalized mixture of the two preferred patterns, and we gradually increase the component along  $I^{A}_1$. Continuous lines: theoretical predictions. Dots: simulations. 
		See \emph{Methods} for details.  
	}
	
	\label{fig:p3}
	
\end{figure}

Having developed an intuitive,  geometric understanding of how a given
unit-rank   connectivity  structure  determines   the  low-dimensional
dynamics in a network, we now reverse our approach to ask how a given
computation   can  be  implemented   by  choosing   appropriately  the
structured  part   of  the   connectivity.   We start with  the
computation  underlying  one  of   the  most  basic  and  most  common
behavioral tasks,  Go-Nogo stimulus  discrimination. In this  task, an
animal has to produce a specific  motor output, e.g.~press a lever or
lick a spout, in response to a stimulus $I^{A}$ (the Go stimulus), and
ignore another stimuli $I^{B}$ (Nogo stimuli). This computation can be
implemented in  a straightforward  way in a  recurrent network  with a
unit-rank connectivity structure. While  such a simple computation does
not in  principle require a  recurrent network, the  implementation we
describe  here illustrates  in a  transparent manner  the relationship
between connectivity, dynamics  and computations in low-rank networks,
and   leads  to   non-trivial  and   directly   testable  experimental
predictions.   It also  provides  the basic  building  block for  more
complex tasks, which we turn to in the next sections.

We model the  sensory stimuli as random patterns
of external inputs to the network,  so that the two stimuli are  represented by two fixed,
randomly-chosen $N$-dimensional  vectors $I^{A}$ and $I^{B}$. To  model the motor
response,  we  supplement  the  network  with an  output  unit,  which
produces  a linear readout  $z(t)=\frac{1}{N}\sum_i w_i  \phi(x_i(t))$ of
network activity (Fig.~\ref{fig:p3} \textsf{A}).   The readout weights $w_i$ are  chosen randomly and
form also  a fixed $N$-dimensional  vector $w$.  The task of the network is to produce an output that
is selective to the Go stimulus:  the readout $z$ at the end of stimulus presentation needs to be non-zero
for the input pattern $I^{A}$ that corresponds to the Go stimulus, and
zero for the other input $I^{B}$. 
 
The two $N$-dimensional vectors  $m$ and $n$ that
generate the appropriate unit-rank connectivity structure to implement
the task can be directly determined from our description of network dynamics. As  shown in Eq.~\ref{eq:mu_inputs} and Fig.~\ref{fig:p2},
the  response of  the network  to the  input pattern  $I$  is in
general two-dimensional and  lies in the plane spanned  by the vectors
$m$ and  $I$. The  output unit will therefore produce a  non-zero readout
only if the readout vector $w$ has a non-vanishing overlap with either
$m$ or $I$. As $w$  is assumed to be uncorrelated, and therefore
orthogonal,  to all input  patterns, this  implies that  the connectivity
vector $m$  needs to have a  non-zero overlap with  the readout vector
$w$ for the network to  produce a non-trivial output. This output will
depend on the amount of activity along $m$, quantified by the overlap $\kappa$.
As shown  in Fig.~\ref{fig:p2}, the overlap $\kappa$  will be non-zero
only  if  $n$ has  a  non-vanishing  overlap  with the  input  pattern. Altogether, implementing the Go-Nogo task therefore requires that the right-connectivity vector $m$ is correlated with the readout vector $w$, and that the left-connectivity vector $n$ is correlated with the Go stimulus $I^{A}$.

Choosing  $m=w$   and  $n=I^{A}$,  therefore   provides  the  simplest
unit-rank  connectivity  that   implements  the  desired  computation.
Fig.~\ref{fig:p3}  illustrates  the   activity  in  the  corresponding
network.   At the  level  of individual  units,  by construction  both
stimuli  elicit large  and heterogeneous  responses (Fig.~\ref{fig:p3}
\textsf{B})   that   display   mixed  selectivity   (Fig.~\ref{fig:p3}
\textsf{D}). As predicted by the  theory, the response to stimulus $B$
is dominantly one-dimensional  and organized  along the input  direction $I^{B}$,
while the response to stimulus  $A$ is two-dimensional and lies in the
plane  defined by  the  right-connectivity vector  $m$  and the  input
direction $I^{A}$ (Fig.~\ref{fig:p3} \textsf{C}). The readout from the
network corresponds  to the  projection of the  activity onto  the $m$
direction,  and   is  non-zero  only  in  response   to  stimulus  $A$
(Fig.~\ref{fig:p3} \textsf{E}), so  that the network indeed implements
the  desired  Go-Nogo task.   Our  framework  therefore  allows us  to
directly link  the connectivity, the low-dimensional  dynamics and the
computation performed by the  network, and leads to two experimentally
testable   predictions.   The   first   one  is   that  performing   a
dimensionality-reduction  separately on responses  to the  two stimuli
should lead  to larger dimensionality of the  trajectories in response
to the Go stimulus. The second prediction is that for the Go stimulus,
the  dominant   directions  of   activity  depend  on   the  recurrent
connectivity  in the  network, while  for  the Nogo  stimulus they  do
not. More specifically, for the  activity elicited by the Go stimulus,
the dominant principal components are combinations of the input vector
$I^{A}$ and  right-connectivity vector  $m$. Therefore if  two neurons
have large principal component weights, they are expected to also have
large  $m$  weights and  therefore  stronger  mutual connections  than
average  (Fig.~\ref{fig:p3}  \textsf{F}  top).  In contrast,  for  the
activity  elicited   by  the  Nogo  stimulus,   the  dominant  principal
components are determined solely by the feed-forward input, so that no
correlation between dominant PC  weights and recurrent connectivity is
expected (Fig.~\ref{fig:p3}  \textsf{F} bottom).  This  prediction can
in  principle   be  directly   tested  in  experiments   analogous  to
\citet{Ho2011}, where  calcium imaging in behaving  animals is combined
with   measurements  of   connectivity   in  a   subset  of   recorded
neurons.  Note  that in this setup  very weak  structured  connectivity  is
sufficient   to   implement  computations,   so   that  the   expected
correlations may be weak if the random part of the connectivity is strong (see Fig.~S5).

The  unit-rank connectivity structure  forms the  fundamental scaffold
for  the desired  input-output transform.  The random  part of  the
connectivity adds variability around the target output, and can induce
additional  chaotic fluctuations. Summing  the activity  of individual  units through  the  readout unit,
however,  averages out this heterogeneity, so  that the  readout error
decreases  with   network  size  as   $1/\sqrt{N}$  (Fig.~S5).
The  present implementation is therefore robust to
noise,  and has desirable  computational   properties in  terms  of
generalization to novel  stimuli. In particular, it can   be  extended   in  a
straightforward way  to the  detection of a  category of  Go stimuli,
rather than a  single stimulus (Fig.~\ref{fig:p3} \textsf{G}).

\subsection*{Detection of a noisy stimulus}

\begin{figure}[h!]

	\centering
	\includegraphics{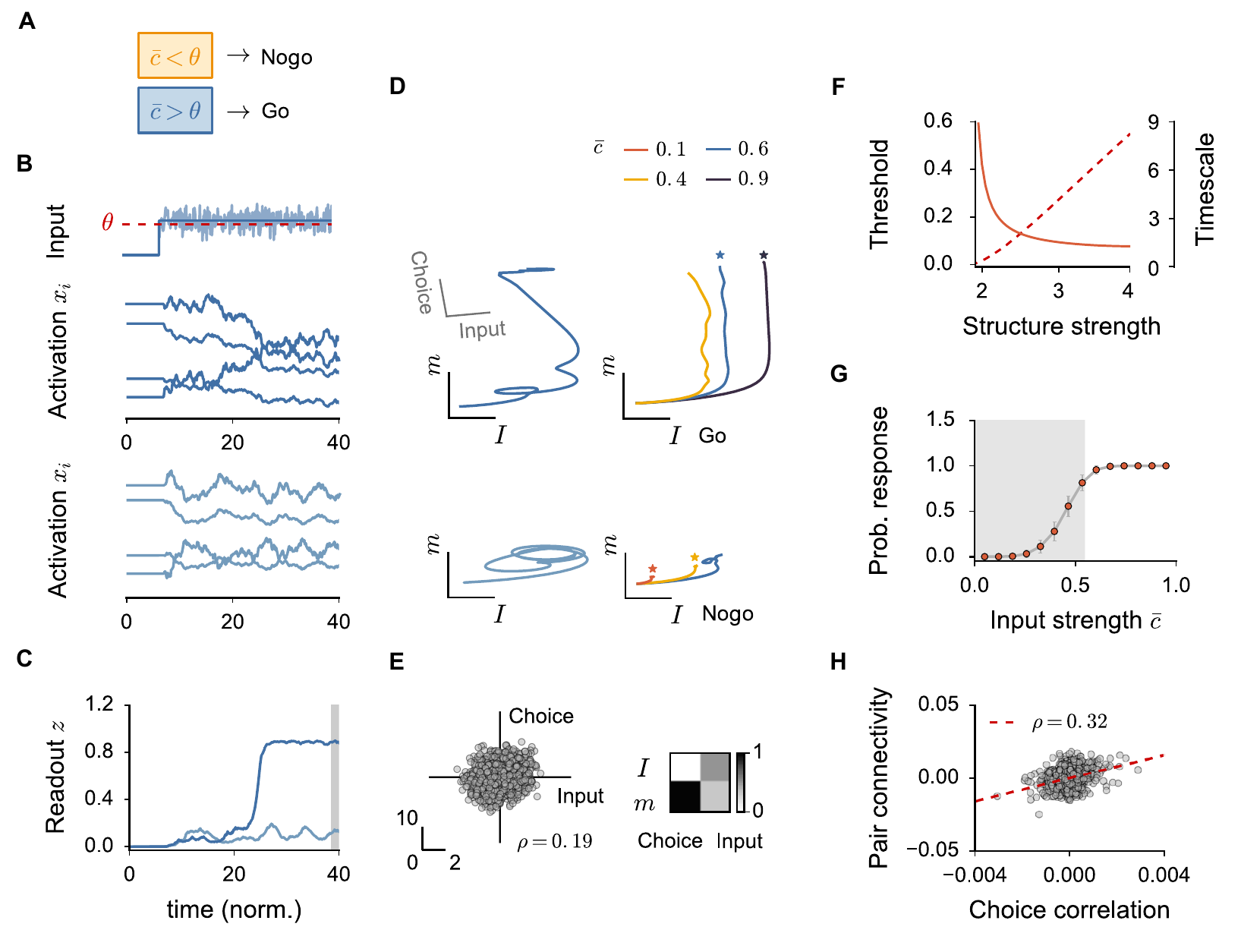}
	
	\caption{{\bf \textsf{Implementing a noisy detection task with a unit-rank  connectivity structure.} }
		\textsf{\bfseries A.}  The network is given a noisy input $c(t)$ along a fixed, random pattern of inputs $I$. The task consists in producing an output if the average input $\bar{c}$ is larger than a threshold $\theta$.
		\textsf{\bfseries B.} Dynamics in a sample network. Top: noisy input and threshold. Bottom: activity traces for four units and two different noise realizations in the stimulus, leading to a Go (dark blue) and a Nogo (light blue) output. 
		\textsf{\bfseries C.} Readout dynamics for the two stimuli.
		\textsf{\bfseries D.} Projections of the population trajectory onto the plane defined by the right-connectivity vector $m$ and the input vector $I$. Left: single-trial trajectories corresponding to \textsf{B}. Right: trial-averaged trajectories, for Go (top) and Nogo (bottom) outputs, and different values of the mean input $\bar{c}$. Stars indicate correct responses. 
		\textsf{\bfseries E.} Left: linear regression coefficients for the input amplitude and the decision outcome. Every dot corresponds to a network unit. Right: correlation coefficients between the vectors $m$ and $I$ and the input and choice regression axes (see \emph{Methods}). Projection directions of the two input and choice regression axes onto the $m - I$ plane are shown in \textsf{D}.
		\textsf{\bfseries F.} Detection threshold (dashed), and time scale of the effective exponential filter (full line) for increasing values of the structure strength. 
		\textsf{\bfseries G.}  Psychometric curve. The shaded area indicates the  bistable region.
		\textsf{\bfseries H.} Average connectivity strength as a function of the product of the linear regression coefficients for the choice variable. Every dot corresponds to a pair of network units. See \emph{Methods} for details. 	
	}
	
	\label{fig:p4}
	
\end{figure}

We  now  turn  to  a  slightly  more  complex  task:  integration  of  a
continuous,  noisy  stimulus.  In contrast  to the  previous discrimination
task, where  the stimuli were completely  different (i.e. orthogonal),
here we  consider a  continuum of stimuli  that differ only  along the
intensity of a  single feature, such as the  coherence of a random-dot
kinetogram \citep{Newsome1989}. In  a given  stimulus presentation, this  feature moreover
fluctuates in time.  We  therefore represent each stimulus as $c(t)I$,
where $I$  is a fixed, randomly  chosen input vector  that encodes the
relevant  stimulus  feature,  and  $c(t)$  is the  amplitude  of  that
feature.  We  consider a  Go-Nogo version of  this task, in  which the
network has to  produce an output only if the average  value of $c$ is
larger than a threshold (Fig.~\ref{fig:p4}
\textsf{A}).

As  for  the  basic  discrimination  task, the  central  requirements  for  a
unit-rank  network   to  implement  this  task  are   that  the  right-connectivity vector $m$ is correlated with the readout vector $w$, and
the left-connectivity vector $n$  is correlated with the input pattern
$I$. A key  novel requirement in the present task  is however that the
response  needs to  be  non-linear to  produce  the Go  output when  the
strength  of the input  along $I$  is larger  than the  threshold.  As
shown in Fig.~\ref{fig:p2}  \textsf{D}, such a non-linearity
can be obtained  when the left- and right-connectivity vectors $n$ and
$m$ have  a strong enough overlap. We  therefore add a  shared component to
$m$ and $n$ along a direction orthogonal to both $w$ and $I$.  In that
setup, if the stimulus intensity $c$  is low, the network will be in a
bistable regime,  in which  the activity along  the direction  $m$ can
take  two  distinct  values  for  the  same  input  (Fig.~\ref{fig:p2}
\textsf{D} right). Assuming that  the lower state represents a
Nogo output, and that the network  is initialized in this state at the
beginning of the trial, increasing  the stimulus intensity $c$ above a
threshold  will lead  to a  sudden  jump, and  therefore a  non-linear
detection of the stimulus.   Because the input amplitude fluctuates noisily in time,
whether such a jump  occurs depends on the integrated estimate of the stimulus intensity. The timescale over which this estimate is integrated is determined by the time-constant of the effective exponential filter describing the network dynamics. In our unit-rank network, this time-constant is set by the connectivity strength, i.e.~the overlap between the left- and
right-connectivity  vectors $m$ and  $n$, which also determines the  value of
the threshold. Arbitrarily large timescales can be obtained by adjusting this overlap close to the bifurcation value, in which case the threshold becomes arbitrarily small (Fig.~\ref{fig:p4}
\textsf{F}). In this section, we fix the structure strength so that the threshold is set to $0.5$, which corresponds to an integration timescale  of the order of the time constant of individual units.

Fig.~\ref{fig:p4}   illustrates    the   activity   in    an   example
implementation of this  network. In a given trial,  as the stimulus is
noisy,  the  activity  of  the individual  units  fluctuates  strongly
(Fig.~\ref{fig:p4}   \textsf{B}).  Our   theory   predicts  that   the
population  trajectory on  average lies  in the  plane defined  by the
connectivity vector  $m$ and the input  pattern $I$ (Fig.~\ref{fig:p4}
\textsf{D}).  Activity  along the  $m$ direction is  picked up  by the
readout, and its value at  the end of stimulus presentation determines
the  output (Fig.~\ref{fig:p4} \textsf{C}).   Because of  the bistable
dynamics in the network, whether the $m$ direction is explored, and an
output  produced, depends  on the  specific noisy  realization  of the
stimulus.  Stimuli  with an  identical average strength  can therefore
either  lead  to  two-dimensional  trajectories  of  activity  and  Go
responses, or  one-dimensional trajectories of  activity corresponding
to Nogo responses  (Fig.~\ref{fig:p4} \textsf{D}).  The probability of
generating  an  output as  function  of  stimulus  strength follows  a
sigmoidal psychometric curve  that reflects the underlying bistability
(Fig.~\ref{fig:p4}  \textsf{G}).  Note  that  the bistability  is  not
clearly apparent  on the level of individual  units. In particular, the  activity of
individual units is always  far from saturation, as their  inputs are distributed
along a zero-centered Gaussian (Eq.~\ref{eq:mu_inputs}).

The responses of individual units are strongly heterogeneous and exhibit mixed
selectivity to  stimulus strength and output choice (Fig.~\ref{fig:p4}
\textsf{E}).  A popular manner
to  interpret such  activity at  the  population level  is a  targeted
dimensional reduction  approach, in which  input and choice
dimensions  are determined through
regression   analyses  \citep{Mante2013}.  
As expected from our theoretical analysis, the two  dimensions  obtained through  regression  are  closely   related  to  $m$  and  $I$;  in
particular, the  choice dimension is  highly correlated with  the right-connectivity  vector  $m$ (Fig.~\ref{fig:p4}
\textsf{E}).  As a result, the plane in which network  activity dominantly lies 
corresponds to the plane defined by the choice and the input dimensions (Fig.~\ref{fig:p4}
\textsf{D}).
Our  framework therefore  directly  links
recurrent connectivity  and effective output choice direction  through the low-dimensional
dynamics.  A  resulting experimentally testable prediction  is  that
neurons with strong choice regressors have stronger mutual connections (Fig.~\ref{fig:p4}
\textsf{H}).

\subsection*{A context-dependent discrimination task}

\begin{figure}[h!]
	
	\centering
	\includegraphics{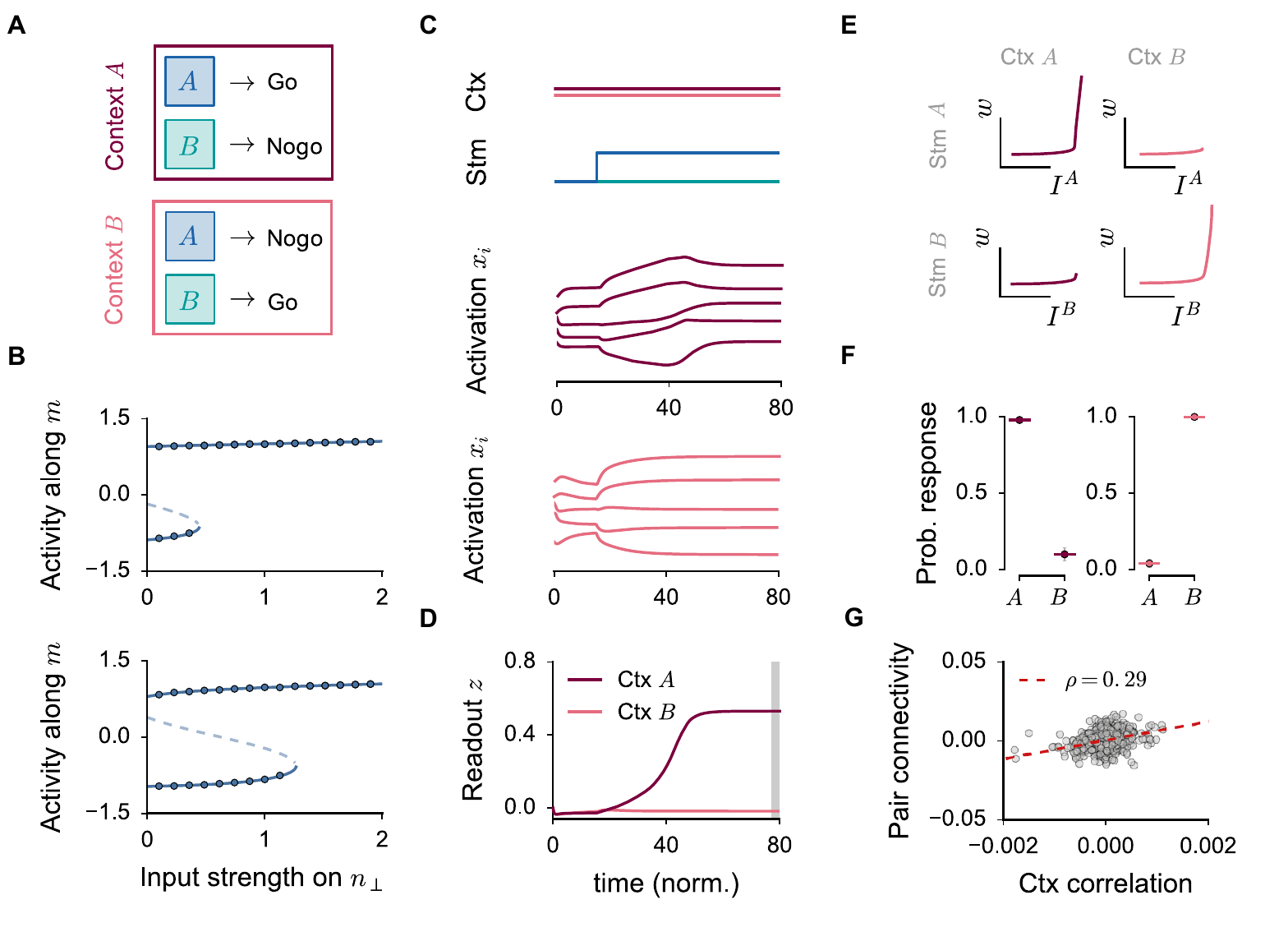}
	
	\caption{{\bf \textsf{Implementing a context-dependent Go-Nogo discrimination task with a rank-two  connectivity structure.} }
		\textsf{\bfseries A.} As in Fig.~\ref{fig:p3}, two stimuli $A$ and $B$ are presented to the network. The task consists in producing an output in response to the Go stimulus, which is determined by the contextual cue ($A$ in Context $A$, $B$ in Context $B$), modeled as inputs along random directions $I_{ctx A}$ and $I_{ctx B}$.
		\textsf{\bfseries B.} Inputs along the overlap direction between the left- and the right-connectivity vectors modulate the response threshold of the network (see also Fig.~S5). 
		\textsf{\bfseries C.} Dynamics in a sample network in response  to the stimulus $A$. Top: stimulus and contextual input. Bottom:  activity for five units in contexts $A$ (crimson) and $B$ (pink).
		\textsf{\bfseries D.} Readout dynamics in the two contexts.
		\textsf{\bfseries E.} Projections of the average population trajectories onto the planes spanned by vectors $w$, $I^A$ and $I^B$. 
		\textsf{\bfseries F.} Network performance in the two contexts.
		\textsf{\bfseries G.} Average connectivity strength between pairs of units as a function of the product between the regression coefficients for context. Every dot corresponds to a pair of network units. See \emph{Methods} for details. 
	}
	
	\label{fig:p5}
	
\end{figure}

We next consider a context-dependent discrimination task, in which the
relevant response  to a stimulus  depends on an  additional, explicit
contextual  cue.   Specifically,  we  focus  on the  task  studied  in
\citet{Saez2015}  where in  one context  (referred  to as  Context $A$),  the
stimulus $A$ requires a Go  output, and the stimulus $B$ a Nogo, while
in the other context (referred to as Context $B$), the associations are
reversed (Fig.~\ref{fig:p5} \textsf{A}).   This  task is  a  direct  extension  of the  basic  binary
discrimination   task   introduced   in   Fig.~\ref{fig:p3},  yet   it   is
significantly more  complex as it  represents a hallmark  of cognitive
flexibility:  a  non-linearly separable,  $XOR$-like computation  that a
single-layer feed-forward network cannot solve \citep{Rigotti2010, Fusi2016}. We will show that this
task can be  implemented in a rank-two recurrent  network that is a
direct extension  of the unit-rank network used for  the discrimination  task in
Fig.~\ref{fig:p4}.

This context-dependent task can be  seen as a combination of two basic,
opposite Go-Nogo  discriminations, each of which  can be independently
implemented by  a unit-rank structure with  the right-connectivity
vector  $m$  correlated to  the readout,  and  the left-connectivity vector  correlated to the Go  input ($I^A$
for Context $A$, $I^B$ for Context $B$).  Combining two such unit-rank
structures, with left-connectivity  vectors $n^{(1)}$ and $n^{(2)}$ correlated
respectively with $I^A$ and $I^B$,  leads to a rank-two connectivity structure that
serves as a  scaffold for the present task.  The  cues for context $A$
and  $B$ are represented  by additional  inputs along  random vectors
$I_{ctx A}$ and $I_{ctx B}$, presented  for the full length of the trial \citep{Remington2018}
(Fig.~\ref{fig:p5} \textsf{C}).  These inputs are the only contextual information
incorporated in the  network. In particular, the readout  vector $w$ is
fixed and independent of the context \citep{Mante2013}. Crucially, since the readout $w$ needs to produce an output for both input stimuli,  both right-connectivity
vectors $m^{(1)}$ and $m^{(2)}$ need to be correlated with it.

The key  requirement for implementing  context-dependent discrimination is
that  each contextual  input effectively  switches off  the irrelevant
association.   To implement  this  requirement, we  rely  on the  same
non-linearity  as for  the  noisy discrimination  task,  based on  the
overlap  between  the left- and  right-connectivity  vectors (Fig.~\ref{fig:p2} 
\textsf{D}). We  however exploit an  additional property, which is  that the
 threshold of the non-linearity  (i.e. the position  of the transition  from a
bistable    to    a    mono-stable   region    in    Fig.~\ref{fig:p2} 
\textsf{D})  can be controlled by an  additional modulatory input along
the overlap direction  between $m$ and $n$  (Figs.~\ref{fig:p5} \textsf{B} and S4).
Such a modulatory input acts as an effective offset for the bistability at the macroscopic, population level (see Eq.~\ref{eq:kappa_modulation} in \emph{Methods}).
A stimulus of a given strength (e.g.~unit strength in Fig.~\ref{fig:p5} \textsf{B}) may therefore induce a transition from the lower to the upper state (Fig.~\ref{fig:p5} \textsf{B} top), or no transition (Fig.~\ref{fig:p5} \textsf{B} bottom) depending on the strength of the modulatory input that sets the threshold value.
While in  the noisy discrimination  task, the overlap between  $m$ and
$n$ was  chosen in an arbitrary  direction, in the  present setting we
take the  overlaps between  each pair of  left- and  right-connectivity
vectors  to lie along  the direction  of the  corresponding contextual
input (i.e. $m^{(1)}$ and $n^{(1)}$ overlap along $I_{ctx A}$, $m^{(2)}$ and $n^{(2)}$  along $I_{ctx B}$), so that contextual inputs directly modulate the threshold of the non-linearity.  The  final
rank-two setup is described in detail in the \emph{Methods}.

Fig.~\ref{fig:p5}  illustrates  the  activity  in an  example  of  the
resulting network implementation.  The  contextual cue is present from
the very beginning of the trial, and effectively sets the network in a
context-dependent initial  state (Fig.~\ref{fig:p5} \textsf{C}) that corresponds to the lower of the two bistable states. The low-dimensional  response of the
network  to  the following  stimulus  is  determined  by this  initial
state and the sustained contextual input. If  the cue for context  $A$ is present, stimulus  $A$  leads to
the crossing  of the non-linearity, a transition from the lower to the upper state, and therefore  a two-dimensional
response in  the plane determined by  $I^A$ and $w$ (Figs.~\ref{fig:p5} \textsf{E} top left),  generating a  Go output (Fig.~\ref{fig:p5} \textsf{D}).  In  contrast, if the  cue for
context $B$ is present,  the threshold of the underlying non-linearity
is increased in the direction of input $I^A$ (Fig.~\ref{fig:p5} \textsf{B} bottom), so that the presentation
of stimulus $A$ does not induce a transition between the lower and upper states, but leads  only to a  one-dimensional trajectory orthogonal
to  the readout,  and therefore  a  Nogo response (Fig.~\ref{fig:p5} \textsf{E} top right).   The situation  is
totally symmetric in response to stimulus $B$ (Fig.~\ref{fig:p5} \textsf{E} bottom), so that contextual cues
fully  reverse  the   stimulus-response  associations (Fig.~\ref{fig:p5} \textsf{F}).  Overall,  this
context-dependent   discrimination  relies   on   strongly  non-linear
interactions between  the stimulus and contextual inputs,  that on the
connectivity  level are  implemented by  overlaps  between the connectivity
vectors along the contextual inputs. A central, experimentally testable prediction of our framework
is  therefore that,  if a  network is  implementing  this computation,
units   with  strong   contextual  selectivity   have on average  stronger  mutual
connections (Fig.~\ref{fig:p5} \textsf{G}).

\subsection*{A context-dependent evidence integration task}

\begin{figure}[h!]

	\centering
	\includegraphics{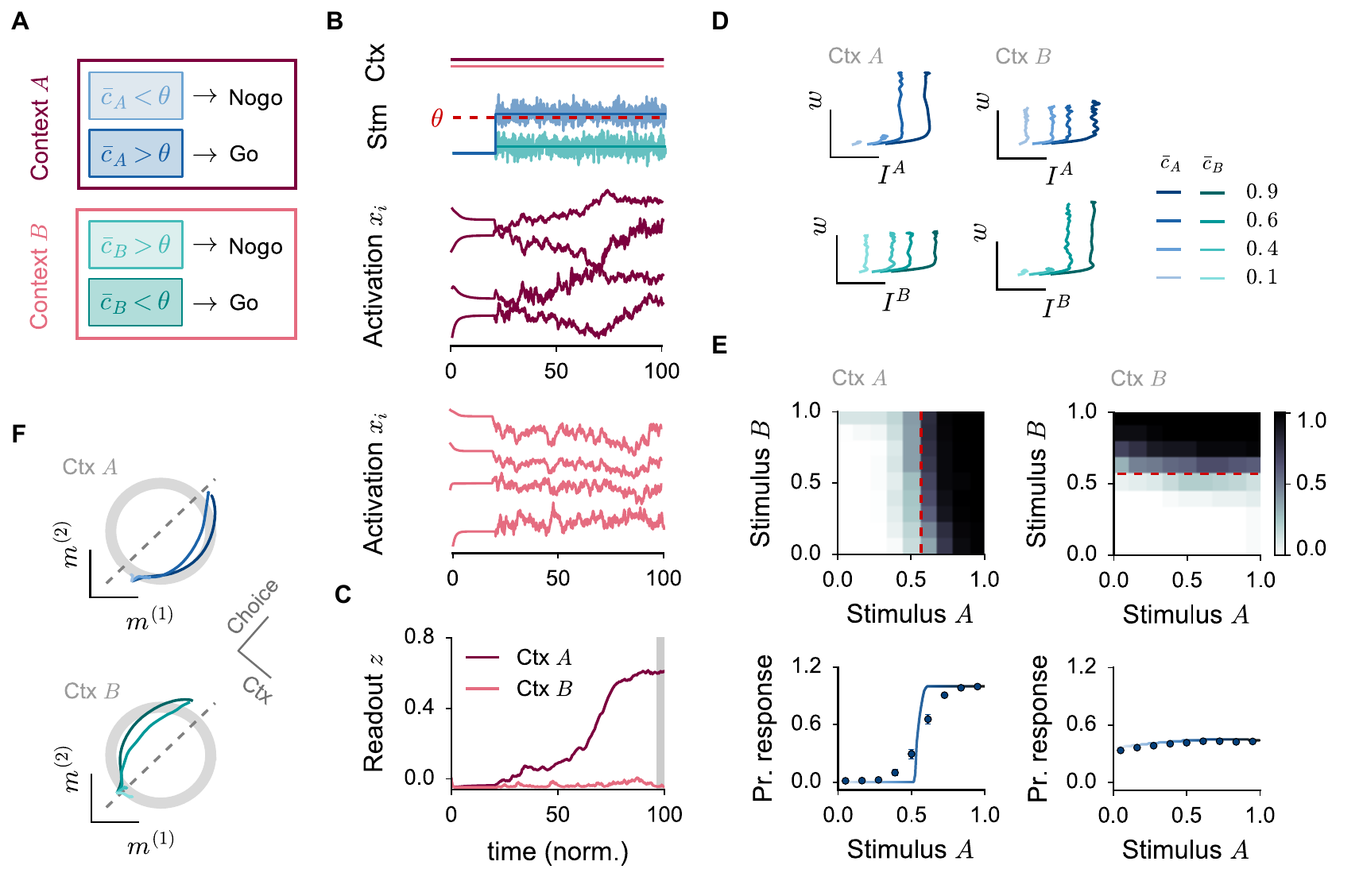}
	
	\caption{   {\bf   \textsf   {Implementing a  context-dependent evidence accumulation  task  using    rank-two connectivity   structure.}}
		\textsf{\bfseries   A.}  The stimuli consist of a superposition of two features $c_A$ and $c_B$ which  fluctuate in time around mean values $\bar{c}_A$ and $\bar{c}_B$. In every trial, a pair of contextual inputs determines the relevant input feature. The task consists in producing an output if the average strength of the relevant feature is larger than a threshold. 
		\textsf{\bfseries   B.} Dynamics in a sample network. Top: stimulus and contextual inputs. Bottom: activity of four  units in contexts A (crimson) and B (pink).
		\textsf{\bfseries C.} Readout dynamics in the two contexts.
		\textsf{\bfseries D.} Average population trajectories projected onto the planes spanned by vectors $w$, $I^A$ and $I^B$. Blue (resp. green) trajectories have been sorted according to the value of the strength of stimulus $A$ (resp. $B$), and averaged across stimulus $B$ (resp. $A$). 
		\textsf{\bfseries E.} Network performance. Top row: probability of response as function of input strengths $\bar{c}_A$ and $\bar{c}_B$ (simulated data). Bottom: probability of response averaged over $\bar{c}_B$. Continuous line: theoretical prediction; dots: simulations.
		\textsf{\bfseries F.}  Projection of the population activity onto the plane defined by the orthogonal components of the vectors $m^{(1)}$ and $m^{(2)}$ , and comparison with the underlying circular attractor (see \emph{Methods}). Trajectories are sorted by the strength of the relevant stimulus, and averaged across the non-relevant one. The direction of the projections of the regression axes for choice and context are indicated in gray. See \emph{Methods} for details. 
	}
	
	\label{fig:p6}
	
\end{figure}

We finally  examine a  task inspired by \citet{Mante2013}  that combines
context-dependent  output and fluctuating,  noisy inputs.  The stimuli
now consist of superpositions of two different features $A$ and $B$, and the
strengths of both features fluctuate  in time during a given trial.  In
\citet{Mante2013},  the stimuli  were random  dot kinetograms,  and the
features $A$ and  $B$ corresponded to the direction  of motion and color
of  these  stimuli.  The  task  consists in  classifying  the  stimuli
according to one  of those features, the relevant  one being indicated
by an explicit contextual cue (Fig.~\ref{fig:p6} \textsf{A}). 

We implemented a  Go-Nogo version of the task, in  which the output is
required to be  non-zero when the relevant feature  is stronger than a
prescribed threshold  (arbitrarily set to $0.5$). The  present task is
therefore  a direct combination  of the  detection task  introduced in
Fig.~\ref{fig:p4}  and the  context-dependent  discrimination task  of
Fig.~\ref{fig:p5}, but the individual stimuli are now two-dimensional,
as they consist of two  independently varied features $A$ and $B$.  In
this task, a significant additional  difficulty is that on every trial
the irrelevant  feature needs  to be ignored,  even if it  is stronger
than the  relevant feature (e.g.~color coherence  stronger than motion
coherence on a motion-context trial).

This  context-dependent evidence integration  task can  be implemented
with exactly the  same rank-two configuration as  the basic context-dependent
discrimination in  Fig.~\ref{fig:p5}, with  contextual gating  relying on
the same  non-linear mechanism as  in Fig.~\ref{fig:p5} \textsf{B}. The
contextual     cue    is     presented     throughout    the     trial
(Fig.~\ref{fig:p6} \textsf{B}), and determines which of the features of
the two-dimensional stimulus leads  to non-linear dynamics along the direction of
connectivity  vectors $m^{(1)}$  and  $m^{(2)}$ (Fig.~\ref{fig:p6} \textsf{D}).
These  directions share a  common component  along the  readout vector
$w$, and the  readout unit picks up the  activity along that dimension.  As  a  consequence, depending  on  the
contextual  cue,  the  same  stimulus  can lead  to  opposite  outputs
(Fig.~\ref{fig:p6} \textsf{C}). Altogether, in  Context $A$, the output
is independent of the values of feature $B$, and conversely in Context
$B$ (Fig.~\ref{fig:p6} \textsf{E}).  The output therefore behaves as if
it were  based on two  orthogonal readout directions, yet  the readout
direction  is unique and  fixed, and  the output  relies instead  on a
context-dependent selection of the relevant input feature \citep{Mante2013}.

An important  additional requirement in the present  task with respect
to the  basic context-dependent integration is that  the network needs
to perform temporal integration  to average out temporal fluctuations
in  the stimulus.  As  illustrated in  Fig.~\ref{fig:p6} \textsf{B-C},
the  network dynamics  in response  to stimuli  indeed exhibit  a slow
timescale,  and progressively integrate  the input.   Strikingly, such
slow  dynamics  do  not  require  additional  constraints  on  network
connectivity;  they   are  a   direct  consequence  of   the  rank-two
connectivity  structure  used  for  contextual  gating  (in  fact  the
dynamics are already slow in the basic contextual discrimination task,
see Fig.~\ref{fig:p5}  \textsf{C-D}). More specifically,  the symmetry
between the  two contexts  implies that two  sets of left-  and right-
connectivity     vectors     have     identical     overlaps     (i.e.
$m^{(1)T}n^{(1)}=m^{(2)T}n^{(2)}$). Without further constraints on the connectivity,
such a symmetric  configuration leads to an emergence  of a continuous
line attractor, with the shape  of a two-dimensional ring in the plane
defined    by    $m^{(1)}$    and    $m^{(2)}$   (see    \emph{Methods}    and
Fig.~S5). In the implementation  of the present task, on top
of symmetric overlaps, the  four connectivity vectors include a common
direction along  the readout  vector. This  additional constraint
eliminates  the ring  attractor, and  stabilizes only  two equilibrium
states that correspond to Go and Nogo outputs. Yet, the ring attractor
is  close  in parameter  space,  and  this  proximity induces  a  slow
manifold in  the dynamics,  so that the  trajectories leading to  a Go
output slowly evolve  along two different sides of  the underlying ring depending
on the  context (Fig.~\ref{fig:p6} \textsf{F}).  As a  result, the two
directions  in the plane  $m^{(1)}-m^{(2)}$ correspond  to choice  and context
axis as  found by regression  analysis (Fig.~\ref{fig:p6} \textsf{F}).
A similiar mechanism  for context-dependent evidence integration based
on a line attractor was previously identified by reverse-engineering a
trained  recurrent network  \citep{Mante2013}.  Whether  the underlying
dynamical structure was a ring as  in our case, or two line attractors
for the two  contexts depended on the details  of the network training
protocol (V.~Mante, Cosyne  2018). Here we show that  such a mechanism
based on a ring attractor can be implemented in a minimal network with
rank-two connectivity structure, but  other solutions can certainly be
found.  Note  that  this  rank-two   network  can  also  serve  as  an
alternative    implementation    for   context-independent    evidence
integration in which the integration timescale and the threshold value
are  fully independent  in  contrast to  the unit-rank  implementation
(Fig.~\ref{fig:p4}).

\section*{Discussion}

Motivated  by  the  observation  that  a  variety  of  approaches  for
implementing computations in recurrent  networks rely on a common type
of connectivity structure,  we studied a class of  models in which the
connectivity matrix consists of a sum  of a fixed, low-rank term and a
random  part. Our  central result  is that  the  low-rank connectivity
structure induces low-dimensional dynamics  in the network, a hallmark
of    population    activity     recorded    in    behaving    animals
\citep{GaoGanguli}. While low-dimensional activity is usually detected
numerically        using        dimensional-reduction       techniques
\citep{CunninghamYu}, we showed that  a mean-field theory allows us to
directly   predict   the  low-dimensional   dynamics   based  on   the
connectivity and input  structure.  This approach led us  to a simple,
geometrical understanding of the relationship between connectivity and
dynamics,    and   enabled    us   to    design   minimal-connectivity
implementations of specific computations.  In particular, we found that
the  dynamical repertoire of  the network  increases quickly  with the
rank  of the  connectivity  structure, so  that rank-two  networks
can already 
implement  a variety  of  computations.  In this  study,  we have  not
explicitly considered  structures with rank  higher than two,  but our
theoretical framework is in principle  valid for arbitrary rank $r \ll N$,
where $N$ is the size of the network.

While other works have examined dynamics in networks with a mixture of
structured    and   random   connectivity    (e.g.~\citet{RoudiLatham, Ahmadian2015}),   the  most   classical   approach  for   implementing
computations  in recurrent  networks has  been  to endow  them with  a
clustered    \citep{Wang2002,AmitBrunel,
  LitwinKumarDoiron}   or
distance-dependent  connectivity \citep{BenYishai1995}.  Such
networks  inherently display low-dimensional  dynamics similar  to our
framework \citep{DoironLitwinKumar2014,     Williamson2016},
as  clustered connectivity  is  in  fact a  special  case of  low-rank
connectivity. Clustered  connectivity, however, is  highly  ordered: each  neuron
belongs to  a single  cluster and therefore  is selective to  a single task
feature (e.g.~a  given stimulus, or  a given output). Neurons  in clustered
networks are  therefore highly specialized and display pure
selectivity \citep{Rigotti2013}.  Here, instead, we  have considered
random low-rank  structures, which generate activity organized along
heterogeneous directions  in state space. As a consequence, stimuli and outputs
are represented in a  random, highly distributed manner and individual
neurons  are  typically responsive  to  several  stimuli, outputs,  or
combinations of the two. Such mixed selectivity is a
ubiquitous property  of cortical neurons \citep{Rigotti2013,   Mante2013,
  ChurchlandShenoy2007},  and confers additional computational  properties  to  our
networks \citep{Kanerva2009}. In
particular,  it allowed  us to  easily extend  to  a context-dependent
situation \citep{Mante2013,Saez2015}  a network implementation of a basic
discrimination task.  This is typically difficult to  do in clustered,
purely selective networks \citep{Rigotti2010}.

The type of  connectivity used in our study is  closely related to the
classical framework of Hopfield networks \citep{Hopfield,Amit1985}. The  aim of  Hopfield networks is  to store in  memory specific
patterns  of activity  by creating  for each  pattern  a corresponding
fixed-point  in the  network dynamics.  This is  achieved by  adding a
unit-rank term  for each item,  and one approach for  investigating the
capacity of  such a  setup has  relied on the  mean-field theory  of a
network with a connectivity that consists  of a sum of a rank-one term
and  a random  matrix \citep{TirozziTsodyks, ShiinoFukai,RoudiLatham}. While this approach is clearly close to
the   one  adopted  in   the  present   study,  there   are  important
differences.  Within  Hopfield   networks,  the  unit-rank  terms  are
symmetric,  so  that the  corresponding  left- and  right-connectivity
vectors are identical for  each pattern. Moreover, the unit-rank terms
that correspond  to different patterns are  generally uncorrelated. In
contrast,  here we  have considered  the more  general case  where the
left- and right-eigenvectors are different, and potentially correlated
between different rank-one terms. Most importantly, our main focus was
on responses to external  inputs and input-output computations, rather
than  memorizing items. In
particular we  showed that  left- and right-connectivity  vectors play
different  roles   with  respect   to  processing  inputs,   with  the
left-connectivity  vector  implementing   input-  selection,  and  the
right-connectivity vector determining the output of the network.

Our  study is  also directly  related to  echo-state networks  (ESN) \citep{JaegerHaas}  and  FORCE  learning  \citep{SussilloAbbott}. In those frameworks,  randomly connected recurrent networks are
trained to  produce specified outputs  using a feedback loop  from a
readout  unit to the  network, which  is mathematically  equivalent to
adding a  rank-one term  to the random  connectivity matrix  \citep{Maass2007}.  In their most  basic implementation, both  ESN and
FORCE  learning train  only the  readout weights.  The
training is performed for a fixed, specified realization of the random
connectivity, so that the  final rank-one structure is correlated with
the random part of the connectivity  and may be strong with respect to
it. In  contrast, the results  presented here rely on  the assumption
that the  low-rank structure is  weak and independent from  the random
part.  Although   ESN  and  FORCE   networks  do  not necessarily  fulfill  this
assumption, in ongoing work we found that our approach describes well
networks  trained using  ESN or  FORCE  to produce  a constant  output
\citep{RivkindBarak}.  Note   that  in  our   framework,  the
computations rely  solely on the structured part  of the connectivity,
but ongoing work suggests that the random part of the connectivity may
play an important role during training.

The  specific network  model used  here is  identical to  most studies
based on trained recurrent  networks \citep{SussilloAbbott,  Mante2013,Sussillo2014}. It  is highly  simplified  and  lacks  many
biophysical  constraints, the  most basic  ones being  positive firing
rates,   the  segregation  between   excitation  and   inhibition  and
interactions through spikes. Recent works have investigated extensions
of  the  abstract  model   used  here  to  networks  with  biophysical
constraints    \citep{Ostojic2014, Kadmon2015, Harish2015,MastrogiuseppeOstojic,   Thalmeier2016}.  Additional work will  be needed to implement  the present
framework in networks of spiking neurons.

Our  results imply novel,  directly testable  experimental predictions
relating  connectivity,  low-dimensional  dynamics  and  computational
properties of individual neurons. Our main result is that the dominant
components   of  low-dimensional   dynamics  are   a   combination  of
feed-forward  input  patterns,  and  vectors specifying  the  low-rank
recurrent  connectivity   (Fig.~\ref{fig:p2}  \textsf{C}).   A  direct
implication is  that, if the  low-dimensional dynamics in  the network
are  generated by  low-rank recurrent  connectivity, two  neurons that
have large loadings in the  dominant principal components will tend to
have  mutual connections stronger than average (Fig.~\ref{fig:p3} \textsf{F} top).  In
contrast,  if  the  low-dimensional  dynamics  are  not  generated  by
recurrent  interactions, but  instead driven  by  feed-forward inputs
alone, no correlation between  principal components and connectivity is
expected    (Fig.~\ref{fig:p3}   \textsf{F}   bottom).     Since   the
low-dimensional  dynamics  based on  recurrent  connectivity form  the
scaffold for computations  in our model, this basic  prediction can be
extended   to   various   task-dependent  properties   of   individual
neurons. For instance, if the recurrent connectivity implements evidence
integration, two units with  strong choice regressors are predicted to
have   mutual   connections stronger than average  (Fig.~\ref{fig:p4}  \textsf{H}).
Analogously,  if  recurrent  connections  implement  context-dependent
associations, two units with strong context regressors are expected to
share   connections stronger than average  (Fig.~\ref{fig:p5} \textsf{G}).  Such  predictions  can in principle be directly  tested in experiments
that combine calcium imaging of neural activity in behaving animals with measurements of
connectivity between a subset of recorded neurons \citep{Ho2011}. It should be noted however that very weak structured connectivity is sufficient to implement computations, so that the expected correlations between connectivity and various selectivity indices may  be weak.

The  class  of recurrent  networks  we  considered  here is  based  on
connectivity matrices  that consist of  an explicit sum of  a low-rank
and a random  part. While this may seem as a  limited class of models,
in fact any arbitrary matrix  can be approximated with a low-rank one,
e.g.~by keeping  a  small  number of  dominant  singular values  and
singular  vectors \citep{Markovsky}  --  this is  the basic  principle
underlying  dimensionality  reduction. A  recurrent  network with  any
arbitrary   connectivity  matrix   can  therefore   in   principle  be
approximated by a low-rank recurrent network. From this point of view,
our theory suggests a simple conjecture: the low-dimensional structure
in connectivity determines  low-dimensional dynamics and computational
properties  of  recurrent networks.   While  more  work  is needed  to
establish under which precise conditions a low-rank network provides a
good  computational approximation  of a  full recurrent  network, this
conjecture provides a simple and practically useful working hypothesis
for reverse-engineering trained neural networks \citep{SussilloBarak},
and  relating  connectivity,   dynamics  and  computations  in  neural
recordings.

\section*{Acknowledgements}
We are grateful to Alexis Dubreuil, Vincent Hakim and Kishore Kuchibhotla  for discussions and feedback on the manuscript. 
This work was funded by the Programme Emergences of City of Paris, and the program
``Investissements d'Avenir'' launched by the French Government and
implemented by the ANR, with the references ANR-10-LABX-0087 IEC and
ANR-11-IDEX-0001-02 PSL* Research University. The funders had no role in study design, data collection and analysis, decision to publish, or preparation of the manuscript.

\section*{Author contributions}
F.M. and S.O. designed the study and wrote the manuscript. F.M. performed model analyses and simulations.

\section*{Declaration of Interests}
The authors declare no competing interests.


\clearpage

\setcounter{secnumdepth}{-2}
\setcounter{tocdepth}{5}
\renewcommand*\contentsname{}

\section{{\textsf \Large STAR Methods}}

\section{Contact for Reagent and Resource Sharing}
Further requests for resources should be directed to and will be fulfilled by the Lead Contact, Srdjan Ostojic (srdjan.ostojic@ens.fr).

\section{Method Details}

\subsection{The network model}

We study large recurrent networks of rate units. Every unit in the network is characterized by a continuous variable $x_i(t)$, commonly interpreted as the total input current. More generically, we also refer to $x_i(t)$ as the \emph{activation} variable. The output of each unit is a non-linear function of its inputs modeled as a  sigmoidal function $\phi(x)$. In line with previous works \citep{Sompolinsky1988, SussilloAbbott, RivkindBarak}, we focus on  $\phi(x)=\tanh(x)$, but we show that qualitatively similar dynamical regimes appear in network models with more realistic, positively defined activation functions (Fig.~S7). The transformed variable $\phi(x_i(t))$ is interpreted as the firing rate of unit $i$, and is also referred to as the \emph{activity} variable.

\noindent The time evolution is specified by the following dynamics:
\begin{equation}\label{eq:dynamics}
	\dot{x}_i(t)=-x_i(t)+\sum_{j=1}^N J_{ij} \phi(x_j(t))+I_i.
\end{equation}

\noindent We considered a particular class of connectivity matrices, which can be written as a sum of two terms: 
\begin{equation}
	J_{ij}=g\chi_{ij}+P_{ij}. 
\end{equation}
Similarly  to   \citep{Sompolinsky1988},  $\chi_{ij}$  is   a  Gaussian
all-to-all random matrix, where every element is drawn from a centered
normal distribution with variance  $1/N$. The parameter $g$ scales the
strength of random connections in the network, and we refer to it also
as the {\em random strength}. The second term $P_{ij}$ is a low-rank
matrix.  In  this  study,  we   consider  the  low-rank  part  of  the
connectivity  fixed, while  the random  part varies  between different
realizations of the connectivity.  Our results rely on two simplifying
assumptions. The  first one is that  the low-rank term  and the random
term  are statistically uncorrelated.  The  second   one  is  that,  as  stated  in
Eq.~\ref{eq:p},  the structured  connectivity is
weak  in the  large $N$  limit, i.e.~it scales as
$1/N$, while  the random  connectivity components $\chi_{ij}$  scale as
$1/\sqrt{N}$.

We  first consider  the simplest  case  where $P_{ij}$  is a  rank-one
matrix, which can generally be written as the external product between
two one-dimensional vectors $m$ and $n$:
\begin{equation}\label{eq:p}
	P_{ij} = \frac{m_in_j}{N}.
\end{equation}
According to our first assumption,  the entries of vectors $m$ and $n$
are   independent   of   the   random   bulk   of   the   connectivity
$\chi_{ij}$. Note that the only non-zero eigenvalue of $P$ is given by
the  scalar product  $m^Tn/N$, and  the corresponding  right  and left
eigenvectors are, respectively, vectors $m$ and $n$. In the following,
we  will   refer  to the eigenvalue $m^Tn/N$  as  the   {\em strength  of  the
	connectivity  structure}, and  to $m$  and $n$  as the  {\em right-  and left-connectivity vectors}. Here we  focus on vectors obtained by generating
the components  from a joint  Gaussian distribution.

More general connectivity structures of rank $r\ll N$ can be written as a sum of unit-rank terms
\begin{equation}
	P_{ij}=\frac{m_i^{(1)}n_j^{(1)}}{N}+\ldots+\frac{m_i^{(r)}n_j^{(r)}}{N},
\end{equation}
and are therefore specified by $r$ pairs of vectors $m^{(k)}$  and
$n^{(k)}$, where different $m$ vectors are linearly independent, and similarly for $n$ vectors.

\subsection{Overview of Dynamical Mean-Field theory}

Our  results rely on a  mathematical  analysis of  network
dynamics    based    on   Dynamical    Mean-Field   (DMF)    theory
\citep{Sompolinsky1988,Rajan2010,Kadmon2015}.     To    help
navigate the analysis, here we provide first a succint overview of the
approach. Full details are given further down in the section {\em Details of Dynamical Mean-Field theory}.

DMF  theory allows  one  to  derive an  effective  description of  the
dynamics by  averaging over the  disorder originating from  the random
part of the connectivity.  Across different realizations of the random
connectivity  matrix $\chi_{ij}$,  the sum  of inputs  to unit  $i$ is
approximated by a Gaussian stochastic process $\eta_i(t)$
\begin{equation}
	\sum_{j=1}^N J_{ij} \phi(x_j(t))+I_i  \approx \eta_i(t),
\end{equation}
so that
each unit obeys a Langevin-like equation:
\begin{equation}\label{eq:dynamics_eta}
	\dot{x}_i(t)=-x_i(t)+\eta_i(t).
\end{equation}
The Gaussian processes $\eta_i$  can in principle have different first
and  second-order   statistics  for  each  unit,   but  are  otherwise
statistically  independent across different  units. As  a consequence,
the activations $x_i$ of different units are also independent Gaussian
stochastic   processes,   coupled  only   through   their  first   and
second-order  statistics.    The  core  of  DMF   theory  consists  of
self-consistent  equations for the  mean $\mu_i$  and auto-correlation
function $\Delta^{I}_i(t)$.

At equilibrium (i.e. in absence of transient dynamics) the  equation for the  mean $\mu_i$  of  $x_i$ is  obtained by directly averaging  Eq.~\ref{eq:dynamics} over the
random part of the connectivity. For a unit-rank connectivity, it reads
\begin{equation}\label{eq:mu_overview}
	\mu_i= \kappa m_i +I_i,  
\end{equation}
where 
\begin{equation}\label{eq:kappa_overview}
	\kappa=\frac{1}{N} \sum_{j=1}^N n_j  \big[\phi_j \big].
\end{equation}
In the last equation, we adopted the short-hand notation $\phi_i(t) \coloneqq \phi(x_i(t))$.
Here  $\big[\phi_j \big]$  is the  average  firing rate  of unit  $j$,
i.e.  $\phi(x_j)$  averaged  over  the Gaussian  variable  $x_j$.  In a geometrical interpretation, the
quantity $\kappa$ represents the overlap between the left-connectivity
vector $n$ and the vector of average firing rates. Equivalently, it is given by a population average of $n_j  \big[\phi_j \big]$, which can also be expressed as
\begin{equation}\label{eq:kappa_overview2}
	\kappa= \int \diff m  \diff n \diff I \: p(m,n,I) \: n \int \mathcal{D}z \phi(m\kappa + I+\sqrt{\Delta_0^I} z)
\end{equation}
where $p(m,n,I)$ is the joint distribution of components of vectors $m$, $n$ and $I$. $\Delta_0^I$ is the variance of $x_i$ (see below), and $ \int \mathcal{D}z =\int_{-\infty}^{+\infty}\frac{e^{-\frac{z^2}{2}}}{\sqrt{2\pi}}\diff z$.

The   auto-correlation    function $\Delta^{I}_i(t)$   quantifies  the
fluctuations of  the activation $x_i$  around the expected  mean. 
Computing this auto-correlation function shows that it is  identical  for  all  units in  the  network,
i.e. independent of  $i$ (see Eq.~\ref{eq:corr_eta}). It can be decomposed  into a static variance, which quantifies the fluctuations of the equilibrium values of $x_i$ across  different realizations of  the random component
of  the  connectivity,  and  an additional  temporal  variance  which
is present  when  the  network  is  in  a temporally  fluctuating,  chaotic
state. In a stationary state, the variance $\Delta^{I}_0 \equiv \Delta^I(t=0)$ can be expressed as
\begin{equation}\label{eq:deltaI_overview}
	\Delta_0^I=g^2 \frac{1}{N} \sum_{j=1}^N [\phi_i^2].
\end{equation}
where $[\phi_i^2]$ is the average of $\phi_i^2(x)$ over the Gaussian  variable  $x_i$.

The                         right-hand-sides                        of
Eqs.~\ref{eq:kappa_overview} and \ref{eq:deltaI_overview}  show  that  both  the
mean $\mu_i$ and variance $\Delta_0^I$ depend on population-averaged,
macroscopic  quantities.   To fully  close  the  DMF description,  the
equations for  single-unit  statistics need  to  be  averaged over  the
population. For static equilibrium dynamics, this leads to two coupled
equations for two macroscopic quantities, the overlap $\kappa$ and the static, population-averaged
variance $\Delta_0$:
\begin{equation}
	\begin{split}
		&\kappa =  F(\kappa, \Delta_0)\\
		&\Delta_0 =  G(\kappa, \Delta_0).
	\end{split}
\end{equation} 
Here $F$  and $G$ are two  non-linear functions, the  specific form of
which  depends  on the  geometrical  arrangement  of the  connectivity
vectors $m$ and $n$ and the input vector $I$.
For   temporally   fluctuating,   chaotic  dynamics   an   additional
macroscopic quantity (corresponding to the temporal variance)  needs to be  taken into
account. In that  case, the full DMF description is  given by a system
of three non-linear equations for three unknowns.
The equilibrium states of  the network dynamics are therefore obtained
by  solving  these  systems  of equations  using  standard  non-linear
methods.

To  describe the  transient dynamics  and assess  the stability  of the
obtained equilibrium states, we determined the spectrum of eigenvalues
at  the  obtained equilibrium fixed  points.  This  spectrum  consists of  two
components: a continuous, random component distributed within a circle
in the complex  plane, and a single outlier  induced by the structured
part  of the  connectivity (Fig.~S1 \textsf{A}, \textsf{D}). The  radius  of the
continuous  component and  the  value  of the  outlier  depend on  the
connectivity parameters.   Although the two quantities  in general are
non-trivially coupled, the value of the radius is mostly controlled by
the strength of the disorder, while the value of the outlier increases
with the  strength $m^Tn/N$ of  the rank-one structure (Fig.~S1 \textsf{F}).  The  equilibrium is stable as  long as the real  part of all
eigenvalues  is less  than  unity.  For  large connectivity  structure
strengths, the  outlier crosses unity, generating  an instability that
leads    to    the    appearance   of    one-dimensional    structured
activity. Increasing the disorder strength  on the other hand leads to
another  instability, corresponding  to the  radius of  the continuous
component crossing unity.  This instability gives
rise to chaotic, fluctuating activity.

When a linear readout with weights $w_i$ is added to the network, its average output is given by
\begin{equation}\label{eq:readout_overview}
	z(t)=\frac{1}{N}\sum_{i=1}^N w_i  \big[\phi_i(t)\big],
\end{equation}
i.e. by the projection of the average network firing rate on the readout vector $w$. This quantity is analogous to $\kappa$, except that the vector $n$ is replaced by the vector $w$, so that similarly to Eq.~\ref{eq:kappa_overview2}, the average readout can also be expressed as
\begin{equation}\label{eq:readout_overview2}
	z= \int \diff m  \diff w \diff I \: p(m,w,I) \: w \int \mathcal{D}y \phi(m\kappa + I+\sqrt{\Delta_0^I} y)
\end{equation}
and therefore directly depends on the joint distribution $p(m,w,I)$ which characterizes the geometric arrangement of vectors $m$, $w$ and $I$.

The DMF theory can be directly extended to connectivity structures of rank $r$ greater than one. The equilibrium mean input to unit $i$ is then given by
\begin{equation}\label{eq:mu_r_overview}
	\mu_i= \sum_{k=1}^{r} \kappa^{(k)} m_i^{(k)} +I_i.  
\end{equation}
The activity therefore lives in an $(r+1)$-dimensional space determined by the $r$ right-connectivity vectors $m^{(k)}$ and the input vector $I$. It is characterized by $r$ overlaps $\kappa^{(k)}$, each of which quantifies the amount of activity along the corresponding direction  $m^{(k)}$. Averaging over the population, the DMF theory then leads
to a system of $r+1$ nonlinear coupled equations for describing stationary dynamics.

\subsection{Details of Dynamical Mean-Field theory}

Here we provide the full details of the mathematical analysis. We start by examining the activity of a network with a rank-one structure in absence of external inputs ($I_i=0$ $\forall i$ in Eq.~\ref{eq:dynamics}).

\subsubsection{Single-unit equations for spontaneous dynamics}

We start by  determining the statistics of the effective noise $\eta_i$ to unit $i$, defined by
\begin{equation}\label{eq:eta}
	\eta_i(t)=g\sum_{j=1}^N \chi_{ij}\phi(x_j(t)) + \frac{m_i}{N} \sum_{j=1}^N n_j \phi(x_j(t)).
\end{equation}

The DMF  theory relies on the hypothesis that a
disordered component  in the  coupling structure, here  represented by
$\chi_{ij}$, efficiently decorrelates  single neuron activity when the
network is  sufficiently large.  We will show that this hypothesis of decorrelated activity is self-consistent for the specific network architecture we study.

As in standard DMF derivations, we characterize self-consistently the distribution of $\eta_i$ by averaging over different realizations of the random matrix $\chi_{ij}$ \citep{Sompolinsky1988, Rajan2010}. In the following, $[.]$ indicates an average over the realizations of the random matrix $\chi_{ij}$, while $\langle . \rangle$ stands for an average over different units of the network. 
Note that the network activity can be equivalently characterized in terms of input current variables $x_i(t)$ or their non-linear transforms $\phi(x_i(t))$. As these two quantities are not independent, the statistics of the distribution of the latter can be written in terms of the statistics of the former.

The mean of the effective noise received by unit $i$ is given by:
\begin{equation}
	[\eta_i(t)]=g\sum_{j=1}^N [\chi_{ij}\phi(x_j(t))] + \frac{m_i}{N} \sum_{j=1}^N n_j [\phi(x_j(t))].
\end{equation}
Under the hypothesis that in large networks, neural activity decorrelates (more specifically, that activity $\phi(x_j(t))$ is independent of its outgoing weights), we have:
\begin{equation}\label{eq:mean_eta}
	[\eta_i(t)]=g\sum_{j=1}^N [\chi_{ij}][\phi(x_j(t))] + \frac{m_i}{N}  \sum_{j=1}^N n_j [\phi(x_j(t))]=m_i \kappa
\end{equation}
as $[\chi_{ij}]=0$. Here we introduced 
\begin{equation}
	\kappa \coloneqq \frac{1}{N}  \sum_{j=1}^N n_j [\phi(x_j(t))] = \langle n_j [\phi_j(t)] \rangle,
\end{equation}
which quantifies the  overlap between the mean population activity vector and the left-connectivity vector $n$. 

Similarly, the noise correlation function is given by
\begin{equation}
	\begin{split}
		[\eta_i(t)\eta_j(t+\tau)]=&g^2\sum_{k=1}^N \sum_{l=1}^N [\chi_{ik}\chi_{jl}][\phi(x_k(t))\phi(x_l(t+\tau))] \\
		&+ \frac{m_im_j}{N^2} \sum_{k=1}^N \sum_{l=1}^N n_k n_l [\phi(x_k(t))\phi(x_l(t+\tau))].
	\end{split}
\end{equation}
Note that every cross-term in the product vanishes since $[\chi_{ij}]=0$. Similarly to standard DMF derivations \citep{Sompolinsky1988}, the first term on the r.h.s.~vanishes for cross-correlations ($i\neq j$) while it survives in the auto-correlation function ($i=j$), as $[\chi_{ik}\chi_{jl}]={\delta_{ij}\delta_{kl}}/{N}$. We get:
\begin{equation}
	[\eta_i(t)\eta_j(t+\tau)]=\delta_{ij}g^2\langle[\phi_i(t)\phi_i(t+\tau)]\rangle + \frac{m_im_j}{N^2} \sum_{k=1}^N \sum_{l=1}^N n_k n_l [\phi(x_k(t))\phi(x_l(t+\tau))].
\end{equation}
We focus now on the second term in the right-hand side. The corresponding sum contains $N$ terms where $k=l$. This contribution vanishes in the large $N$ limit because of the $1/N^2$ scaling. According to our starting hypothesis, when $k\neq l$, activity decorrelates: $[\phi_k(t)\phi_l(t+\tau)]=[\phi_k(t)][\phi_l(t+\tau)]$. To the leading order in $N$, we get:
\begin{equation}
	\begin{split}
		[\eta_i(t)\eta_j(t+\tau)]&=\delta_{ij}g^2\langle[\phi_i(t)\phi_i(t+\tau)]\rangle + \frac{m_im_j}{N^2} \sum_{k} n_k [\phi(x_k(t))] \sum_{l\neq k} n_l  [\phi(x_l(t+\tau))]\\
		&=\delta_{ij}g^2\langle[\phi_i(t)\phi_i(t+\tau)]\rangle + m_i m_j \kappa^2
	\end{split}
\end{equation}
so that:
\begin{equation}\label{eq:corr_eta}
	[\eta_i(t)\eta_j(t+\tau)]-[\eta_i(t)][\eta_j(t)]=\delta_{ij}g^2\langle[\phi_i(t)\phi_i(t+\tau)]\rangle.
\end{equation}
We therefore find that the statistics of the effective input are uncorrelated across different units, so that our initial hypothesis is self-consistent.

To conclude, for every unit $i$, we computed the first- and the second-order statistics of the effective input $\eta_i(t)$. The expressions we obtained show that the individual noise statistics depend on the statistics of the full network activity. In particular, the mean of the effective input depends on the average overlap $\kappa$, but varies from unit to unit through the components of the right-connectivity vector $m$. On the other hand, the auto-correlation of the effective input is identical for all units, and determined by  the population-averaged firing rate auto-correlation $\langle[\phi_i(t)\phi_i(t+\tau)]\rangle$.

Once the statistics of  $\eta_i(t)$ have been determined, a self-consistent solution for the activation variable $x_i(t)$ can be derived by solving the Langevin-like stochastic process from Eq.~\ref{eq:dynamics_eta}. As a first step, we look at its stationary solutions, which correspond to the fixed points of the original network dynamics.

\subsubsection{Population-averaged equations for stationary solutions}\label{par:dmf_stationary}

For any solution that does not depend on time, the mean $\mu_i$ and the variance $\Delta_0^I$ of the variable $x_i$ with respect to different realizations of the random connectivity coincide with the statistics of the effective noise $\eta_i$. From Eqs.~\ref{eq:mean_eta} and \ref{eq:corr_eta}, the mean $\mu_i$ and variance $\Delta_0^I$ of the input to unit $i$ therefore read
\begin{equation}\label{eq:stationary_ind}
	\begin{split}
		&\mu_i \coloneqq [x_i] = m_i \kappa\\
		&\Delta_0^I \coloneqq [x_i^2] - [x_i]^2 = g^2 \langle[\phi_i^2]\rangle
	\end{split}
\end{equation}
while any other cross-variance $[x_ix_j] - [x_i][x_j]$ vanishes.
We conclude that, on average, the structured connectivity $P_{ij}$ shapes the network activity along the direction specified by its right eigenvector $m$. Such a heterogeneous stationary state critically relies on a non-vanishing overlap $\kappa$ between the left eigenvector $n$ and the average population activity vector $[\phi]$. Across different realizations of the random connectivity, the input currents $x_i$  fluctuate around these mean values. The typical size of fluctuations is determined by the individual variance $\Delta_0^I$, equal for every unit in the network.

The r.h.s.~of Eq.~\ref{eq:stationary_ind} contains two population averaged quantities, the overlap $\kappa$ and the second moment of the activity $\langle[\phi_i^2]\rangle$. To close the equations, these quantities need to be expressed self-consistently. Averaging Eq.~\ref{eq:stationary_ind} over the population, we get expressions for the population-averaged mean $\mu$ and variance $\Delta_0 $ of the input:
\begin{equation}\label{eq:stationary}
	\begin{split}
		&\mu \coloneqq \langle [x_i] \rangle = \langle m_i \rangle \kappa\\
		&\Delta_0 \coloneqq  \langle [x_i^2] \rangle -  \langle [x_i] \rangle^2= g^2 \langle [\phi_i^2] \rangle + (\langle m_i^2 \rangle - \langle m_i \rangle ^2 ) \kappa^2.
	\end{split}
\end{equation}
Note that the total population variance $\Delta_0$ is a sum of two terms: the first term, proportional to the strength of the random part of connectivity, coincides with the individual variability $\Delta_0^I$ which emerges from different realizations of $\chi_{ij}$; the second term, proportional to the variance of the right-connectivity vector $m$, coincides with the variance induced at the population level by the spread of the mean values $\mu_i\propto m_i$.
When the vector $m$ is homogeneous ($m_i=\bar{m}$), input currents $x_i$ are centered around the same mean value $\mu$, and the second variance term vanishes.

We next  derive  appropriate expression for the r.h.s.~terms $\kappa$ and $\langle[\phi_i^2]\rangle$.
To start with, we rewrite $[\phi_i]$ by substituting the average over the random connectivity  with the equivalent Gaussian integral:
\begin{equation}
	\begin{split}
		[\phi_i]
		=\int \mathcal{D}z \phi(\mu_i + \sqrt{\Delta_0^I} z)
	\end{split}
\end{equation}
where we used the short-hand notation $ \int \mathcal{D}z =\int_{-\infty}^{+\infty}\frac{e^{-\frac{z^2}{2}}}{\sqrt{2\pi}}\diff z$. To obtain $\kappa$, $[\phi_i]$ needs to be multiplied by $n_i$ and averaged over the population. This average can be expressed by representing the fixed vectors $m$ and $n$ through the joint distribution of their elements over the components:
\begin{equation}
	p(m,n)=\frac{1}{N}\sum_{j=1}^{N}\delta(m-m_j)\delta(n-n_j). 
\end{equation}
This leads to
\begin{equation}\label{eq:kappa}
	\begin{split}
		\kappa&=\langle n_i \int \mathcal{D}z \phi(\mu_i + \sqrt{\Delta_0^I} z)\rangle\\
		&= \int \diff m \int \diff n \: p(m,n) \: n \int \mathcal{D}z \phi(m\kappa + \sqrt{\Delta_0^I} z).
	\end{split}
\end{equation}
Similarly, a suitable expression for the  second-order momentum of the firing rate is given by:
\begin{equation}\label{eq:phisq}
	\langle[\phi_i^2]\rangle=\int \diff m \: p(m) \: \int \mathcal{D}z \phi^2(m\kappa + \sqrt{\Delta_0^I} z).
\end{equation}

Eqs. \ref{eq:kappa} and \ref{eq:phisq}, combined with Eq. \ref{eq:stationary}, provide a closed set of equations for determining $\kappa$ and $\Delta_0$ once the vectors $m$ and $n$ have been specified.

To further simplify the problem, we reduce the full distribution $p(m,n)$ of elements $m_i$ and $n_i$ to their first- and second-order momenta. That is equivalent to substituting the probability density $p(m,n)$ with a bivariate Gaussian distribution. We therefore write:
\begin{equation}\label{eq:mn}
	\begin{split}
		&m = M_m + \Sigma_m\sqrt{1 - \rho}\: x_1 + \Sigma_m\sqrt{\rho}\: y \\
		&n = M_n + \Sigma_n\sqrt{1 -\rho}\: x_2  + \Sigma_n\sqrt{\rho}\: y
	\end{split}
\end{equation}
where $x_1$, $x_2$ and $y$ are three  normal Gaussian processes. Here, $M_m$ (resp.~$M_n$) and $\Sigma_m$ (resp.~$\Sigma_n$) correspond to the mean and the standard deviation of $m$ (resp.~$n$), while the covariance between $m$ and $n$ is given by $ \langle m_i n_i\rangle - M_mM_n = \Sigma_m\Sigma_n\rho$. Within a geometrical interpretation, $M_m$ and $M_n$ are the projections of $N-$dimensional vectors $m$ and $n$ onto the unitary vector $u=(1,1,...1)/{N}$, $\Sigma_m\sqrt{\rho}$ and  $\Sigma_n\sqrt{\rho}$ are the projections onto a  direction orthogonal to $u$ and common to $m$ and $n$, and $\Sigma_m\sqrt{1 - \rho}$ and $\Sigma_n\sqrt{1 -\rho}$ scale the parts of $m$ and $n$ that are mutually orthogonal.

The expression for $\kappa$ becomes:
\begin{equation}
	\begin{split}
		\kappa = &\int \mathcal{D}y\int \mathcal{D}x_2 \: (M_n + \Sigma_n\sqrt{1 -\rho} x_2 + \Sigma_n\sqrt{\rho} y)  \\
		& \times \int \mathcal{D}z\int \mathcal{D}x_1 \: \phi( \kappa (M_m + \Sigma_m\sqrt{1 - \rho} x_1 + \Sigma_m\sqrt{\rho} y) + \sqrt{\Delta_0^I}z)
	\end{split}
\end{equation}
which gives rise to three terms when expanding the sum $M_n + \Sigma_n\sqrt{1 -\rho} x_2 + \Sigma_n\sqrt{\rho} y$. 
The first term can be rewritten as:
\begin{equation}
	\begin{split}
		&M_n \int \mathcal{D}z \: \phi(M_m \kappa + \sqrt{\Delta_0^I + \Sigma_m^2 \kappa^2}z)\\
		& =M_n \int \mathcal{D}z \: \phi(\mu+\sqrt{\Delta_0}z)\\
		&=M_n \langle[\phi_i]\rangle,
	\end{split}
\end{equation}
which coincides with the overlap between vectors $n$ and $[\phi]$ along the unitary direction $u=(1,1,...1)/{N}$.
In the last step, we rewrote our expression for $\kappa$ in terms of the population averaged statistics $\mu$ and $\Delta_0$ (Eq.~\ref{eq:stationary}).

\noindent The second term vanishes, while the third one gives:
\begin{equation}
	\begin{split}
		&\Sigma_n\sqrt{\rho}  \int \mathcal{D}y \: y  \int \mathcal{D}z\int \mathcal{D}x_1 \: \phi( \kappa (M_m + \Sigma_m \sqrt{1 - \rho} x_1 + \Sigma_m\sqrt{\rho} y) + \sqrt{\Delta_0^I}z) \\
		&= \kappa \rho \Sigma_m \Sigma_n \langle[\phi_i']\rangle
	\end{split}
\end{equation}
which coincides with the overlap between $n$ and $[\phi]$ in a direction orthogonal to $u$.
Here we used the equality:
\begin{equation}\label{eq:trick}
	\int \mathcal{D}z \:  z f(z) = \int \mathcal{D}z  \:  \frac{\diff f(z)}{\diff z}
\end{equation}
which is obtained by integrating by parts. 

Through a similar reasoning we obtain:
\begin{equation}\label{eq:phisq_gaussian}
	\langle[\phi_i^2]\rangle = \int \mathcal{D}z \: \phi^2(\mu+\sqrt{\Delta_0}z)
\end{equation}
as in standard DMF derivations.

To conclude, the mean-field description of stationary solutions reduces to the system of three implicit equations for $\mu$, $\kappa$ and $\Delta_0$:
\begin{equation}\label{eq:stationary_general}
	\begin{split}
		&\mu = M_m \kappa \\
		&\Delta_0 =g^2 \langle [\phi_i^2] \rangle + \Sigma_m^2\kappa^2 \\
		&\kappa = M_m \langle[\phi_i]\rangle + \kappa \rho \Sigma_m \Sigma_n \langle[\phi_i']\rangle.
	\end{split}
\end{equation}
Both averages $\langle[.]\rangle$ are performed with respect to a Gaussian distribution of mean $\mu$ and variance $\Delta_0$.
Once $\mu$, $\Delta_0$ and $\kappa$ have been determined, the single unit mean $\mu_i$ and the individual variance $\Delta_0^I$ are obtained from Eq. \ref{eq:stationary_ind}.

The dynamical mean-field equations given in Eq. \ref{eq:stationary_general} can be fully solved to determine stationary solutions. Detailed descriptions of these solutions are provided further down for two particular cases: (i) overlap between $m$ and $n$ only along the unitary direction $u$ ($M_m\neq 0$, $M_n\neq 0$, $\rho=0$); (ii) overlap between $m$ and $n$ only in a direction orthogonal to  $u$ ($M_m= M_n=0$,  $\rho \neq 0$).

\subsubsection*{Transient dynamics and stability of stationary solutions}\label{par:dmf_stability}\label{par:lin_stability}

We now turn to transient dynamics around fixed points, and to the related problem of evaluating whether the stationary solutions found within DMF are stable with respect to the original network dynamics (Eq.~\ref{eq:dynamics}).

For any given realization of the connectivity matrix, the network we consider is completely deterministic. We can then study the local, transient dynamics by linearizing the dynamics around any stationary solution. We therefore look at the time evolution of a small displacement away from the fixed point: $x(t)=x_i^0+x_i^1(t)$.  For any generic stationary  solution $\{x_i^0\}$ the linearized dynamics are given by the stability matrix $S_{ij}$ which reads:
\begin{equation}\label{eq:stab_mtx}
	S_{ij}=\phi'(x_j^0)\left(g\chi_{ij}+\frac{m_in_j}{N}\right).
\end{equation}
If the real part of every eigenvalue of $S_{ij}$ is smaller than unity, the perturbation decays in time  and thus the stationary solution is stable.

\paragraph{Homogeneous stationary solutions}
We first consider homogeneous stationary solutions, for which $x_i^0=\bar{x}$ for all units. A particular homogeneous solution is the trivial solution $x_i=0$, which the network admits for all parameter values when the transfer function is $\phi(x)=\tanh(x)$. Other homogeneous solutions can be obtained when the vector $m$ is homogeneous, i.e. $m_i=M_m$ for all $i$.

For homogeneous  solutions, the stability matrix reduces to a scaled version of the connectivity
matrix $J_{ij}$:
\begin{equation}
	S_{ij}=\phi'(\bar{x})J_{ij}.
\end{equation}
We  are thus left with
the problem of evaluating the eigenspectrum of the global connectivity
matrix  $J_{ij}$.   The  matrix   $J_{ij}$  consists  of  a  full-rank
component $\chi_{ij}$,  the entries of which  are drawn at  random, and  of a
structured component of small dimensionality  with fixed entries. We focus on the  limit of large networks; in that limit, an
analytical  prediction for  the  spectrum of  its  eigenvalues can  be
derived.

Because of the $1/N$ scaling, the matrix norm of $P_{ij}$ is bounded as $N$ increases. We can then apply results from random matrix theory \citep{Tao2013} which predict that, in the large $N$ limit, the eigenspectra of the random and the structured parts do not interact, but sum together.
The eigenspectrum of $J_{ij}$ therefore consists of two separated components, inherited respectively from the random and the structured terms (Fig.~S1 \textsf{A}). Similarly to \citep{Girko}, the random term $\chi_{ij}$ returns a set of $N-1$ eigenvalues which lie on the complex plane in a compact circular region of radius $g$. In addition to this component, the eigenspectrum of $J_{ij}$ contains the non-zero eigenvalues of $P_{ij}$: in the case of a rank-one matrix, one single outlier eigenvalue is centered at the position $\sum_i m_i n_i / N=\langle m_i n_i\rangle$. In Fig.~S1 \textsf{B} we measure both the outlier position and the radius of the compact circular component. We show that deviations from the theoretical predictions are in general small and decay to zero as the system size is increased. 

Going        back         to        the        stability        matrix
$S_{ij}=\phi'(\bar{x})J_{ij}$,   we   conclude   that   a
homogeneous stationary  solution can  lose stability in  two different
ways,   when   either   $m^T n/N$    or   $g$   become   larger   than
$1/\phi'(\bar{x})$.   We expect  different kinds  of  instabilities to
occur in the two cases. When $g$ crosses the instability line, a large
number of  random directions become unstable  at the same  time. As in
\citep{Sompolinsky1988}, this  instability is  expected to lead  to the
onset of irregular temporal activity.  When the instability is lead by
the outlier, instead, the trivial  fixed point becomes unstable in one
unique direction  given by the corresponding  eigenvector. When $g=0$,
this eigenvector coincides exactly with  $m$. For finite values of the
disorder  $g$,  the  outlier  eigenvector fluctuates depending on the random part of the connectivity, but remains strongly correlated with $m$  (Fig.~S1 \textsf{C}), which therefore determines the average direction of the instability. Above the instability, as the network dynamics is completely symmetric with respect to a change of sign of the input variables, we expect the non-linear boundaries to generate two symmetric stationary solutions. 

\paragraph{Heterogeneous stationary solutions}
A second type of possible stationary solutions are heterogeneous fixed points, in which different units reach different equilibrium values. For such fixed points, the linearized stability matrix $S_{ij}$ is obtained by multiplying each column of the connectivity matrix $J_{ij}$ by a different gain value (see Eq.~\ref{eq:stab_mtx}), so that the eigenspectrum of $S_{ij}$ is not trivially related to the spectrum of $J_{ij}$.

Numerical investigations reveal that, as for $J_{ij}$, the eigenspectrum of $S_{ij}$ consists of  two discrete components: one compact set of $N-1$ eigenvalues contained in a circle on the complex plane, and a single isolated outlier eigenvalue (Fig.~S1 \textsf{D}). 

As previously noticed in \citep{Harish2015}, the radius of the circular compact set $r$ can be computed as in \citep{Rajan2006, Aljadeff2015} by summing the variances of the distributions in every column of $S_{ij}$. To the leading order in $N$:
\begin{equation}
	r = g \sqrt{\frac{1}{N}\sum_{j=1}^N \phi'^2(x_j^0)}
\end{equation}
which, in large networks, can be approximated by the mean-field average:
\begin{equation}\label{eq:radius_het}
	r = g \sqrt{\langle [\phi_i'^2 ]\rangle}.
\end{equation}
Note that, because of the weak scaling in $P_{ij}$, the structured connectivity term does not  appear explicitly in the expression for the radius. As  the structured part of the connectivity determines the heterogeneous fixed point, the value of $r$ however depends implicitly on the structured connectivity term through $\langle [\phi_i'^2 ]\rangle$, which is computed as a Gaussian integral over a distribution with mean $\mu$ and variance $\Delta_0$ given by Eq.~\ref{eq:stationary_general}.
In Fig.~S1 \textsf{D-F}, we show that Eq.~\ref{eq:radius_het}  approximates well the radius of finite-size, numerically computed eigenspectra.
Whenever the mean-field theory predicts instabilities led by $r$, we expect the network dynamics to converge to irregular non-stationary solutions. Consistently, at the critical point, where $r=1$,  the DMF equations predict the onset of temporally fluctuating solutions (see later on in \emph{Methods}).

We now turn to the problem of evaluating the position of the outlier eigenvalue. In the case of heterogeneous fixed points, the structured and the random components of the matrix $S_{ij}$ are strongly correlated, as they both scale with the multiplicative factor $\phi'(x_j^0)$, which correlates with the particular realization of the random part of the connectivity $\chi_{ij}$. As a consequence, $\chi_{ij}$ cannot be considered as a truly random matrix with respect to ${m_i\phi'(x_j^0)n_j}/{N}$, and in contrast to the case of homogeneous fixed points, results from \citep{Girko} do not hold.

We determined numerically the position of the outlier in finite-size eigenspectra (Fig.~S1 \textsf{D-F}). We found that its value indeed significantly deviates from the only non-zero eigenvalue of the rank-one structure ${m_i\phi'(x_j^0)n_j}/{N}$, which can be computed in the mean-field framework (when $\rho=0$, it corresponds to $M_mM_n\langle[\phi'_i]\rangle + M_n\kappa\Sigma_m^2\langle[\phi''_i]\rangle$).
On the other hand, the value of the outlier coincides exactly with the eigenvalue of ${m_i\phi'(x_j^0)n_j}/{N}$ whenever the random component $\chi_{ij}$ is shuffled (black dots in Fig.~S1 \textsf{F}). This observation confirms that the position of the outlier critically depends on the correlations existing between the rank-one structure ${m_i\phi'(x_j^0)n_j}/{N}$ and its specific realization of the random bulk $\chi_{ij}$.

\paragraph{Mean-field analysis of transient dynamics and  stability of stationary solutions}
As  for heterogeneous  fixed  points  we were  not  able to assess the position of the  outlying
eigenvalue  using random  matrix  theory, we  turned  to a  mean-field
analysis  to determine  transient  activity. This
analysis  allowed  us to  determine  accurately  the  position of  the
outlier, and therefore the stability of heterogeneous fixed points. The approach exploited here is based on \citep{Kadmon2015}.

We consider the stability of the single unit activation  $x_i$ when averaged across different realizations of the random connectivity and its random eigenmodes. Directly averaging across realizations the network dynamics defined in Eq.~\ref{eq:dynamics} yields the time evolution of the mean activation $\mu_i$ of unit $i$:
\begin{equation}\label{eq:mui}
	\dot{\mu}_i(t) = -\mu_i(t) + m_i\kappa(t).
\end{equation}
We observe that we can write: $\mu_i(t)=m_i \tilde{\kappa}(t)$, where $\tilde{\kappa}$ is the low-pass filtered version of $\kappa$: $(1+\diff/\diff t) \tilde{\kappa}(t) = \kappa(t)$.
Small perturbations around the fixed point solution read: $\mu_i(t)=\mu_i^0 + \mu_i^1(t)$. The equilibrium values $\mu_i^0$ correspond to the DMF stationary solution computed from Eq.~\ref{eq:stationary_ind} and \ref{eq:stationary_general}: $\mu_i^0 = m_i \kappa^0$. The first-order perturbations thus obey:
\begin{equation}
	\dot{\mu_i}^1(t) = -\mu_i^1(t) + m_i\kappa^1(t),
\end{equation}
indicating that the decay time scale of the mean activity is inherited by the decay time constant of $\kappa^1$. 
An additional equation for the time evolution of $\kappa^1$ thus needs to be derived.

When activity is perturbed, the firing activity $\phi_i$ of unit $i$ can be evaluated at the first order: $\phi_i^0 \rightarrow \phi_i^0 +\phi_i^1(t) = \phi(x_i^0) +\phi'(x_i^0)x_i^1(t)$. As a consequence, the first-order in $\kappa$ reads:
\begin{equation}\label{eq:kappa1}
	\kappa^1(t) = \langle n_i [\phi'(x_i^0)x_i^1(t)]\rangle.
\end{equation}
Summing Eq.~\ref{eq:kappa1} to its time-derivative, we get:
\begin{equation}
	\dot{\kappa}^1(t)=-\kappa^1(t)+(1+\frac{\diff}{\diff t}) \langle n_i [\phi'(x_i^0)x_i^1(t)]\rangle.
\end{equation}

In order to simplify the r.h.s., we start by considering the average with respect to the random part of the connectivity for a single unit $i$. In order to compute $[\phi'(x_i^0)x_i^1]$, we explicitly build $x_i^0$ and $x_i^t \coloneqq x_i(t)$ as Gaussian variables centered respectively in $\mu_i^0 $ and $\mu_i^t$.
We will call $\Delta_0^{I0}$ and $\Delta_0^{It}$ the variances of the two variables, and $\Delta^{I,t:0}$ their two-times correlation defined by $\Delta^{I,t:0} = [x_i^tx_i^0] -  [x_i^t][x_i^0]$. We can then write the two variables as
\begin{equation}
	\begin{split}
		&x_i^0 = \mu_i^0 + \sqrt{\Delta_0^{I0} - \Delta^{I,t:0}} x_1 + \sqrt{\Delta^{I,t:0}} y\\
		&x_i^t =  \mu_i^t + \sqrt{\Delta_0^{It} - \Delta^{I,t:0}} x_2 + \sqrt{\Delta^{I,t:0}} y\\
	\end{split}
\end{equation}
The first-order response of $x_i$ is given by the difference between $x_i^t$ and $x_i^0$, and reads:
\begin{equation}
	x_i^1 = \mu_i^1 + \sqrt{\Delta_0^{It} - \Delta^{I,t:0}} x_2 - \sqrt{\Delta_0^{I0} - \Delta^{I,t:0}} x_1 .
\end{equation}
As in classical DMF derivations \citep{Sompolinsky1988, Rajan2010, Kadmon2015}, $x_1$, $x_2$ and $y$ are standard normal variables. By integrating over their distributions we can write:
\begin{equation}
	\begin{split}
		[\phi'(x_i^0)x_i^1] = & \int\mathcal{D}x_1 \int\mathcal{D}x_2 \left( \mu_i^1 + \sqrt{\Delta_0^{It} - \Delta^{I,t:0}} x_2 - \sqrt{\Delta_0^{I0} - \Delta^{I,t:0}} x_1 \right) \\
		& \times  \int\mathcal{D}y \phi' \left( \mu_i^0 + \sqrt{\Delta_0^{I0} - \Delta^{I,t:0}} x_1 + \sqrt{\Delta^{I,t:0}} y \right).
	\end{split}
\end{equation}
Integrating by parts as in Eq.~\ref{eq:trick} we  get:
\begin{equation}\label{stab_interm}
	[\phi'(x_i^0)x_i^1] = \mu_i^1 [\phi_i'] + \left( \Delta^{I,t:0} - \Delta_0^{I0} \right)[\phi_i''] 
\end{equation}
where the Gaussian integrals $[\phi_i']$ and  $[\phi_i'']$ are evaluated using the fixed point statistics.

Note that, at the fixed point, $\Delta^{I,t:0} = \Delta_0^{I0}$. As a consequence, $\Delta^{I,t:0} - \Delta_0^{I0}$ gives a first-order response:
\begin{equation}
	\Delta^{I, 1:0} \coloneqq \Delta^{I,t:0} - \Delta_0^{I0}= [x_i^1x_i^0]- [x_i^1][x_i^0] = [x_i^1x_i^0] - \mu_i^0\mu_i^1
\end{equation}
which can be rewritten as a function of the global second-order statistics $\Delta^{1:0}=\langle[x_i^1x_i^0]\rangle - \langle [x_i^1]\rangle \langle[x_i^0]\rangle$ as: 
\begin{equation}\label{eq:delta0Idelta0}
	\begin{split}
		\Delta^{I, 1:0} &=\Delta^{1:0} - \{ \langle \mu_i^1\mu_i^0 \rangle - \langle \mu_i^1\rangle \langle\mu_i^0 \rangle \}\\
		&= \Delta^{1:0} - \Sigma_m^2 \tilde{\kappa}^0 \tilde{\kappa}^1.
	\end{split}
\end{equation}

Eq.~\ref{eq:delta0Idelta0} can be rewritten in terms of the first-order perturbation for the global equal-time variance: $\Delta_0^1 = \Delta_0^t - \Delta_0^0$. We consider that, by definition:
\begin{equation}
	\begin{split}
		&\Delta^{1:0} = \sum_{j=1}^N x_j^1 \frac{\partial \Delta^{t:0}}{\partial x_j^t}\Bigr|_0\\
		&\Delta_0^{1} = \sum_{j=1}^N x_j^1 \frac{\partial \Delta_0^{t}}{\partial x_j^t}\Bigr|_0.
	\end{split}
\end{equation}
We then observe that, when the derivatives are evaluated at the fixed point, we have:
\begin{equation}
	\frac{\partial \Delta^{t:0}}{\partial x_j^t}\Bigr|_0 = \frac{1}{2}\frac{\partial \Delta_0^{t}}{\partial x_j^t}\Bigr|_0,
\end{equation}
and we conclude that:
\begin{equation}\label{eq:trick_delta0}
	\Delta^{1:0} = \frac{1}{2} \Delta_0^1
\end{equation}.

Eq. \ref{stab_interm} thus becomes:
\begin{equation}
	[\phi'(x_i^0)x_i^1] = m_i \ \tilde{\kappa}^1 [\phi_i'] + \left(  \frac{\Delta_0^1}{2} - \Sigma_m^2  \tilde{\kappa}^0 \tilde{\kappa}^1 \right)[\phi_i''] .
\end{equation}
In a second step, we perform the average across different units of the population, by writing $m$ and $n$ as in Eq.~\ref{eq:mn}.
After some algebra, we get:
\begin{equation}
	\begin{split}
		\langle n_i [\phi'(x_i^0)x_i^1(t)]\rangle&  = \ \tilde{\kappa}^1 \left[  (M_mM_n+\rho\Sigma_m\Sigma_n)\langle[\phi_i']\rangle + \rho\kappa^0M_m\Sigma_m\Sigma_n \langle[\phi_i'']\rangle \right] \\
		& \: \: \: \: + \frac{\Delta_0^1}{2}  \left[ M_n \langle[\phi_i'']\rangle + \rho \kappa^0\Sigma_m\Sigma_n \langle[\phi_i''']\rangle \right]\\
		&\coloneqq  \tilde{\kappa}^1 a + \Delta_0^1 \: b
	\end{split}
\end{equation}
where constants $a$ and $b$ were defined as:
\begin{equation}
	\begin{split}
		&a = (M_mM_n+\rho\Sigma_m\Sigma_n)\langle[\phi_i']\rangle + \rho\kappa^0M_m\Sigma_m\Sigma_n \langle[\phi_i'']\rangle \\
		&b = \frac{1}{2}\left\{ M_n \langle[\phi_i'']\rangle + \rho \kappa^0 \Sigma_m\Sigma_n \langle[\phi_i''']\rangle \right\} .
	\end{split}
\end{equation}

The time evolution of $\kappa$ can be finally rewritten as:
\begin{equation}\label{eq:kappa_ev}
	\begin{split}
		\dot{\kappa}^1 (t) = - \kappa^1 (t) + (1+\frac{\diff}{\diff t}) \left\{  \tilde{\kappa}^1 a + \Delta_0^1 b \right\},
	\end{split}
\end{equation}
so that the time evolution of the perturbed variance must be considered as well. 

In order to isolate the evolution law of $\Delta_0$, we rewrite the activation variable $x_i(t)$ by separating the uniform and the heterogeneous components: $x_i(t)=\mu(t)+\delta x_i(t)$.
The time evolution for the residual $\delta x_i(t)$ is given by:
\begin{equation}
	\dot{\delta x_i}(t)=-\delta x_i(t) + g\sum_{j=1}^N \chi_{ij}\phi(x_j(t)) + (m_i-M_m)\kappa(t)
\end{equation}
so that, squaring:
\begin{equation}
	\begin{split}
		&\left(\frac{\diff \delta x_i(t)}{\diff t}\right)^2+ 2 \delta x_i(t)\frac{\diff \delta x_i(t)}{\diff t} + \delta x_i(t)^2=g^2 \sum_{j=1}^N\sum_{k=1}^N \chi_{ij}\chi_{ik}\phi(x_j(t))\phi(x_k(t)) \\
		&+ {(m_i-M_m)^2}\kappa(t)^2+g{(m_i-M_m)}\kappa(t)\sum_{k=1}^N \chi_{ij}\phi(x_k(t)).
	\end{split}
\end{equation}
Averaging over $i$ and the realizations of the disorder yields:
\begin{equation}\label{eq:dyn_delta0}
	\begin{split}
		\frac{\diff \Delta_0(t)}{\diff t} & = - \Delta_0(t) + g^2 \langle [\phi_i^2(t)]\rangle + \Sigma_m^2 \kappa(t)^2 - \langle\left[\left(\frac{\diff \delta x_i(t)}{\diff t}\right)^2\right]\rangle\\
		&\coloneqq  - \Delta_0(t) +G(\mu, \Delta_0,\kappa) - \langle\left[\left(\frac{\diff \delta x_i(t)}{\diff t}\right)^2\right]\rangle
	\end{split}
\end{equation}
as by definition we have: $\langle[\delta x^2_i(t)]\rangle=\Delta_0(t)$.

Expanding the dynamics of $\Delta_0$ to the first order, we get:
\begin{equation}
	\dot{\Delta^1_0}(t)=-\Delta_0^1(t) + \mu^1 \frac{\partial G}{\partial \mu}\Bigr|_0 + \Delta_0^1 \frac{\partial G}{\partial \Delta_0}\Bigr|_0 + \kappa^1 \frac{\partial G}{\partial \kappa}\Bigr|_0.
\end{equation}
Note that we could neglect the contributions originating from the last term of Eq.~\ref{eq:dyn_delta0} because they do not enter at the leading order. Indeed we have:
\begin{equation}
	\frac{\partial}{\partial \mu}\langle\left[\left(\frac{\diff \delta x_i(t)}{\diff t}\right)^2\right]\rangle \Bigr|_0 =2\langle\left[ \frac{\diff \delta x_i(t)}{\diff t}\frac{\partial}{\partial \mu}  \frac{\diff \delta x_i (t)}{\diff t} \right]\rangle\Bigr|_0  =0
\end{equation}
since temporal derivatives for every $i$ vanish when evaluated at the fixed point.

A little algebra returns the last three linear coefficients:
\begin{equation}
	\begin{split}
		&\frac{\partial G}{\partial \mu}\Bigr|_0  =2 g^2 \langle [\phi_i\phi_i']\rangle \\
		&\frac{\partial G}{\partial \Delta_0}\Bigr|_0= g^2\left\{\langle [\phi_i'^2]\rangle+ \langle [\phi_i\phi_i'']\rangle \right\}\\
		&\frac{\partial G}{\partial \kappa}\Bigr|_0= 2\Sigma_m^2\kappa^0.
	\end{split}
\end{equation}

Collecting all the results together in Eq.~\ref{eq:kappa_ev} we obtain:
\begin{equation}
	\begin{split}
		\dot{\kappa}^1 (t)  = - \kappa^1 (t) + a \kappa^1 (t) + b \: \left\{ \mu^1 \frac{ \partial G }{\partial \mu}\Bigr|_{0} + \Delta_0^1 \frac{ \partial  G }{\partial \Delta_0}\Bigr|_{0}+\kappa^1 \frac{ \partial  G }{\partial \kappa}\Bigr|_{0} \right\}.
	\end{split}
\end{equation}

By averaging Eq.~\ref{eq:mui} we furthermore obtain:
\begin{equation}\label{eq:dyn_mu}
	\begin{split}
		\dot{\mu^1}(t)=-\mu^1(t)+M_m\kappa^1.
	\end{split}
\end{equation}

We finally obtained that the perturbation time scale is determined by the population-averaged dynamics:
\begin{equation}\label{eq:kappa_dyn}
	\frac{\diff}{\diff t}\begin{pmatrix}  \mu^1 \\ \Delta_0^1 \\ \kappa^1\end{pmatrix} = - \begin{pmatrix} \mu^1 \\ \Delta_0^1 \\ \kappa^1 \end{pmatrix} + \mathcal{M} \begin{pmatrix} \mu^1 \\ \Delta_0^1 \\ \kappa^1 \end{pmatrix} 
\end{equation}
where the evolution matrix $\mathcal{M}$ is defined as:
\begin{equation}\label{eq:M}
	\mathcal{M} = 
	\begin{pmatrix} 0 & 0 & M_m\\ 
		2 g^2 \langle [\phi_i\phi_i']\rangle & g^2\left\{\langle [\phi_i'^2]\rangle + \langle [\phi_i\phi_i'']\rangle \right\} & 2\Sigma_m^2\kappa^0\\
		2 b g^2 \langle [\phi_i\phi_i']\rangle & b g^2\left\{\langle [\phi_i'^2]\rangle + \langle [\phi_i\phi_i'']\rangle \right\} & b  2\Sigma_m^2\kappa^0 + a
	\end{pmatrix} .
\end{equation}

Note that one eigenvalue of matrix $\mathcal{M}$, which corresponds to the low-pass filtering between $\kappa$ and $\mu$, is always fixed to zero.

Eqs. \ref{eq:kappa_dyn} and \ref{eq:M} reveal that, during the relaxation to equilibrium, the transient dynamics of the first- and second-order statistics of the activity are tightly coupled. 
Diagonalizing $\mathcal{M}$ allows to retrieve the largest decay timescale of the network, which indicates the average, structural stability of stationary states.

When an outlier eigenvalue is present in the  eigenspectrum of the stability matrix $S_{ij}$, the largest decay time scale from $\mathcal{M}$ predicts its position.
The corresponding eigenvector $\hat{e}$ contains indeed a structured component along $m$, which is not washed out by averaging across different realizations of $\chi_{ij}$. 

The second non-zero eigenvalue of $\mathcal{M}$, which vanishes at $g=0$, measures a second and smaller effective timescale, which derives from averaging across the remaining $N-1$ random modes.

Varying $g$, we computed the largest eigenvalue of $\mathcal{M}$ for corresponding stationary solutions of mean-field equations. In Fig.~S1 \textsf{F} we show that, when the stability eigenspectrum includes an outlier eigenvalue, its position is correctly predicted by the largest eigenvalue of $\mathcal{M}$.  The mismatch between the two values is small and can be understood as a finite-size effect (Fig.~S1 \textsf{E}, gray). 

To conclude, we found that the stability of arbitrary stationary solutions can be assessed by evaluating, with the help of  mean-field theory, both the values of the radius (Eq.~\ref{eq:radius_het}) and the outlier (Eq.~\ref{eq:M}) of the stability eigenspectrum. Instabilities led by the two different components are expected to reshape activity into two qualitatively different classes of dynamical regimes, which are discussed in detail, further in \emph{Methods}, for two specific classes of structures.

\subsubsection*{Dynamical Mean Field equations for chaotic solutions}\label{par:DMF_chaotic}
When a stationary state loses stability due to the compact component of the stability eigenspectrum, the network activity starts developing irregular temporal fluctuations. Such temporally fluctuating states can be described within the DMF theory by taking into account the full temporal auto-correlation function of the effective noise $\eta_i$ \citep{Sompolinsky1988}.
For the sake of simplicity, here we derive directly the mean-field equations for population-averaged statistics, and we eventually link them back to single unit quantities.

By differentiating twice Eq.~\ref{eq:dynamics_eta}, and by substituting the appropriate expression for the statistics of the noise $\eta_i$, we derive that the auto-correlation function $\Delta(\tau)=\langle [x_i(t+\tau)x_i(t) ]\rangle-\langle[x_i(t)]\rangle^2$ obeys the second-order differential equation:
\begin{equation}\label{eq:deltadot}
	\ddot{\Delta}(\tau)=\Delta(\tau)-g^2\langle[\phi_i(t)\phi_i(t+\tau)]\rangle-\Sigma_m^2\kappa^2.
\end{equation}

In this context, the activation variance $\Delta_0$ coincides with the peak of the full auto-correlation function: $\Delta_0=\Delta(\tau=0)$. We expect the total variance to include a temporal term, coinciding with the amplitude of chaotic fluctuations, and a quenched one, representing the spread across the population due to the disorder in $\chi_{ij}$ and the structure imposed by the right-connectivity vector $m$.

In order to compute the full rate auto-correlation function $\langle[\phi_i(t)\phi_i(t+\tau)]\rangle$, we need to explicitly build two correlated Gaussian variables $x(t)$ and $x(t+\tau)$, such that:
\begin{equation}
	\begin{split}
		&\langle[x_i(t)]\rangle=\langle[x_i(t+\tau)]\rangle=\mu\\
		&\langle[x_i^2(t)]\rangle-\langle[x_i(t)]\rangle^2 = \langle[x_i^2(t+\tau)]\rangle-\langle[x_i(t)]\rangle^2 =\Delta_0\\
		& \langle[x_i(t+\tau)x_i(t)]\rangle-\langle[x_i(t)]\rangle^2=\Delta(\tau).
	\end{split}
\end{equation}
Following previous studies \citep{Sompolinsky1988, Rajan2010}, we obtain:
\begin{equation}
	\langle[\phi_i(t)\phi_i(t+\tau)]\rangle=\int \mathcal{D}z  \left[ \int \mathcal{D}x \phi (\mu+\sqrt{\Delta_0-\Delta}x+\sqrt{\Delta}z) \right]^2
\end{equation}
where we used the short-hand notation $\Delta\coloneqq\Delta(\tau)$ and we assumed for simplicity $\Delta>0$. As we show later, this requirement is satisfied by our final solution.

In order to visualize the dynamics of the solutions of Eq.~\ref{eq:deltadot}, we study the equivalent problem of a classical particle moving in a one-dimensional potential \citep{Sompolinsky1988, Rajan2010}:
\begin{equation}
	\ddot{\Delta}(\tau)=-\frac{\partial V}{\partial \Delta}
\end{equation}
where the potential $V$ is given by an integration over $\Delta$:
\begin{equation}
	V(\Delta, \Delta_0)=-\frac{\Delta^2}{2}+g^2\langle[\Phi_i(t)\Phi_i(t+\tau)]\rangle+\Sigma_m^2\kappa^2\Delta
\end{equation}
and $\Phi(x)=\int_{-\infty}^{x}\phi(x')\diff x'$. As the potential $V$ depends self-consistently on the initial condition $\Delta_0$, the shape of the auto-correlation function $\Delta(\tau)$ depends parametrically on the value of $\Delta_0$. Similarly to previous works, we isolate the solutions that decay monotonically from $\Delta_0$ to an asymptotic value $\Delta(\tau\rightarrow \infty)\coloneqq\Delta_{\infty}$, where $\Delta_{\infty}$ is determined by $\diff V / \diff \Delta|_{\Delta=\Delta_{\infty}}=0$. This translates into a first condition to be imposed. A second equation comes from the energy conservation condition: $V(\Delta_0, \Delta_0)=V(\Delta_{\infty}, \Delta_0)$. Combined with the usual equation for the mean $\mu$ and the overlap $\kappa$, the system of equations to be solved becomes:
\begin{equation}\label{eq:DMF_chaotic}
	\begin{split}
		&\mu=M_m\kappa\\
		&\kappa = M_n \langle[\phi_i]\rangle + \rho \kappa  \langle[\phi_i']\rangle\\
		&\frac{\Delta_0^2-\Delta_{\infty}^2}{2}=g^2\left\{\int \mathcal{D}z \Phi^2(\mu+\sqrt{\Delta_0}z) -\int \mathcal{D}z  \left[ \int \mathcal{D}x \Phi (\mu+\sqrt{\Delta_0-\Delta_{\infty}}x+\sqrt{\Delta_{\infty}}z) \right]^2\right\} +\Sigma_m^2\kappa^2 (\Delta_0 - \Delta_{\infty})\\
		&\Delta_{\infty}=g^2\int \mathcal{D}z  \left[ \int \mathcal{D}x \phi (\mu+\sqrt{\Delta_0-\Delta_{\infty}}x+\sqrt{\Delta_{\infty}}z) \right]^2 + \Sigma_m^2\kappa^2.
	\end{split}
\end{equation}

The temporally fluctuating state is therefore described by a closed set of equations for the mean activity $\mu$, the overlap $\kappa$, the zero-lag variance $\Delta_0$ and the long-time variance $\Delta_{\infty}$. The difference $\Delta_0-\Delta_{\infty}$ represents the amplitude of temporal fluctuations. If temporal fluctuations are absent, $\Delta_0=\Delta_{\infty}$, and the system of equations we just derived reduces to the DMF description for stationary solutions given in Eq.~\ref{eq:stationary_general}.

A similar set of equations can be derived  for single unit activity. As for static stationary states, the mean activity of unit $i$ is given by
\begin{equation}
	\mu_i=m_i\kappa.
\end{equation}
The static variance around this mean activity is identical for all units and given by
\begin{equation}
	\Delta_{\infty}^I=g^2\int \mathcal{D}z  \left[ \int \mathcal{D}x \phi (\mu+\sqrt{\Delta_0-\Delta_{\infty}}x+\sqrt{\Delta_{\infty}}z) \right]^2 = \Delta_{\infty} -  \Sigma_m^2\kappa^2
\end{equation}
while the temporal component $\Delta_T^I$ of the variance is identical to the population averaged temporal variance 
\begin{equation}
	\Delta_T^I=\Delta_0-\Delta_{\infty}.
\end{equation}

To conclude,  similarly to static stationary states, the structured connectivity $P_{ij}$ shapes network activity in the direction defined by its right eigenvector $m$ whenever the overlap $\kappa$ does not vanish.
For this reason, the mean-field theory predicts in some parameter regions the existence of more than one chaotic solution. A formal analysis of the stability properties of the different solutions has not been performed. We nevertheless observe from numerical simulations that chaotic solutions tend to inherit the stability properties of the stationary solution they develop from. Specifically, when an homogeneous solution generates two heterogeneous bistable ones, we notice that the former loses stability in favor of the latter.

We finally observe that the critical coupling at which the DMF theory predicts the onset of chaotic fluctuations can be computed by imposing that, at the critical point, the concavity of the potential function $V(\Delta)$ is inverted \citep{Sompolinsky1988, Harish2015}:
\begin{equation}
	\frac{\diff^2 V (\Delta, \Delta_0)}{\diff \Delta^2}\Bigr|_{\Delta_{\infty}} = 0
\end{equation}
and the temporal component of the variance vanishes: $\Delta_0=\Delta_{\infty}$. These two conditions are equivalent to the expression: $1=g^2  \langle [\phi_i'^2] \rangle$ where, as we saw, $g^2  \langle [\phi_i'^2] \rangle$ coincides with the squared value of the radius of the compact component of the stability eigenspectrum (Eq.~\ref{eq:radius_het}). In the phase diagram of Fig.~\ref{fig:p1}  \textsf{B}, we solved this equation for $g$ to derive the position of the instability boundary from stationary to chaotic regimes.

\subsubsection{Spontaneous dynamics: structures overlapping on the unitary direction}\label{par:independent}
In this section, we analyze in detail a specific case, in which the connectivity vectors $m$ and $n$ overlap solely along the unitary direction $u=(1,1,...1)/{N}$. Within the statistical description of vector components, in this situation the joint probability density $p(m,n)$ can be replaced by the product two normal distributions (respectively, $\mathcal{N}(M_m, \Sigma_m^2)$ and $\mathcal{N}(M_n, \Sigma_n^2)$). 
The mean values $M_m$ and $M_n$ represent the projections of $m$ and $n$ on the common direction $u$, and the overlap between $m$ and $n$ is given by  $M_mM_n$. The components $m$ and $n$ are otherwise independent, the fluctuations representing the remaining parts of $m$ and $n$ that lie along mutually orthogonal directions. In this situation, the expression for $\kappa$ simplifies to
\begin{equation}
	\begin{split}  \label{eq:kappa_unitary}
		\kappa &= \langle n_i [\phi_i]\rangle\\
		&=M_n \langle[\phi_i]\rangle
	\end{split}
\end{equation}
so that a  non-zero overlap $\kappa$ can be obtained  only if the mean
population  activity  $\langle[\phi_i]\rangle$  is non-zero.   Choosing
independently drawn $m$ and $n$ vectors thus slightly simplifies the mean-field
network description.  The main qualitative features resulting from the
interaction  between the structured  and the  random component  of the
connectivity  can  however  already   be  observed,  and  more  easily
understood, within this simplified setting.

\paragraph{Stationary solutions}

The DMF description for stationary solutions reduces to a  system of two non-linear equations for the population averaged mean $\mu$  and variance $\Delta_0$:
\begin{equation}\label{eq:stationary_final}
	\begin{split}
		&\mu = M_m M_n \langle[\phi_i]\rangle \coloneqq F(\mu, \Delta_0)\\
		&\Delta_0 = g^2 \langle[\phi_i^2]\rangle + \Sigma_m^2 M_n^2 \langle[\phi_i]\rangle^2\coloneqq G(\mu, \Delta_0).
	\end{split}
\end{equation}
The population averages $\langle[\phi_i]\rangle$ and $\langle[\phi_i^2]\rangle$ are computed as Gaussian integrals similarly to Eq.~\ref{eq:phisq_gaussian}. Eq. \ref{eq:stationary_final} can be solved numerically for $\mu$ and $\Delta_0$  by iterating the equations up to convergence, which is equivalent to numerically simulating the two-dimensional dynamical system given by  
\begin{equation}\label{eq:stationary_final_dyn}
	\begin{split}
		&\dot{\mu}(t) = -\mu + F(\mu, \Delta_0)\\
		&\dot{\Delta}_0(t) = -\Delta_0 + G(\mu, \Delta_0),
	\end{split}
\end{equation}
since the fixed points of this dynamical system correspond to solutions of Eq.~\ref{eq:stationary_final}. 
Gaussian integrals in the form of $\langle[\phi_i]\rangle$ are evaluated numerically through Gauss-Hermite quadrature with a sampling over 200 points. 
Unstable solutions can be computed by iterating the same equations after having inverted the sign of the time variable in the first equation.

As the system of equations in Eq.~\ref{eq:stationary_final} is two-dimensional, we can investigate the number and the nature of stationary solutions through a simple graphical approach (Fig.~S1 \textsf{G}). We plot on the $\mu-\Delta_0$ plane the loci of points where  the two individual equations
\begin{equation}
	\begin{split}
		&\mu=F(\mu, \Delta_0) \\
		&\Delta_0=G(\mu, \Delta_0)
	\end{split}
\end{equation}
are satisfied. In analogy with dynamical systems approaches, we refer to the two corresponding curves as the DMF \emph{nullclines}. The solutions of Eq.~\ref{eq:stationary_final} are then given by the intersections of the two nullclines.

To begin with, we focus on the nullcline defined by the first equation
(also  referred to  as the  $\mu$ nullcline).  With respect  to $\mu$,
$F(\mu, \Delta_0)$  is an odd  sigmoidal function whose  maximal slope
depends    on     the    value    of     $\Delta_0$    and    $M_mM_n$.   When $g=0$
and  $\Sigma_m=0$, the  input  variance $\Delta_0$  vanishes. In  this
case, the points of the  $\mu$ nullcline trivially reduce to the roots
of  the  equation:  $\mu=M_mM_n\phi(\mu)$,  which  admits  either  one
($M_mM_n<1$), or three solutions ($M_mM_n>1$).  Non-zero values of $g$
and  $\Sigma_m$ imply  finite and  positive values  of  $\Delta_0$. As
$\Delta_0$    increases,    the     solutions    to    the    equation
$\mu=M_mM_n\langle[\phi_i]\rangle$  vary smoothly,  delineating  the full
nullcline  in  the  $\mu-\Delta_0$  plane.   As in  the  case  without
disorder  ($g=0$  and   $\Sigma_m=0$),  for  low  structure  strengths
($M_mM_n<1$), the $\mu$ nullcline consists of a unique branch: $\mu=0$
$\forall \Delta_0$. At
high  structure strengths  ($M_mM_n>1$), instead,  its  shape smoothly
transforms   into  a   symmetric   pitchfork.

The $\Delta_0$ nullcline is given by the solutions of $\Delta_0=G(\mu,
\Delta_0)$ for $\Delta_0$ as  function of $\mu$. As $G(\mu, \Delta_0)$
depends quadratically on $\mu$, the $\Delta_0$ nullcline
has a symmetric~\emph{V}-shape centered in $\mu=0$. The ordinate of its vertex is controlled by the parameter $g$, as the second term of the second equation in \ref{eq:stationary_final} vanishes at $\mu=0$. For $\mu=0$, the slope of $G(\mu, \Delta_0)$ in $\Delta_0=0$ is equal to $g^2$. As a consequence, for $g<1$, the vertex of the $\Delta_0$ nullcline is fixed in $(0,0)$, while for  $g>1$, the vertex is located at $\Delta_0>0$ and an isolated point remains at $(0,0)$.

The stationary  solutions of the  DMF equations are determined  by the
intersections  between the  two  nullclines.  For  all  values of  the
parameters,  the  nullclines   intersect    in    $\mu=0$,   $\Delta_0=0$,
corresponding  to the trivial,  homogeneous stationary  solution. The
existence  of  other  solutions  are  determined  by  the  qualitative
features of the individual nullclines, that depend on whether $M_mM_n$
and $g$  are smaller  or greater than  one (Fig.~S1 \textsf{G}).
The  following qualitative  situations can  be distinguished:  (i) for
$M_mM_n<1$  and $g<1$,  only  the trivial  solutions  exist; (ii)  for
$M_mM_n>1$,  two additional, symmetric  solutions exist  for non-zero
values   of  $\mu$   and  $\Delta_0$,   corresponding   to  symmetric,
heterogeneous  stationary  states; (iii)  for  $g>1$, an  additional
solution  exist  for  $\mu=0$  and $\Delta_0>0$,  corresponding  to  a
heterogeneous  solution  in   which  individual  units  have  non-zero
stationary  activity,   but  the  population-average   vanishes.  For
$M_mM_n>1$,   this   solution   can   co-exist  with   the   symmetric
heterogeneous  ones, but  in the  limit of  large $g$  these solutions
disappear (Fig.~S1 \textsf{G}).

The next step is to assess  the stability of the various solutions. As
explained  earlier on,  the stability  of the  trivial  state $\mu=0$,
$\Delta_0=0$ can  be readily assessed using random  matrix theory arguments (Fig.~S1 \textsf{A-B}).  This  state  is  stable  only  for  $M_mM_n<1$  and
$g<1$. At $M_mM_n=1$, it  loses stability due to the outlying
eigenvalue of  the stability matrix, leading to  the bifurcation already
observed at the level of  nullclines. At $g=1$, the instability is due
to the  radius of the  bulk of the  spectrum. This leads to  a chaotic
state,  not   predicted  from  the  nullclines   for  the  stationary
solutions.

The  stability  of heterogeneous  stationary  states  is assessed  by
determining separately the radius of  the bulk of the spectrum and the
position of the outlier (Fig.~S1 \textsf{D-F}).  The radius is
determined  from Eq.~\ref{eq:radius_het}.   The  outlier is  instead
computed as  the leading eigenvalue  of the stability matrix  given in
Eq.~\ref{eq:M}. Note that in the present framework, where the overlap is defined along the unitary direction, it is possible to show that the latter is equivalent  to computing the
leading  stability  eigenvalue   of  the  effective  dynamical  system
introduced  in Eq.~\ref{eq:stationary_final_dyn}, linearized  around the
corresponding fixed  point. The bifurcation obtained  when the outlier
crosses  unity is  equivalent to  the bifurcation  predicted  from the
nullclines  when the  symmetric solutions  disappear in  favor  of the
heterogeneous  solution of  mean zero  (Fig.~S1 \textsf{G}). For
$M_mM_n>1$, we  however find that as  $g$ is increased,  the radius of
the bulk of the spectrum  always leads to a chaotic instability before
the  outlier  becomes   unstable.  Correspondingly,  the  $\mu=0$  and
$\Delta_0>0$  stationary state  that  exist for  large  $g$ is  never
stable.

\paragraph{Chaotic solutions} For large $g$, the instabilities of the stationary points generated by the bulk of the spectrum are expected to give rise to chaotic dynamics. We therefore turn to the DMF theory for chaotic states, which are described by an additional variable that quantifies temporal fluctuations. For the case studied here of connectivity vectors $m$ and $n$ overlapping only along the unitary direction, Eq.~\ref{eq:DMF_chaotic} become

\begin{equation}\label{eq:DMF_chaotic_unitary}
	\begin{split}
		&\mu=F(\mu, \Delta_0, \Delta_{\infty})=M_mM_n\int \mathcal{D}z \phi(\mu+\sqrt{\Delta_0}z)\\
		&\Delta_0=G(\mu, \Delta_0, \Delta_{\infty})=\left[\Delta_{\infty}^2+2g^2\left\{\int \mathcal{D}z \Phi^2(\mu+\sqrt{\Delta_0}z) \right.\right. \\
		&\left. -\int \mathcal{D}z  \left[ \int \mathcal{D}x \Phi (\mu+\sqrt{\Delta_0-\Delta_{\infty}}x+\sqrt{\Delta_{\infty}}z) \right]^2\right\} +\biggl. M_n^2\Sigma_m^2\langle[\phi_i]\rangle^2 (\Delta_0 - \Delta_{\infty})\biggr]^{\frac{1}{2}}\\
		&\Delta_{\infty}=H(\mu, \Delta_0, \Delta_{\infty})=g^2\int \mathcal{D}z  \left[ \int \mathcal{D}x \phi (\mu+\sqrt{\Delta_0-\Delta_{\infty}}x+\sqrt{\Delta_{\infty}}z) \right]^2 + M_n^2\Sigma_m^2\langle[\phi_i]\rangle^2.
	\end{split}
\end{equation}

As the system to be solved is now three-dimensional, graphical approaches have only limited use. Similarly to the stationary state, a practical and stable way to find numerically the solutions is to iterate the dynamical system given by
\begin{equation}\label{eq:macro_ev_2}
	\begin{split}
		&\dot{\mu}=-\mu+F(\mu, \Delta_0,\Delta_{\infty})\\
		&\dot{\Delta}_0=-\Delta_0+G(\mu, \Delta_0,\Delta_{\infty})\\
		&\dot{\Delta}_{\infty}=-\Delta_{\infty}+H(\mu, \Delta_0,\Delta_{\infty}).
	\end{split}
\end{equation}
where the double Gaussian integrals from Eq.~\ref{eq:DMF_chaotic_unitary} can be evaluated numerically as two nested Gauss-Hermite quadratures.
Note that stationary states simply correspond to solutions for which $\Delta_0=\Delta_{\infty}$. 

As  for stationary  solutions, different  types of  chaotic solutions
appear depending on the values  of the structure strength $M_mM_n$ and
the    disorder    strength   $g$.    If    $g>1$   and    $M_mM_n<1$, a  single  chaotic  state  exists
corresponding  to   $\mu=0$ and  $\Delta_{\infty}=0$, meaning  that  the temporally  averaged
activity of all units vanishes, so that fluctuations are only temporal
(Fig.~\ref{fig:p1} {B} red).  As  $M_mM_n$ crosses unity,  two symmetric  states  appear  with
non-zero values of $\mu$ and $\Delta_{\infty}$. These states correspond to bistable heterogeneous chaotic states (Fig.~\ref{fig:p1} {B} orange) that are analogous to bistable heterogeneous stationary states. 

The critical  disorder strength  $g_B$ at which  heterogeneous chaotic
states   emerge    (gray   boundary   in   the    phase   diagram   of
Fig.~\ref{fig:p1}) is  computed by evaluating the  linear stability of
the dynamics in \ref{eq:macro_ev_2} around the central solution $(0,
\Delta_0, 0)$.   A long but  straightforward algebra reveals  that the
stability matrix, evaluated in $(0, \Delta_0, 0)$, is simply given by
\begin{equation}
	\begin{pmatrix}
		M_m M_n \langle \phi' \rangle & 0 & 0\\
		0 & {\frac{g^2\left(\langle \phi^2 \rangle  +\langle \Phi \phi' \rangle  - \langle \Phi \rangle  \langle \phi' \rangle \right)}{\Delta_0}} & 0\\
		0 & 0 & g^2 \langle \phi' \rangle ^2
	\end{pmatrix},
\end{equation}
such that $g_{B}$ corresponds to the value of the random strength $g$ for which the largest of its three eigenvalues crosses unity.

\subsubsection{Spontaneous dynamics: structures overlapping on an arbitrary direction}

In the previous section, we focused on the simplified scenario where  the connectivity vectors $m$ and $n$ overlapped only in the unitary direction. Here, we briefly turn to the opposite case where the overlap along the unitary direction $u$ vanishes (i.e. $M_m=0$, $M_n=0$), but the overlap $\rho$ along a direction orthogonal to $u$ is non-zero.
As we will show, although the equations describing the network activity present some formal differences, they lead to qualitatively similar regimes. The same qualitative results apply as well to the general case, where an overlap exists on both the unitary and an orthogonal direction.

The network dynamics can be studied by solving the DMF equations \ref{eq:stationary_general} and \ref{eq:DMF_chaotic} by setting $\mu = 0$.
Stationary solutions are now determined by:
\begin{equation}\label{eq:static_arbitrary}
	\begin{split}
		&\kappa = \rho \kappa \Sigma_m\Sigma_n\langle [\phi'_i(0, \Delta_0)]\rangle \coloneqq F(\kappa, \Delta_0)\\
		&\Delta_0 = g^2  \langle [\phi_i^2(0, \Delta_0)]\rangle+\Sigma_m^2 \kappa^2 \coloneqq G(\kappa, \Delta_0).
	\end{split}
\end{equation}
Note that, in this more general case, the relevant first-order statistics of network activity is given by the overlap $\kappa$, which now can take non-zero values even when the population-averaged activity $\langle[\phi_i]\rangle$ vanishes.

As in the previous case, the stationary solutions can be analyzed in terms of nullclines (Fig.~S2 \textsf{A}). The main difference lies in the $\kappa$ nullcline given by $\kappa=\rho \kappa \Sigma_m\Sigma_n \langle [\phi_i'(0, \Delta_0)]\rangle$.
As both sides of the first equation are linear and homogeneous in $\kappa$, two classes of solutions exist: a trivial solution ($\kappa=0$ for any $\Delta_0$), and a non-trivial one ($\Delta_0=\tilde{\Delta}_0$ for any $\kappa$), with $\tilde{\Delta}_0$ determined by:
\begin{equation}\label{eq:delta0tilde}
	\langle [\phi_i'(0, \tilde{\Delta}_0)]\rangle=1/(\rho\Sigma_m\Sigma_n).
\end{equation}
Because $0<\phi'(x)<1$, Eq.~\ref{eq:delta0tilde} admits non-trivial solutions only for sufficiently large overlap values: $\rho>1/\Sigma_m\Sigma_n$. In consequence, the $\kappa$ nullcline takes qualitatively different shapes depending on the value of $\rho$: (i) for $\rho<1/\Sigma_m\Sigma_n$, it consists only of a vertical branch $\kappa=0$; (ii) for $\rho>1/\Sigma_m\Sigma_n$ an additional horizontal branch $\Delta_0=\tilde{\Delta}_0$ appears (Fig.~S2 \textsf{A}).

The $\Delta_0$ branch is qualitatively similar to the previously studied case of $m$ and $n$ overlapping along the unitary direction, with a qualitative change when the disorder parameter $g$ crosses unity.

The stationary  solutions are given by the  intersections between the
two  nullclines. Although  the  shape of  the  $\kappa$ nullcline  is
distinct from the shape of the $\mu$ nullcline studied in the previous
case, qualitatively similar regimes are found. The trivial stationary
state $\kappa=0$,  $\Delta_0=0$ exists for all  parameter values. When
the   structure   strength  $\rho\Sigma_m\Sigma_n$   exceeds unity,  two   symmetric
heterogeneous  states  appear   with  non-zero  $\kappa$  values of opposite signs  (but
vanishing  mean $\mu$).  Finally for  large $g$  an  additional state
appears with $\kappa=0$, $\Delta_0>0$.

Similarly to Fig.~\ref{fig:p1}, the solutions of Eq.~\ref{eq:static_arbitrary}, which correspond to stationary activity states, are shown in blue in Fig.~S2 \textsf{B-D}. 

In Fig.~S2 \textsf{B} we address their stability properties: again we find that when non-centered stationary solutions exist, the central fixed point becomes unstable. The instability is led by the outlier eigenvalue of the stability eigenspectrum.
Similarly to Fig.~\ref{fig:p1}, furthermore, the DMF theory predicts an instability to chaotic phases for high $g$ values. As for stationary states, both heterogeneous and homogeneous chaotic solutions are admitted (Fig.~S2 \textsf{C-D}); heterogeneous chaotic states exist in a parameter region where the values of $g$ and $\rho$ are comparable.

\subsubsection{Response to external inputs}\label{par:dmf_input}

In  this section,  we  examine the  effect  of non-vanishing  external
inputs on  the network  dynamics. We consider  the situation  in which
every unit receives  a potentially different input $I_i$,  so that the
pattern  of  inputs at  the  network  level  is characterized  by  the
$N$-dimensional  vector $I=\{I_i\}$. The  network dynamics  in general
depend on the  geometrical arrangement of the vector  $I$ with respect
to  the  connectivity  vectors  $m$  and  $n$.   Within  the  statistical
description  used  in  DMF  theory,  the input  pattern  is  therefore
characterized  by the  first-  and second-order  statistics $M_I$  and
$\Sigma_I$  of   its  elements,  as   well  as  by the value of   the  correlations $\Sigma_{mI}$  and $\Sigma_{nI}$ with the  vectors $m$  and  $n$.  In
geometric terms, $M_I$ quantifies the  component of $I$ along the unit
direction  $u$,  while $\Sigma_{mI}$  and  $\Sigma_{nI}$ quantify  the
overlaps with $m$ and $n$ along directions orthogonal to $u$.  For the
sake of simplicity, here we consider two connectivity vectors  $m$ and $n$ that overlap solely on the unitary direction ($\rho=0$). The two vectors thus read (see Eq.~\ref{eq:mn}):
\begin{equation}
	\begin{split}
		&m = M_m + \Sigma_m x_1 \\
		&n = M_n + \Sigma_n  x_2.
	\end{split}
\end{equation}
The input pattern can overlap with the connectivity vectors on the common ($u$) and on the orthogonal directions ($x_1$ and $x_2$). It can moreover include further orthogonal components of strength $\Sigma_{\perp}$. The most general expression for the input vector can thus be written as:
\begin{equation}\label{eq:input}
	I = M_I + \frac{\Sigma_{mI}}{\Sigma_m} x_1 +  \frac{\Sigma_{nI}}{\Sigma_n} x_2 + \Sigma_{\perp} h
\end{equation}
where  $h$  is  a standard  normal  vector.   We  first focus  on  the
equilibrium response  to constant inputs,  and then turn  to transient
dynamics.

The mean-field equations  in presence of external inputs can be derived in a straightforward fashion 
by following the same steps as in the input-free case. We start by considering the statistics of the effective coupling term, which is given by $\xi_i(t)=\eta_i(t)+I_i(t)$, with $\eta_i(t)$ defined as in Eq.~\ref{eq:eta}. We can then exploit the statistics of  $\eta_i(t)$ which have been computed in the previous paragraphs to obtain the equation for the mean activity:
\begin{equation}\label{eq:mui_input}
	\mu_i=[x_i]=m_i \kappa+I_i.
\end{equation}
Eq.~\ref{eq:mui_input} indicates that the direction of the average network activity is determined by a combination of the structured recurrent connectivity and the external input pattern. The final direction of the activation vector in the $N$-dimensional population space is controlled by the value of the overlap $\kappa$, which depends on the relative orientations of $m$, $n$ and $I$. Its value is given by the self-consistent equation:
\begin{equation}\label{eq:kappa_inp}
	\begin{split}
		\kappa &=\langle n_i [\phi_i]\rangle\\
		& = \langle n_i \int \mathcal{D}z \phi(m_i\kappa + I_i + \sqrt{\Delta_0^I}z_i) \rangle\\
		&= M_n\langle[\phi_i]\rangle + \Sigma_{nI}\langle[\phi'_i]\rangle,
	\end{split}
\end{equation}
as both vectors $m$ and $I$ share non-trivial overlap directions with $n$.

The second-order statistics of the noise are given by:
\begin{equation}
	[\xi_i(t) \xi_j(t+\tau)]=\delta_{ij}g^2 \langle [\phi_i(t)\phi_i(t+\tau)] \rangle  + m_i m_j \kappa^2 + (m_i I_j + m_j I_i) \kappa +I_i I_j .
\end{equation}
Averaging across the population we obtain:
\begin{equation}
	\langle [\xi_i(t) \xi_i(t+\tau)] \rangle-\langle[\xi_i(t)]\rangle^2=g^2 \langle [\phi_i^2] \rangle +\Sigma_m^2 \kappa ^2 + 2 \Sigma_{mI}  \kappa  + \Sigma_I^2.
\end{equation}
The first term of the r.h.s.~represents the quenched variability inherited from the random connectivity matrix, while $\Sigma_{\mu}^2 = \Sigma_m^2 \kappa^2  + 2 \Sigma_{mI}  \kappa  + \Sigma_I^2$ represents the variance induced by the structure, which is inherited from both vectors $m$ and $I$ (Eq.~\ref{eq:mui_input}). From Eq.~\ref{eq:input}, the variance of the input reads:
\begin{equation}
	\Sigma_I^2 = \frac{\Sigma_{mI}^2}{\Sigma_m^2} + \frac{\Sigma_{nI}^2}{\Sigma_n^2} + \Sigma_{\perp}^2.
\end{equation}

The final DMF equations to be solved are given by the following system:
\begin{equation}\label{eq:mf_input}
	\begin{split}
		&\mu=M_m \kappa+M_I\\
		&\ddot{\Delta}=\Delta- \left\{ g^2 \langle [\phi_i(t)\phi(t+\tau)] \rangle+\Sigma_m^2 \kappa ^2 + 2 \Sigma_{mI}  \kappa  + \Sigma_I^2 \right\}\\
		&\kappa = M_n\langle[\phi_i]\rangle + \Sigma_{nI}\langle[\phi'_i]\rangle
	\end{split}
\end{equation}
which, similarly to the cases we examined in detail so far, admits both stationary and chaotic solutions.
As for spontaneous dynamics, the instabilities to chaos are computed by evaluating the radius of the eigenspectrum of the stability matrix $S_{ij}$ (Eq.~\ref{eq:radius_het}). The stability matrix can admit an outlier eigenvalue as well, whose value can be predicted with a mean-field stability analysis.
Extending the arguments already presented in the previous paragraphs allows to show that the effective stability matrix $\mathcal{M}$ is given by:
\begin{equation}\label{eq:input_M}
	\mathcal{M} = 
	\begin{pmatrix} 0 & 0 & M_m\\ 
		2 g^2 \langle [\phi_i\phi_i']\rangle & g^2\left\{\langle [\phi_i'^2]\rangle + \langle [\phi_i\phi_i'']\rangle \right\} & 2\Sigma_m^2\kappa^0+2\Sigma_{mI}\\
		2 b g^2 \langle [\phi_i\phi_i']\rangle & b g^2\left\{\langle [\phi_i'^2]\rangle + \langle [\phi_i\phi_i'']\rangle \right\} & b  (2\Sigma_m^2\kappa^0+2\Sigma_{mI}) + a
	\end{pmatrix},
\end{equation}
with:
\begin{equation}
	\begin{split}
		a = M_m M_n  \langle [\phi_i']\rangle + M_m \Sigma_{nI}  \langle [\phi_i'']\rangle\\
		b = \frac{1}{2} \left\{ M_n  \langle [\phi_i'']\rangle +  \Sigma_{nI}  \langle [\phi_i''']\rangle \right\}.
	\end{split}
\end{equation}
As in the input-free case, when the stability eigenspectrum contains one outlier eigenvalue, its position is well predicted by the largest eigenvalue of $\mathcal{M}$. 

In the following, we refer to Fig.~\ref{fig:p2} and analyse in detail the contribution of every input direction to the final network dynamics.

In Fig.~\ref{fig:p2} \textsf{D} (left), we consider a unit-rank structure whose vectors $m$ and $n$ are orthogonal: $M_m=M_n=0$. The input direction is orthogonal to the connectivity vectors: $\Sigma_{mI} = \Sigma_{nI} = 0$, so that the input strength is quantified by the amplitude of the component along $h$ ($\Sigma_{\perp}$). In this configuration, because of Eq.~\ref{eq:kappa_inp}, the amount of structured activity quantified by $\kappa$ systematically vanishes.

In Fig.~\ref{fig:p2} \textsf{D} (center), we consider again orthogonal connectivity vectors, but we take an input pattern which overlaps with $n$ along $x_2$. We keep $\Sigma_{\perp}=1$ fixed and we vary the component of the input along $n$ by increasing $\Sigma_{nI}$.
As  can be seen from the equation for $\kappa$ (Eq.~\ref{eq:mf_input}), the overlap $\Sigma_{nI}$ between the input and the left vector $n$ has the effect of increasing the value of $\kappa$, which would otherwise vanish since the structure has null strength ($M_n=0$). In response to the input, a structured state emerges.
From the same equation, furthermore, one can notice that the $\Sigma_{nI}$ term has the effect of breaking  the sign reversal symmetry ($x \rightarrow\ -x$) that characterizes the mean-field equations in the case of spontaneous dynamics.

In Fig.~\ref{fig:p2} \textsf{D} (right), we include strong non-vanishing structure strengths ($M_m M_n = 3.5$). In absence of external activity, the network dynamics thus admit two bistable solutions (Fig.~\ref{fig:p1}). We consider an input pattern that correlates with $n$ but is orthogonal to the structure overlap direction ($M_I=0$, $\Sigma_{nI}>0$). In this configuration, the external input has the effect of disrupting the symmetry between the two stable solutions. 
For sufficiently strong input values, one of the two stable solutions disappears by annihilating with the unstable one. 

In Fig.~S4 \textsf{C}, we show that the value of the critical input strength for which one of the two stable solution disappears can be controlled by an additional external input that overlaps with $n$ on a different, orthogonal direction. Specifically, in Fig.~S4 \textsf{C}, we tune the additional input along the direction of the structure overlap $u$. This input component can be thought as a modulatory signal which controls the way the network dynamics process the input stimulus along $x_2$. 
In models of computational tasks that employ non-linear input responses (Fig.~\ref{fig:p4}), a modulatory input along the structure overlap can regulate the  threshold value of the input strength that the network has learnt to detect. Similarly, in Figs.~\ref{fig:p5} and \ref{fig:p6}, modulatory inputs are used to completely block the response to the non-relevant input stimulus, so that the readout can produce context-dependent outputs.

\paragraph{Asymmetric solutions}

A major effect of external inputs is that they break the sign reversal symmetry ($x \rightarrow\ -x$) present in the network dynamics without inputs.
As a consequence, in the parameter regions where the network dynamics admit bistable structured states,  the two stable solutions are characterized by different statistics and stability properties. 

To illustrate this effect, we focus on the simple case where the external input pattern $I$ overlaps with the connectivity vectors $m$ and $n$ solely on the unitary direction ($M_I \neq 0$, $\Sigma_{mI} = \Sigma_{nI}=0$).
The solutions of the system of equations corresponding to stationary states can be visualized with the help of the graphical approach, which unveils the symmetry breaking of network dynamics induced by external inputs (Fig.~S4 \textsf{D}).

Similarly to the input-free case, the $\Delta_0$ nullcline consists of a symmetric \emph{V}-shaped curve. In contrast to before, however, the vertex of the nullcline is no longer fixed in $(0,0)$, but takes positive ordinate values also at low $g$ values. The value of $G(0, \Delta_0)$, indeed, does not vanish, because of the finite contribution from the input pattern $\Sigma_I^2$.

The nullcline curves of $\mu$ are instead strongly asymmetric.
For low $M_mM_n$ values, one single $\mu$ nullcline exists. In contrast to the input-free case, this nullcline is no longer centered in zero. As a consequence, it intersects the $\Delta_0$ nullclines in one non-zero point, corresponding to a unique heterogeneous stationary solution.
As $M_mM_n$ increases, a second, separated branch can appear. In contrast to the input-free case, the structure strength at which the second branch appears is not always equal to unity, but depends on the mean value of the input.
If $M_mM_n$ is strong enough, the negative branch of the nullcline can intersect the $\Delta_0$ nullcline in two different fixed points, while a third solution is built on the positive $\mu$ nullcline. As $g$ increases, the two intersections on the negative branch become closer and closer and they eventually collapse together. At a critical value $g_B$, the network activity discontinuously jumps from negative to positive mean solutions.

As they are no longer symmetrical, the stability of the positive and the negative fixed points has to be assessed separately, and gives rise to different instability boundaries. 
Computing the position of the outlier reveals that, when more than one solution is admitted by the mean-field system of equations, the centered one is always unstable.

As the stability boundaries of different stationary solutions do not necessarily coincide, in presence of external input patterns the phase diagram of the dynamics are in general more complex (Fig.~S4 \textsf{A-C}). Specifically, hybrid dynamical regimes, where one static solution co-exists with a chaotic attractor, can be observed.

\paragraph{Transient dynamics}
We now turn to transient dynamics evoked by a temporal step in the external input (Fig.~\ref{fig:p2} \textsf{B}).
We specifically examine the projection of the activation vector and its average onto the two salient directions spanned by vectors $m$ and $I$.

The transient dynamics of relaxation to a stationary solution can be assessed by linearizing the mean-field dynamics. We compute the time course of the average activation vector $\{\mu_i\}$, and we finally project it onto the two orthogonal directions which are indicated in the small insets of Fig.~\ref{fig:p2} \textsf{B}.

Similarly to Eq.~\ref{eq:mui}, the time evolution of $\mu_i$ is governed by:
\begin{equation}
	\dot{\mu}_i(t) = -\mu_i(t) + m_i \kappa (t) + I_i (t)
\end{equation}
so that, at every point in time:
\begin{equation}
	\mu_i(t) =  m_i \tilde{\kappa} (t) + \tilde{I_i} (t),  \label{eq:mu_temp}
\end{equation}
where $\tilde{\kappa} (t)$ and $\tilde{I_i} (t)$ coincide with the low-pass filtered versions of $\kappa(t)$ and $I(t)$.

When the network activity is freely decaying back to an equilibrium stationary state, $\tilde{I_i} (t)$ coincides with a simple exponential relaxation to the pattern $I_i$. The decay time scale is set by the time evolution of activity (Eq.~\ref{eq:dynamics}), which  is taken here to be equal to unity:
\begin{equation}\label{eq:I_transient}
	\tilde{I_i} (t) = I_i + (I^{ic}_i - I_i) e^{-t}.
\end{equation}

The time scale of $\tilde{\kappa}(t)$ is inherited from the dynamics of $\kappa(t)$.
We thus refer to our mean-field stability analysis, and  compute the relaxation time of the population statistics $\kappa(t)$ as the largest eigenvalue of the stability matrix $\mathcal{M}$. The eigenvalue predicts a time constant $\tau_r$, which is in general larger than unity. As a consequence, the relaxation of $\kappa(t)$ obeys, for small displacements:
\begin{equation}\label{eq:kappa_transient}
	{\kappa} (t) = \kappa^0 + (\kappa^{ic} - \kappa^0 ) e^{-\frac{t}{\tau_r}},
\end{equation}
where the asymptotic value of $\kappa^0$ is determined from the equilibrium mean-field equations (Eqs.~\ref{eq:mf_input}).
Finally, the time course of $\tilde{\kappa}(t)$ is derived as the low-pass filter version of Eq.~\ref{eq:kappa_transient} with unit decay time scale.

\subsubsection{Rank-two connectivity structures}

In the following paragraphs, we provide the detailed analysis for network models with rank-two connectivity structures. The structured component of the connectivity can be written as:
\begin{equation}
	P_{ij}=\frac{m_i^{(1)}n_j^{(1)}}{N}+\frac{m_i^{(2)}n_j^{(2)}}{N},
\end{equation}
where the vector pairs $m^{(1)}$ and $m^{(2)}$, $n^{(1)}$ and $n^{(2)}$ are assumed to be linearly independent.

As in the case of unit-rank structures, we determine the network statistics by exploiting the link between linear stability analysis and mean-field description. The study of the properties of eigenvalues and eigenvectors for the low-dimensional matrix $P_{ij}$ helps to predict the complex behavior of activity above the instability and to restrict our attention to the cases of interest.

The mean activity of the network in response to a fixed input pattern $I_i$ is given by:
\begin{equation}
	\mu_i=  \kappa_1 m_i^{(1)}+\kappa_2 m_i^{(2)} +I_i.
\end{equation}
The final direction of the population activity is thus determined by the overlap values $\kappa_1 = \langle n^{(1)}_i [\phi_i]\rangle$ and $\kappa_2 = \langle n^{(2)}_i [\phi_i]\rangle$.

The expression of the mean-field equations for the first- and second-order statistics are determined by the geometrical arrangement of the connectivity and the input vectors. 
Similarly to the unit-rank case, the simplest mean-field solutions correspond to stationary states, which inherit the structure of the most unstable eigenvectors of the connectivity matrix $J_{ij}$.
The stability of the heterogeneous stationary states can be assessed as before by evaluating separately the value of the radius (Eq.~\ref{eq:radius_het}) and the position of the outliers of the linear stability matrix $S_{ij}$.

Similarly to the unit-rank case, it is possible to compute the position of the outlier eigenvalues by studying the linearized dynamics of the network statistics close to the fixed point, that is given by:
\begin{equation}
	\frac{\diff}{\diff t}\begin{pmatrix}  \mu^1 \\ \Delta_0^1 \\ \kappa_1^{1} \\ \kappa_2^{1}\end{pmatrix} = - \begin{pmatrix} \mu^1 \\ \Delta_0^1 \\ \kappa_1^{1} \\ \kappa_2^{1} \end{pmatrix} + \mathcal{M} \begin{pmatrix} \mu^1 \\ \Delta_0^1 \\\kappa_1^{1} \\ \kappa_2^{1} \end{pmatrix} .
\end{equation}
Note that, in $\kappa_k^{l}$, the subscript $k = 1, 2$ refers to the left vector $n^{(k)}$ with which the overlap is computed, while the superscript $l = 0, 1$ indicates the order of the perturbation away from the fixed point. 

In order to compute the elements of the linear stability matrix $\mathcal{M}$, we follow and extend the reasoning discussed in details for the unit-rank case.
We start by considering the time evolution of the linearized activity $\mu_i^1$, which similarly to Eq. \ref{eq:mui} reads:
\begin{equation}
	\dot{\mu}_i^1(t) = -\mu_i^1 + m_i^{(1)} \kappa_1^{1} + m_i^{(2)} \kappa_2^{1}.
\end{equation}
At every point in time, we can write: $\mu_i^t = m_i^{(1)} \tilde{\kappa}_1^{t} + m_i^{(2)} \tilde{\kappa}_2^{t}$, where $\tilde{\kappa}_k^{t}$ is the low-pass filtered version of ${\kappa}_k^{t}$: $(1+\diff/\diff t) \tilde{\kappa}_k^{ t} = {\kappa}_k^{ t}$.

In the case of orthogonal (zero mean), random connectivity vectors, we get:
\begin{equation}
	\dot{\mu}^1(t) = -\mu^1,
\end{equation}
so that the elements in the first row of $\mathcal{M}$ vanish.
In analogy with Eq.~\ref{eq:dyn_delta0}, the linearized dynamics of $\Delta_0$ gives instead:
\begin{equation}
	\dot{\Delta_0^1} = -\Delta_0^1 + 2g^2 \langle[\phi_i\phi_i']\rangle \mu^1 + g^2 \{  \langle[\phi_i'^2]\rangle +  \langle[\phi_i\phi_i'']\rangle\} \Delta_0^1 + 2\Sigma_m^2 \kappa_1^{0} \kappa_1^{1} + 2\Sigma_m^2 \kappa_2^{0} \kappa_2^{1}.
\end{equation}

Similarly to the unit-rank case (Eq.~\ref{eq:kappa1}), in order to determine the linear response of $\kappa_1$ we need to compute:
\begin{equation}\label{eq:tobeapplied}
	\kappa_1^1 = \langle n^{(1)}_i [x_i^1 \phi'(x_i^0)] \rangle = \langle n^{(1)}_i \mu_i[\phi_i']\rangle + \left( \frac{\Delta_0^1}{2} - \langle \mu_i^1 \mu_i^0\rangle -\langle \mu_i^1 \rangle \langle \mu_i^0\rangle  \right) \langle n^{(1)}_i [\phi_i'']\rangle
\end{equation}
A similar expression can be derived for  $\kappa_2^{1}$.

In general, the integrals in the r.h.s.~can be expressed in terms of the perturbations $\tilde{\kappa}_1^1$, $\tilde{\kappa}_2^1$ and $\Delta_0^1$, leading to expressions of the form:
\begin{equation}
	\begin{split}
		\kappa_1^1 = a_{11} \tilde{\kappa}_1^{1}+ a_{12} \tilde{\kappa}_2^{1} + b_1 \Delta_0^1\\
		\kappa_2^1 = a_{21} \tilde{\kappa}_1^{1}+ a_{22} \tilde{\kappa}_2^{1} + b_2 \Delta_0^1.
	\end{split}
\end{equation}
Applying the operator $(1+\diff/\diff t)$ to the Eq.~\ref{eq:tobeapplied} allows to reshape the results in the final matrix form:
\begin{equation}\label{eq:M_rank2}
	\mathcal{M} = 
	\begin{pmatrix} 0 & 0 & 0 & 0\\ 
		2 g^2 \langle [\phi_i\phi_i']\rangle & g^2\left\{\langle [\phi_i'^2]\rangle + \langle [\phi_i\phi_i'']\rangle \right\} & 2\Sigma_m^2\kappa_1^{0} & 2\Sigma_m^2\kappa_2^{0}\\
		2 b_1 g^2 \langle [\phi_i\phi_i']\rangle & b_1 g^2\left\{\langle [\phi_i'^2]\rangle + \langle [\phi_i\phi_i'']\rangle \right\} & 2 b_1  \Sigma_m^2\kappa_1^{0} + a_{11} &  2 b_1  \Sigma_m^2\kappa_2^{0} + a_{12} \\
		2 b_2 g^2 \langle [\phi_i\phi_i']\rangle & b_2 g^2\left\{\langle [\phi_i'^2]\rangle + \langle [\phi_i\phi_i'']\rangle \right\} & 2 b_2  \Sigma_m^2\kappa_1^{0} + a_{21} &  2 b_2  \Sigma_m^2\kappa_2^{0} + a_{22} \\
	\end{pmatrix},
\end{equation}
where the values of the constants $a$ and $b$ depend on the geometric arrangement of the structure and the input vectors.

In the following, we consider several specific cases of interest.
Note that the non-linear network dynamics is determined by the relative orientation of the structure and input vectors, but also by the characteristics of the statistical distribution of their elements. In contrast to the cases we analyzed so far, the precise shape of the distribution of the entries in the connectivity vectors can play an important role when the rank of $P_{ij}$ is larger than unity. In the following, we focus on the case of broadly, normally distributed patterns.

\paragraph{Rank-two structures with null overlap}

The simplest case we consider consists of rank-two matrices whose four connectivity vectors $m^{(1)}$, $m^{(2)}$, $n^{(1)}$ and $n^{(2)}$  are mutually orthogonal. From the point of view of responses to inputs, networks with this structure behave as superpositions of two independent unit-rank structures.

Similarly to the unit-rank case, if the connectivity vectors are orthogonal, the network is silent in absence of external inputs: $\kappa^1 = \kappa^2 = 0$.  A single homogeneous state -- stationary or chaotic -- is the unique stable attractor of the dynamics. Consistently, the eigenspectrum of $J_{ij}$ does not contain any outlier, since every eigenvalue of $P_{ij}$ vanishes.

In order to compute the eigenspectrum of $P_{ij}$, we can rotate the matrix onto a basis defined by an orthonormal set of vectors, and compute its eigenvalues in the transformed basis. For simplicity, we consider an orthonormal set whose first four vectors are built from the connectivity vectors:
\begin{equation}
	\begin{split}
		&u_1 = \alpha_1 m^{(1)}\\
		&u_2 = \alpha_2 m^{(2)}\\
		&u_3 = \alpha_3 n^{(1)}\\
		&u_4 = \alpha_4 n^{(2)},\\
	\end{split}
\end{equation}
where the coefficient $\alpha_k$ ($k=1, ..., 4$) denote the normalization factors. In this basis, the first four rows and columns of the rotated matrix $P'_{ij}$ read:
\begin{equation}P'_{ij}=
	\frac{1}{N}
	\begin{pmatrix} 
		0 & 0 & \frac{1}{\alpha_1\alpha_3} & 0\\\ 
		0 & 0 & 0 & \frac{1}{\alpha_2\alpha_4} \\
		0 & 0 & 0 & 0 \\
		0 & 0 & 0 & 0
	\end{pmatrix},
\end{equation}
all the remaining entries being fixed to 0. From the present matrix form, it easy to verify that all the eigenvalues of $P'_{ij}$, and thus all the eigenvalues of $P_{ij}$, vanish. 
Note that rewriting $P_{ij}$ in an orthonormal basis simplifies the search for its eigenvalues also in more complex cases where the connectivity vectors share several overlap directions. In those cases, a proper basis needs to be built starting from the connectivity vectors through a Gram-Schmidt orthonormalization process.

As a side note we observe that, even though $P_{ij}'$ (and thus $P_{ij}$) admits only vanishing eigenvalues, its rank is still equal to two. Indeed, the rank can be computed as $N$ minus the dimensionality of the kernel associated to $P_{ij}'$, defined by any vector $x$ obeying $P'x=0$. As $P'_{ij}$ contains $N-2$ empty rows, the last equations impose two independent contraints on the components of $x$. As a consequence, the dimensionality of the kernel equals $N-2$, and the rank is equal to two.

We turn to  responses that are obtained in presence of external inputs.
We examine the network dynamics in response to a normalized input $I$ which partially correlates with one of the left-connectivity vectors, here $n^{(1)}$:
\begin{equation}
	{I} = n^{(1)} \frac{\Sigma_{nI} }{\Sigma_n^2} +x\sqrt{\Sigma_I^2 - \frac{\Sigma_{nI}^2}{\Sigma_n^4}}.
\end{equation}

Similarly to the unit-rank case, we find that ${I}$ elicits a network response in the plane ${I} - m^{(1)}$. The overlap values are given by: 
\begin{equation}\label{eq:selective_two}
	\begin{split}
		&\kappa_1 =\Sigma_{nI}  \langle[\phi_i']\rangle\\
		&\kappa_2 = 0,
	\end{split}
\end{equation}
and they can be used  to close the mean-field equations together with the equation for the first ($\mu=0$) and second-order statistics. In the case of stationary states we have:
\begin{equation}
	\Delta_0 = g^2 \langle [\phi_i^2] \rangle + \Sigma_m^2 \left( \kappa_1^2 + \kappa_2^2 \right) + \Sigma_I^2.
\end{equation}
Similar arguments allow to derive the two equations needed for the chaotic states.

In order to assess the stability of the stationary states, we evaluate the position of the outliers in the stability eigenspectrum by computing the eigenvalues of $\mathcal{M}$ (Eq.~\ref{eq:M_rank2}).
In the case of orthogonal structures and correlated input patterns ${I}$, a little algebra reveals that all the $a$ values vanish, while we have:
\begin{equation}
	\begin{split}
		&b^1 = \frac{1}{2}\Sigma_{nI} \langle [\phi_i'']\rangle\\
		&b^2 = 0.
	\end{split}
\end{equation}

We conclude that the first and the last row of $\mathcal{M}$ always vanish. Furthermore, the second and the third rows are proportional one to the other. As a consequence, the stability analysis predicts at most one outlier eigenvalue, which is indeed observed in the spectrum (not shown). The outlier is negative, as the effect of introducing inputs in the direction of the left vector $n^{(1)}$ is to further stabilize the dynamics.
As it will be shown, more than one outlier can be observed in the case where the low-dimensional structure involves overlap directions.

\paragraph{Rank-two structures with internal pairwise overlap}

As a second case, we consider structured matrices where the two connectivity pairs $m^{(1)}$ and $n^{(1)}$, $m^{(2)}$ and $n^{(2)}$ share two different overlap directions, defined by vectors $y_1$ and $y_2$. We set:
\begin{equation}
	\begin{split}
		&m^{(1)} = \sqrt{\Sigma^2 - \rho_1^2} \: x_1 + \rho_1 y_1\\
		&m^{(2)} = \sqrt{\Sigma^2 - \rho_2^2} \: x_2 + \rho_2 y_2\\
		&n^{(1)} = \sqrt{\Sigma^2 - \rho_1^2} \: x_3 + \rho_1 y_1\\
		&n^{(2)} = \sqrt{\Sigma^2 - \rho_2^2} \: x_4 + \rho_2 y_2.
	\end{split}
\end{equation}
where $\Sigma^2$ is the variance of the connectivity vectors and $\rho_1^2$ and $\rho_2^2$ quantify the overlaps along the directions $y_1$ and $y_2$. 

By rotating $P_{ij}$ onto the orthonormal basis that can be built from $m^{(1)}$ and $m^{(2)}$ by orthogonalizing the left vectors $n^{(1)}$ and $n^{(2)}$, one can easily check that the two non-zero eigenvalues of $P_{ij}$ are given by $\lambda_1 = \rho_1^2$ and $\lambda_2 = \rho_2^2$. They correspond, respectively, to the two right-eigenvectors $m^{(1)}$ and $m^{(2)}$. In absence of external inputs, an instability is thus likely to occur in the direction of the $m^{(k)}$ vector which corresponds to the strongest overlap.

We specifically focus on the degenerate condition where the two overlaps are equally strong, $\rho_1=\rho_2=\rho$, and any combination of $m^{(1)}$ and $m^{(2)}$ is a right-eigenvector.
The mean-field equations for the first-order statistics read:
\begin{equation}
	\begin{split}
		& \kappa_1 = \rho^2 \kappa_1  \langle [\phi_i'] \rangle\\
		& \kappa_2 = \rho^2 \kappa_2  \langle [\phi_i'] \rangle.
	\end{split}
\end{equation}
Similarly to Eq.~\ref{eq:static_arbitrary}, the two equations admit a silent ($\kappa_1=\kappa_2=0$) and a non-trivial state, determined by two identical conditions which read: 
\begin{equation}
	1 = \rho^2  \langle [\phi_i '(0, \Delta_0)] \rangle.
\end{equation}
The equation above determines the value of $\Delta_0$. Note that the non-trivial state exists only for $\rho>1$.

A second condition is imposed by the equation for the second-order momentum which reads, for stationary solutions:
\begin{equation}\label{eq:delta0_ring}
	\Delta_0 = g^2 \langle [\phi_i^2] \rangle + \Sigma^2 \left( \kappa_1^2 + \kappa_2^2 \right).
\end{equation}
As the value of $\Delta_0$ is fixed, the mean-field set of equations fixes only the sum $\kappa_1^2 + \kappa_2^2$, but not each single component. The mean-field thus returns a one-dimensional continuum of solutions, the shape of which resembles a ring of radius $\sqrt{\kappa_1^2 + \kappa_2^2}$ in the $m^{(1)}-m^{(2)}$ plane (see Fig.~S5 	\textsf{D-E}).
Similarly to the unit-rank case, the value of the radius can be computed explicitly by solving numerically the two mean-field equations (three in the case of chaotic regimes), and depends on the relative magnitude of $\rho^2$ compared to $g$ (Fig.~S5 \textsf{F}). Highly disordered connectivities have the usual effect of suppressing non-trivial structured solutions in favour of homogeneous and unstructured states. For sufficiently high $g$ values, furthermore, structured solution can display chaotic dynamics (Fig.~S5 \textsf{E} and Fig.~S5 \textsf{F}, red).

A linear stability analysis reveals that the one-dimensional solution consists of a continuous set of marginally stable states.
Similarly to the orthogonal vectors case, the position of the outliers in the eigenspectra of $S_{ij}$ can be evaluated by computing the reduced stability matrix $\mathcal{M}$, which reads:
\begin{equation}\label{eq:M_ring}
	\mathcal{M} = 
	\begin{pmatrix} 0 & 0 & 0 & 0\\ 
		2 g^2 \langle [\phi_i\phi_i']\rangle & g^2\left\{\langle [\phi_i'^2]\rangle + \langle [\phi_i\phi_i'']\rangle \right\} & 2\Sigma_m^2\kappa_1^{0} & 2\Sigma_m^2\kappa_2^{0}\\
		2 b_1 g^2 \langle [\phi_i\phi_i']\rangle & b_1 g^2\left\{\langle [\phi_i'^2]\rangle + \langle [\phi_i\phi_i'']\rangle \right\} & 2 b_1  \Sigma_m^2\kappa_1^{0} + a_{11} &  2 b_1  \Sigma_m^2\kappa_2^{0}  \\
		2 b_2 g^2 \langle [\phi_i\phi_i']\rangle & b_2 g^2\left\{\langle [\phi_i'^2]\rangle + \langle [\phi_i\phi_i'']\rangle \right\} & 2 b_2  \Sigma_m^2\kappa_1^{0}  &  2 b_2  \Sigma_m^2\kappa_2^{0} + a_{22} \\
	\end{pmatrix} ,
\end{equation}
with:
\begin{equation}
	\begin{split}
		&a_{11} = \rho^2 \langle [\phi_i'] \rangle\\
		&b_1 = \frac{1}{2} \rho^2 {\kappa}_1^{0} \langle [\phi_i'''] \rangle
	\end{split}
\end{equation}
and
\begin{equation}
	\begin{split}
		&a_{22} = \rho^2 \langle [\phi_i'] \rangle\\
		&b_2 = \frac{1}{2} \rho^2 {\kappa}_2^{0} \langle [\phi_i'''] \rangle.
	\end{split}
\end{equation}

As shown in Fig.~S5 \textsf{G}, diagonalizing the stability matrix $\mathcal{M}$ returns the values of two distinct outlier eigenvalues. The third non-zero eigenvalue of $\mathcal{M}$ lies instead systematically inside the compact component of the spectrum, and corresponds to an average measure of the time scales inherited by the random modes. One of the two outliers is tuned exactly to the stability boundary for every value of the parameters which generate a ring solution. This marginally stable eigenvalue is responsible for the slow dynamical time scales which are observed in numerical simulations of the network activity (Fig.~S5 	\textsf{D-E}). 

The  DMF predictions  formally  hold in  the  limit of  infinite-size
networks; in simulations of finite-size networks, the dynamics instead
always converge  on a small  number of equilibrium  spontaneous states
located  on  the   ring  (see  Fig.~S5  \textsf{D-E}).   The
equilibrium  reached  in  a  given  situation  is  determined  by  the
corresponding realization of the  random part of the connectivity, and
the  initial   conditions.   Different  realizations   of  the  random
connectivity lead  to different equilibrium states,  which all however
lie on the predicted ring (see Fig.~S5 \textsf{D-E}).  For a
given  realization  of  the  random connectivity,  transient  dynamics
moreover  show a  clear signature  of the  ring structure.  Indeed the
points on the ring are close  to stable and form a slow manifold.  The
convergence to  the equilibrium activity  is therefore very  slow, and
the temporal dynamics explore the ring structure.

We next examine how the structured, ring-shaped solution is perturbed by the injection of external input patterns.

We consider  an input pattern ${I}$ of variance $\Sigma_I^2$. 
When ${I}$ does not share any overlap direction with the left vectors $n^{(1)}$ and $n^{(2)}$, the mean-field equations are affected solely by an extra term $\Sigma_I$ which needs to be included in the equation for the second-order statistics (Eq.~\ref{eq:delta0_ring}). As the equations for the first-order statistics do not change, the one-dimensional degeneracy of the solution persists. The extra term $\Sigma_I^2$  however decreases the value of the radius of the ring.

When the input contains a component which overlaps with one or both left vectors $n^{(1)}$ and $n^{(2)}$, the degeneracy in the two equations for $\kappa_1$ and $\kappa_2$ is broken. As a consequence, the one-dimensional solution collapses onto a unique stable point.
Consider for example an input pattern of the form:
\begin{equation}\label{eq:selective_input_ring}
	{I} =\Sigma_I \left(\sqrt{1-\alpha} \: x_3 + \sqrt{\alpha} \: x_4\right).
\end{equation}
The equations for the first order become:
\begin{equation}
	\begin{split}
		& \kappa_1 = \left(\rho^2 \kappa_1 + \Sigma_I \sqrt{1-\alpha} \sqrt{\Sigma^2 - \rho^2}\right)  \langle [\phi_i'] \rangle\\
		& \kappa_2 = \left(\rho^2 \kappa_2 + \Sigma_I \sqrt{\alpha} \sqrt{\Sigma^2 - \rho^2}\right)   \langle [\phi_i'] \rangle
	\end{split}
\end{equation}
or, alternatively:
\begin{equation}
	\begin{split}
		\kappa_1 = \frac{ \Sigma_I  \sqrt{1-\alpha} \sqrt{\Sigma^2 - \rho^2} \langle [\phi_i']\rangle } {1-\rho^2 \langle [\phi_i'] \rangle }\\
		\kappa_2 = \frac{ \Sigma_I  \sqrt{\alpha} \sqrt{\Sigma^2 - \rho^2} \langle [\phi_i']\rangle } {1-\rho^2 \langle [\phi_i'] \rangle }.
	\end{split}
\end{equation}
The values of $\kappa_1$ and $\kappa_2$ are thus uniquely specified, and can be computed by iterating the two equations together with the expression for the second-order statistics:
\begin{equation}
	\Delta_0 = g^2 \langle [\phi_i^2] \rangle + \Sigma^2 \left( \kappa_1^2 + \kappa_2^2 \right)+\Sigma_I^2.
\end{equation}

In a similar way, the presence of correlated external inputs affect the values of the entries of the reduced stability matrix $\mathcal{M}$:
\begin{equation}
	\begin{split}
		&b_1 = \frac{1}{2} \left( \rho^2 {\kappa}_1^{0}  +  \Sigma_I  \sqrt{1-\alpha} \sqrt{\Sigma^2 - \rho^2} \right) \langle [\phi'''_i] \rangle\\
		&b_2 = \frac{1}{2} \left( \rho^2 {\kappa}_2^{0}  +  \Sigma_I  \sqrt{\alpha} \sqrt{\Sigma^2 - \rho^2} \right) \langle [\phi'''_i] \rangle.
	\end{split}
\end{equation}

In Fig.~S5 \textsf{H-I}, we focus on the case of an external input pattern aligned with $x_3$ (and thus $n^{(1)}$) . We fix $\alpha=0$, that implies $\kappa_2 = 0$.

Solving the mean-field equations reveals that, according to the strength of the input $\Sigma_I$, one or three fixed points exist.
When the input is weak with respect to the structure overlap $\rho^2$, two fixed points appear in the proximity of the ring, along the direction defined by the axis $\kappa^2 = 0$ (Fig.~S5 \textsf{H} top). In particular, when ${I}$ positively correlates with  $n^{(1)}$, only the fixed point with positive value of $\kappa_1$ gets stabilized. The remaining two solutions are characterized by one outlier eigenvalue which lays above the instability boundary, and are thus unstable.
On the other hand, when the input is sufficiently strong, solely the stable fixed point survives (Fig.~S5 \textsf{H} bottom). Activity is then robustly projected in the direction defined by the right vector $m^{(1)}$.

\paragraph{Rank-two structures for oscillations}

We finally consider the following configuration:
\begin{equation}
	\begin{split}
		&m^{(1)} = \alpha x_1 + \rho y_1\\
		&m^{(2)} = \alpha x_2 + \rho y_2\\
		&n^{(1)} = \alpha x_3 +  \rho y_2 + \gamma\rho y_1\\
		&n^{(2)} = \alpha x_4 - \rho y_1,
	\end{split}
\end{equation}
where the right- and the left-connectivity vectors share two cross-overlap directions $y_1$ and $y_2$. Note that the vectors in one of the two pairs, $m^{(1)}-n^{(2)}$, are negatively correlated. 
A second overlap is introduced internally to the $m^{(1)}-n^{(1)}$ pair, and scales with the parameter $\gamma$. The directions $x_j$, with $k=1,...,4$, represent uncorrelated terms. Note that different values of $\alpha$ affect quantitatively the network statistics, but they do not change the phase diagram in Fig.~S8  \textsf{A}.

By rotating $P_{ij}$ on a proper orthonormal basis, one can check that its eigenvalues are given by:
\begin{equation}\label{eq:lambda_osc}
	\lambda_{\pm} = \frac{\gamma \rho^2}{2} \left(1 \pm \sqrt{1-\frac{4}{\gamma^2}}\right),
\end{equation}
and they are complex conjugate for $\gamma<2$. In this case, the internal overlap $\gamma$ has the effect of returning a non-vanishing real part. The two complex conjugate eigenvectors are given by:
\begin{equation}\label{eq:complex_ep}
	e^{\pm} = \left(-\frac{\gamma}{2}m^{(1)} + m^{(2)}\right) \pm i \sqrt{\left| 1 - \frac{4}{\gamma^2} \right|} m^{(1)}.
\end{equation}

The eigenspectrum of $J_{ij} = g\chi_{ij} + P_{ij}$ inherits the pair of non-zero eigenvalues of $P_{ij}$. When $g<1$ and $\gamma<2$, the trivial fixed point thus undergoes a Hopf bifurcation when the real part of $\lambda$ crosses unity (Fig.~S8 \textsf{A}, blue). 
When $\gamma>2$, instead, the two eigenvalues are real. One bifurcation to bistable stationary activity occurs when the largest eigenvalue $\lambda_+$ crosses unity (Fig.~S8 \textsf{A}, gray).

On the boundary corresponding to the Hopf bifurcation, the frequency of instability $\omega_H$ is determined by the imaginary part of Eq.~\ref{eq:lambda_osc}. 
At the instability, the oscillatory activity of unit $i$ can be represented as a point on the complex plane. Since close to the bifurcation we can write:
\begin{equation}\label{eq:mui_complex}
	\mu_i = e^+_i e ^{i\omega_Ht} + c.c. \: ,
\end{equation}
its coordinates are given by the real and the imaginary part of the $i$th component of the complex eigenvector $e^+$.
The phase of oscillation can then be computed as the angle defined by this point with respect to the real axis.
Note that the disorder in the elements of the eigenvector $e^+$, which is inherited by the random distribution of the entries of the connectivity vectors $m^{(1)}$ and $m^{(2)}$, tends to favor a broad distribution of phases across the population.

In the limit case where the real and the imaginary parts of the complex amplitude of the oscillators are randomly and independently distributed, the population response resembles a circular cloud in the complex plane. In this case, the phase distribution across the population is flat. Note that a completely flat phase distribution can be obtained for arbitrary frequency values by adopting a rank-two structure where an internal overlap of magnitude $\gamma \rho^2$ exists between vectors $m^{(2)}$ and $n^{(2)}$ as well.

In the present case, for every finite value of $\gamma$, the real and the imaginary part of $e^+_i$ are anti-correlated through $m^{(1)}$ (Eq.~\ref{eq:complex_ep}). 
Correlations tend to align the network response on two main and opposite phases, as shown in the phase histograms of Fig.~S8 \textsf{C-D}. The distribution of phases becomes sharper and sharper in the $\gamma \rightarrow 2$ limit, as the distribution in the complex plane collapses on the real axis.

The phase distribution across the population is reflected in the shape of the closed orbit defined by activity on the $m^{(1)}-m^{(2)}$ plane, whose components are given by $\kappa_1$ and $\kappa_2$. The phase of the oscillations in $\kappa_1$ (resp. $\kappa_2$) can be computed by projecting the eigenvector $e^+$ on the right-connectivity vectors $n^{(1)}$ and $n^{(2)}$:
\begin{equation}
	\begin{split}
		\kappa_1 = |\kappa_1| e ^{i(\Phi_1 + \omega_Ht)} + c.c. &= \langle n_i^{(1)} [\phi_i] \rangle \\
		\kappa_2 = |\kappa_2| e ^{i(\Phi_2 + \omega_Ht)} + c.c. &= \langle n_i^{(2)} [\phi_i] \rangle \\
	\end{split}
\end{equation}

By using Eqs.~\ref{eq:complex_ep} and \ref{eq:mui_complex} we get, in the linear regime:
\begin{equation}
	\begin{split}
		\kappa_1 & =  \left[ \langle n_i^{(1)} m_i^{(2)} \rangle - \frac{\gamma}{2} \langle n_i^{(1)} m_i^{(1)} \rangle + i \langle n_i^{(1)} m_i^{(1)} \rangle \sqrt{\left| 1 - \frac{4}{\gamma^2} \right|} \right]  e ^{i\omega_Ht} + c.c.\\
		& = \left[ \rho^2 \left(1 - \frac{\gamma^2}{2} \right) + i \gamma \rho^2 \sqrt{\left| 1 - \frac{4}{\gamma^2} \right|} \right] e ^{i\omega_Ht} + c.c.
	\end{split}
\end{equation}
while:
\begin{equation}
	\begin{split}
		\kappa_2 & =  \left[ \langle n_i^{(2)} m_i^{(2)} \rangle - \frac{\gamma}{2} \langle n_i^{(2)} m_i^{(1)} \rangle + i \langle n_i^{(2)} m_i^{(1)} \rangle \sqrt{\left| 1 - \frac{4}{\gamma^2} \right|} \right]  e ^{i\omega_Ht} + c.c.\\
		& = \left[ \rho^2 \frac{\gamma}{2} - i \rho^2 \sqrt{\left| 1 - \frac{4}{\gamma^2} \right|} \right] e ^{i\omega_Ht} + c.c.
	\end{split}
\end{equation}
When $\gamma$ is close to 2, the complex amplitudes of $\kappa_1$ and $\kappa_2$ vanish. However, their real parts have different signs. We thus get: $\Phi_2 = 0$, $\Phi_1 = \pi$.
As a consequence, at large $\gamma$ values, the oscillatory activity in $\kappa_1$ and $\kappa_2$ tends to be strongly in anti-phase. 

Stationary solutions can be instead easily analyzed with the standard mean-field approach.
The equations for the first order statistics read:
\begin{equation}
	\begin{split}
		& \kappa^1 = (\gamma\rho^2 \kappa^1 + \rho^2 \kappa^2)  \langle [\phi_i'] \rangle\\
		& \kappa^2 = - \rho^2 \kappa^1 \langle [\phi_i'] \rangle.
	\end{split}
\end{equation}
The two equations can be combined together to give the following condition on $\langle [\phi_i'] \rangle$, which in turn determines the value of $\Delta_0$:
\begin{equation}
	\rho^4  \langle [\phi_i'] \rangle^2 - \gamma \rho^2  \langle [\phi_i'] \rangle + 1 = 0.
\end{equation}
The mean-field equations thus admit two solutions, given by:
\begin{equation}
	\langle [\phi_i'] \rangle_{\pm}= \frac{\gamma}{2\rho^2} \left(1+\pm \sqrt{1 - \frac{4}{\gamma^2}}\right)
\end{equation}
which, similarly to Eq.~\ref{eq:lambda_osc}, take real values for $\gamma>2$. Because of the constraints on the sigmoidal activation function, the mean-field solutions are acceptable only if $|\langle [\phi_i'] \rangle|<1$. 
As it can be easily checked, the condition $\langle [\phi_i'] \rangle_{-}<1$ coincides with imposing $\lambda_+>1$. We conclude that two stationary solutions exist above the instability boundary of the trivial fixed point (Fig.~S8 \textsf{A}, gray).
A second pair of solutions appears for $\langle [\phi_i'] \rangle_{+}<1$, which coincide with $\lambda_->1$ (Fig.~S8 \textsf{A}, dashed), where the second outlier of $J_{ij}$ becomes unstable. This second pair of solutions is however always dynamically unstable, as it can be checked by evaluating the outliers of their stability matrix through Eq.~\ref{eq:M_rank2}. The coefficients of the reduced matrix $\mathcal{M}$ read:
\begin{equation}
	\begin{split}
		&a_{11} = \gamma \rho^2 \langle [\phi_i'] \rangle\\
		&a_{12} =  \rho^2 \langle [\phi_i'] \rangle\\
		&b_1 = \frac{1}{2} \rho^2 ({\kappa}^{20} + \gamma {\kappa}^{10}) \langle [\phi_i''] \rangle
	\end{split}
\end{equation}
and
\begin{equation}
	\begin{split}
		&a_{21} = - \rho^2 \langle [\phi'] \rangle\\
		&a_{22} = 0\\
		&b_2 = - \frac{1}{2} \rho^2 {\kappa}^{10} \langle [\phi''] \rangle.
	\end{split}
\end{equation}

On the phase diagram boundary corresponding to $\gamma = 2$, the stable and the unstable pair of stationary solutions annihilate and disappear. At slightly smaller values of $\gamma$ ($\gamma\lesssim 2$), the network develops highly non-linear and slow oscillations which can be thought of as smooth jumps between the two annihilation points (Fig.~S8 \textsf{D}).

\subsection{Implementation of computational tasks}

\subsubsection{Go-Nogo discrimination}

Here we describe and analyze the unit-rank implementation of the Go-Nogo discrimination task  (Fig.~\ref{fig:p3}).

The network receives inputs specified by  $N$-dimensional
vectors $I^{k}$. In every trial, the input vector coincides with one among the two vectors $I^A$ and $I^B$, representing respectively the Go and the Nogo stimuli.
The components of the two input patterns
are generated independently from  a Gaussian distribution of mean zero
and  variance  $\Sigma_I$.  As   the  components  of  the  inputs  are
uncorrelated,  the  two  vectors  are
mutually orthogonal in the limit of large $N$.

The network activity is  read-out linearly through a vector $w$ generated from a Gaussian distribution of mean zero and variance $\Sigma_w^2$. The readout value is given by:
\begin{equation}\label{eq:readout_general}
	z =\frac{1}{N}\sum_{i=1}^N w_i \phi(x_i).
\end{equation}

We fix the connectivity vectors $m$ and $n$ such that: (i) the readout is selective, i.e.~$z\neq 0$ if the input is $I^A$ and $z=0$ for the input $I^B$; (ii) the readout is specific to the vector $w$, i.e.~it is zero for any readout vector uncorrelated with $w$.
The simplest network architecture which satisfies these requirements is given by:
\begin{equation}
	\begin{split}
		&m=w\\
		&n=I^A,
	\end{split}
\end{equation}
i.e.~the right-connectivity vector $m$ corresponds to the readout vector, and the left-connectivity vector corresponds to the  preferred stimulus $I^A$.

The response of the network can be analyzed by referring to the stationary and chaotic solutions of Eq.~\ref{eq:mf_input}. In the case analyzed here, the connectivity vectors have no overlap direction, so we set $M_m = M_n = M_I = \Sigma_{mI} = 0$, which implies $\mu=0$.
The first-order network statistics are  determined by the overlap $\Sigma_{nI}$ between the left-connectivity vector and the input vector. As the left-connectivity is given by $I^A$, $\Sigma_{nI}$ is the overlap between the current input pattern $I$ and the preferred pattern $I^A$, and it takes values $\Sigma_{nI} = \Sigma_I^2$ during the Go stimulus presentation and $\Sigma_{nI} = 0$ otherwise. From Eq.~\ref{eq:kappa_inp} we have:
\begin{equation}
	\begin{split}
		\kappa &= \langle n_i[\phi_i] \rangle\\
		&= \langle I_i^A [\phi_i] \rangle.
	\end{split}
\end{equation}
As a consequence, when the Go stimulus is presented ($I=I^A$):
\begin{equation}
	\kappa =\Sigma_I^2 \langle [\phi_i'] \rangle,
\end{equation}
while the first-order statistics $\kappa$ vanishes in response to any orthogonal pattern $I^B$.

When activity is read out by the specific decoding vector $w$, the readout value is:
\begin{equation}\label{eq:readout}
	\begin{split}
		z & = \langle w_i [\phi_i] \rangle \\
		& = \langle w_i \int \mathcal{D}z \phi(m_i \kappa + I_i + \sqrt{\Delta_0^I} z) \rangle\\
		& = \langle w_i \int \mathcal{D}z \phi(w_i \kappa + I_i + \sqrt{\Delta_0^I} z) \rangle\\
		& = \kappa \Sigma_w^2 \langle [\phi_i'] \rangle,
	\end{split}
\end{equation}
while we trivially obtain $z=0$ for any decoding set orthogonal to both connectivity vectors $m$ and $n$.

In Fig.~\ref{fig:p3} \textsf{C}, we display the transient dynamics predicted by the mean-field theory within the $m-I$ plane. In order to compute the predicted trajectory, we use Eqs.~\ref{eq:I_transient} and \ref{eq:kappa_transient}, where the slowest time-scale of $\kappa$ is computed by diagonalizing the reduced stability matrix in Eq.~\ref{eq:input_M}.

In Fig.~\ref{fig:p3} \textsf{G}, we test the generalization properties of a network which responds to two Go patterns $I^A_1$ and $I^A_2$. We examine the response to a normalized mixture input defined as:
\begin{equation}
	I = \sqrt{\alpha} I^A_1 + \sqrt{1-\alpha}I^A_2,
\end{equation}
so that the variance of the total input is fixed and equal to $\Sigma_I^2$. We set $n=I_1^A+I_2^A$, so that the equation for the first-order statistics reads:
\begin{equation}
	\begin{split}
		\kappa & = \langle I_{1i}^A[\phi_i] \rangle + \langle I_{2i}^A[\phi_i] \rangle\\
		& = (\sqrt{\alpha} +\sqrt{1-\alpha}) \Sigma_I^2 \langle [\phi_i'] \rangle.
	\end{split}
\end{equation}

\subsubsection{Detection of a continuous noisy stimulus}

In Fig.~\ref{fig:p4}, we construct a network model which performs a Go-Nogo detection task on a one-dimensional continuous stimulus.

The stimulus consists of an input of time-varying amplitude $c(t)I$. As in Fig.~\ref{fig:p3}, the input direction $I$ is a centered Gaussian vector of variance $\Sigma_I^2$. The strength value $c(t)$ includes a stationary component $\bar{c}$ together with additive white noise of standard deviation $\sigma$. Less importantly, we include in the input an orthogonal component of quenched noise of unitary variance. The network output is defined at the level of an orthogonal readout as in Eq.~\ref{eq:readout_general}, and the task consists in responding to the stimulus when the strength of the input $c$ is larger than a given threshold.

We obtain highly non-linear readout responses by considering non-vanishing overlaps between the connectivity vectors $m$ and $n$. The simplest setup consists of taking:
\begin{equation}  \label{eq:threshold_overlaps}
	\begin{split}
		&m = w + \rho_m y \\
		&n = I+ \rho_n y,  
	\end{split}
\end{equation}
where $y$ is a standard gaussian vector which defines a direction common to $m$ and $n$, but orthogonal both to $w$ and $I$. 

For this configuration, as in Eq.~\ref{eq:kappa_inp}, the mean-field equation for the first-order statistics includes two terms, generated respectively by the input and the rank-one structure:
\begin{equation}
	\kappa = (\rho_m\rho_n \kappa + \bar{c} \: \Sigma_I^2)  \langle [\phi_i'] \rangle.
\end{equation}

Before the stimulus presentation ($\bar{c}=0$, $\sigma=0$), the structure overlap $\rho_m\rho_n$  is strong enough to generate two bistable solutions (Fig.~\ref{fig:p1}). We set the negative $\kappa$ solution to represent the Nogo condition, and we initialize the network in this state. To have a zero output in this condition, we add an offset to the readout.

When an input along the preferred direction is presented ($\bar{c}>0$), two asymmetric solutions exist only when the strength of the input $\bar{c}$ is not too large (Fig.~\ref{fig:p2}  \textsf{D} right). 
When the correlation $\bar{c}$ is large, instead, only the positive branch of the solution is retrieved (Fig.~\ref{fig:p2}  \textsf{D} right). As a consequence, the average value of $\kappa$ (and thus the readout signal) jumps to positive values, which define the Go output condition. 

More generally, in order to compute the network performance (Fig.~\ref{fig:p4} \textsf{G}), the network is said to respond to the stimulus if the readout $z$ at the end of the stimulus presentation takes values larger than one half of the readout value expected for the upper state.

The threshold value for $\bar{c}$ at which the bistability disappears  is mostly determined by the strength of the structure overlap,  but depends also the input and readout parameters $\Sigma_I$ and $\Sigma_w$.
For practical purposes, in order to obtain the model implementation illustrated in Fig.~\ref{fig:p4}, we first fix the values of $\Sigma_I=1.2$, $\Sigma_w=1.2$ and $\rho_n=2$. We then tune the value of $\rho_m$ in order to obtain a threshold value for $c$ close to 0.5. This leads to $\rho_m=2$.

In Fig.~\ref{fig:p4} \textsf{F} we vary $\rho_m$ and we show that the value of the threshold decreases to zero as the structure strength $\rho_m\rho_n$ decreases from its original value ($\rho_m\rho_n=4$). Rank-one structures characterized by different strengths thus correspond to different thresholds, but also induce different dynamical time-scales in the network. As a rough estimate of this time-scale, we compute the inverse of the outlier eigenvalue from the stability matrix of the fixed point corresponding to the Go resting state ($\bar{c}=0$). The value of the outlier can be computed from the linearized mean-field equations (Eq.~\ref{eq:M}). We show that arbitrarily large time-scales are only obtained by decreasing the value of the structure strength to the critical point where the two bistable branches of the solution emerge from the trivial fixed point. In this configuration, the threshold detected by the network is arbitrarily small.

\subsubsection{Contextual modulation of threshold value}

Here we briefly illustrate how the threshold of detection can be controlled by an additional modulatory input (Fig.~\ref{fig:p5} B). Modulatory inputs are used in Figs.~\ref{fig:p5} and \ref{fig:p6} to implement more complex tasks which require context-dependent responses to stimuli.
Any input direction which overlaps with the left-connectivity vector $n$ and is orthogonal to the stimulus axis $I$ can serve as modulatory input. For simplicity, we consider modulatory inputs which are aligned with the overlap direction $y$ (see Eq.~\ref{eq:threshold_overlaps}). The total external input to the network contains the modulatory component $\gamma y$ together with the stimulus term $c(t)I$, where  $\gamma$ is a scalar which controls the strength of the modulation. The mean-field equation for the first-order statistics reads:
\begin{equation}\label{eq:kappa_modulation}
    \kappa = (\rho_m\rho_n \kappa + \rho_n \gamma + \bar{c} \: \Sigma_I^2)  \langle [\phi_i'] \rangle.
\end{equation}

Eq.~\ref{eq:kappa_modulation} indicates that the modulatory component of the input acts as a constant offset to the stimulus strength. Its net effect is to shift the response curve of the network along the $x$ axis (Fig.~\ref{fig:p5} B) by an amount directly regulated by the parameter $\gamma$. Varying $\gamma$ thus results in network models which detect variable threshold values.

\subsubsection{Rank-two structures for context-dependent computations}

Here  we  provide  details  on  the rank-two  implementation  of  the
context-dependent   tasks. The same model has been used for both tasks in Figs.~\ref{fig:p5} and  \ref{fig:p6}. 

The  stimuli   consist  of
combinations  of two  different features  $A$  and $B$  that correspond  to
inputs along two directions $I^{A}$  and $ I^{B}$, generated as Gaussian random vectors of variance $\Sigma_I^2$. Contextual cues are
represented  as additional  inputs along  directions $I_{ctx  A}$ and
$I_{ctx B}$ of unit variance. The total input pattern to the network on a given trial is therefore given by:
\begin{equation}
	I(t) = c_A(t) I^A + c_B(t)I^B + \gamma_A I_{ctx A} + \gamma_B I_{ctx B} .
\end{equation}
The values $c_A$ and $c_B$ express the strength of the stimulus along the two feature directions. They are given by the sum of stationary average values ($\bar{c}_A$, $\bar{c}_B$), and temporary fluctuating components generated from  independent realizations of white noise with standard deviation $\sigma$. In the simple discrimination version of the task (Fig.~\ref{fig:p5}), inputs are noise-free ($\sigma=0$) and consist of a single feature in each trial ($\bar{c}_A=1$ and $\bar{c}_B=0$ or vice versa). In the evidence integration version of the task (Fig.~\ref{fig:p6}), inputs are noisy ($\sigma>0$) and include non-zero average components along both feature directions.
Finally, the parameters $\gamma_A$ and $\gamma_B$ control the two modulatory inputs which are taken in the directions defined by  $I_{ctx A}$ and $I_{ctx B}$.

In order to implement context-dependent computations,  we define a unique readout signal $z(t)$ by using a common readout set $w$ of unit variance (Eq.~\ref{eq:readout_general}), to which we add an offset so that the baseline Nogo output is set to zero.  The network is said to respond to the stimulus if the value of the total readout at the end of the stimulus presentation takes values larger than one half of the largest predicted value for the upper state.

The rank-two connectivity matrix we consider is given by:
\begin{equation}\label{eq:structure_context}
	\begin{split}
		& m^{(1)} = y_A + \rho_m I_{ctx A}  + \beta_m w\\
		& n^{(1)} = I^A + \rho_n I_{ctx A}  + \beta_n w \\
		& m^{(2)} = y_B + \rho_m I_{ctx B}  + \beta_m w \\
		& n^{(2)} = I^B + \rho_n I_{ctx B}  + \beta_n w,
	\end{split}
\end{equation}
where vectors $y_A$ and $y_B$ represent the orthogonal components of the right-connectivity vectors and are generated as Gaussian vectors of fixed variance (for simplicity, we set $\Sigma_y = \Sigma_I$).

For our choice of the parameters, the network solves the two different tasks by relying on the strongly non-linear responses generated by the interplay between the recurrent connectivity and the feed-forward inputs (details  given below). 

For weak input values, the network dynamics is characterized by two stable attractors (Fig.~\ref{fig:p6} \textsf{F}). 
As in Fig.~\ref{fig:p4}, we initialize the network in the state characterized by negative $\kappa_1$ and $\kappa_2$ values before the stimulus presentation. This dynamical attractor corresponds to the Nogo state. For strong input strengths, the network can jump to the Go state, defined as the stable attractor characterized by positive $\kappa_1$ and $\kappa_2$ values.

The rank-two connectivity matrix has been designed as an extension of the unit-rank recurrent connectivity employed in Fig.~\ref{fig:p4}.
We started by setting:
\begin{equation}\label{eq:structure_context_first}
	\begin{split}
		& m^{(1)} = y_A + \rho_m I_{ctx A} \\
		& n^{(1)} = I^A + \rho_n I_{ctx A} \\
		& m^{(2)} = y_B + \rho_m I_{ctx B} \\
		& n^{(2)} = I^B + \rho_n I_{ctx B} .
	\end{split}
\end{equation}
Note that, because the only overlap directions ($I_{ctx A} $ and $I_{ctx B}$) are internal to the $m^{(1)}-n^{(1)}$ and $m^{(1)}-n^{(1)}$ pairs, Eq.~\ref{eq:structure_context_first} describes a rank-two structure which generates a continuous ring attractor as in Fig.~S5 \textsf{D-I} (gray circles in Fig.~\ref{fig:p6} \textsf{F}).

The readout $z(t)$ should detect the presence of both stimuli directions. As a consequence, it should be sensitive to both overlap values $\kappa_1$ and $\kappa_2$. For this reason, we introduce a common term in the four connectivity vectors that is aligned to the common readout (Eq.~\ref{eq:structure_context}).

Introducing a common overlap direction has the effect of destabilizing the continuous attractor dynamics along the direction $\kappa_1=\kappa_2$ (dashed line in Fig.~\ref{fig:p6} \textsf{F}), where two stable and symmetric fixed points are generated.
The equations for the first-order spontaneous dynamics read indeed:
\begin{equation}
	\begin{split}
		\kappa_1 = \langle n^{(1)}[\phi_i]\rangle = \rho_m\rho_n \kappa_1 \langle [\phi_i'] \rangle +  \beta_m\beta_n (\kappa_1 + \kappa_2)\langle [\phi_i'] \rangle\\
		\kappa_2 =  \langle n^{(2)}[\phi_i]\rangle = \rho_m\rho_n \kappa_2 \langle [\phi_i'] \rangle +  \beta_m\beta_n (\kappa_1 + \kappa_2)\langle [\phi_i'] \rangle
	\end{split}
\end{equation}
from which the value of $\kappa_1=\kappa_2=\bar{\kappa}$ can be derived by dividing and multiplying together the two equations. The final readout signal contains a contribution from both first-order statistics:
\begin{equation}\label{eq:context_readout}
	z(t) = \langle w_i [\phi_i] \rangle = \beta_m  (\kappa_1 + \kappa_2) \langle [\phi_i'] \rangle.
\end{equation}

The input-driven dynamics of the network are determined by the interplay between the structure strength and the contextual and  stimulus inputs.
Crucially, the modulatory inputs along $I_{ctx A}$ and  $I_{ctx B}$ are used to gate a context-dependent response. Similarly to Fig.~\ref{fig:p5} \textsf{B}, a strong and negative gating variable along $I_{ctx A}$ can completely suppress the response to stimulus $I^A$, so that the readout signal is left free to respond to $I^B$. 

The overall effects of the inputs on the dynamics can  be quantified by solving the mean-field equations.
For the first-order statistics, we obtain:
\begin{equation}
	\begin{split}
		\kappa_1 =  \langle [\phi_i'] \rangle  \left\{ \rho_m\rho_n \kappa_1 +  \beta_m\beta_n (\kappa_1 + \kappa_2) + \bar{c}_A\Sigma_I^2 + \rho_n \gamma_A \right\}  \\
		\kappa_2 =  \langle [\phi_i'] \rangle  \left\{ \rho_m\rho_n \kappa_2 +  \beta_m\beta_n (\kappa_1 + \kappa_2) + \bar{c}_B\Sigma_I^2 + \rho_n \gamma_B \right\} 
	\end{split}
\end{equation}
while the second-order gives, in the case of stationary regimes:
\begin{equation}
	\Delta_0 = g^2  \langle [\phi_i^2] \rangle + \Sigma_w^2 (\kappa_1^2 + \kappa_2^2) + \beta_m^2 (\kappa_1^2 + \kappa_2^2) + \Sigma_I^2 (\bar{c}_A^2 + \bar{c}_B^2) + (\rho_m \kappa_1 + \gamma_A)^2 + (\rho_m \kappa_2 + \gamma_B)^2.
\end{equation}

Fig.~S5 \textsf{L-M} displays the values of the first-order statistics and the readout response in the two contexts. 
Note that, when the response to $I^A$ (resp.~$I^B$) is blocked at the level of the readout, the relative first-order statistics $\kappa_1$ (resp. $\kappa_2$) does not vanish, but actively contributes to the final network response.

The average activation variable of single neurons contains entangled contributions from the main directions of the dynamics, which are inherited both from the external inputs and the recurrent architecture:
\begin{equation}
	\begin{split}
		\mu_i = [x_i] = & (y_{A,i} + \rho_m I_{ctx A,i} + \beta_m w_i ) \kappa_1 +  (y_{B,i} + \rho_m I_{ctx B,i} + \beta_m w_i ) \kappa_2 \\
		& + \bar{c}_A I^A_i + \bar{c}_B I^B_i + \gamma_1 I_{ctx A,i} + \gamma_2 I_{ctx B,i}.
	\end{split}
\end{equation}
In Figs.~\ref{fig:p5} \textsf{E} and \ref{fig:p6} \textsf{D}, we project the averaged activation $\mu_i$ in the directions that are more salient to the task. The projection along $w$, which reflects the output decision, is proportional to the readout value (Eq.~\ref{eq:context_readout}). The input signals affect instead the average activity through the values of $\kappa_1$ and $\kappa_2$, but can be also read out directly along the input directions.
Note that the projection on the input direction $I^A$ (resp.~$I^B$) is proportional to the signal $\bar{c}_A$ (resp. $\bar{c}_B$) regardless of the configuration of the modulatory inputs selecting one input channel or the other.

In  practical terms, in order to obtain the network architecture that has been used in Figs.~\ref{fig:p5} and \ref{fig:p6}, we fixed the parameters step by step. 
We first considered input patterns only along $I^A$ ($\bar{c}_B=0$), and we fixed two arbitrary values of $\beta_m$ and $\beta_n$. In particular, we considered intermediate values of $\beta$. Large values of $\beta$ tend to return large activity variance, which requires  evaluating with very high precision the Gaussian integrals present in the mean-field equations. Small values of $\beta$ bring instead the network activity closer to a continuous-attractor structure, and turn into larger finite-size effects.
In a second step, we fix $\rho_m$ and $\rho_n$ such that the network detects normalized input components along $I^A$ only when they are larger than a threshold value, that is taken around 0.5. We then looked for a pair of gating variables strengths $[\gamma_A, \gamma_B]$ which completely suppresses the response to $I^A$ by extending the range of bistable activity. The opposite pattern can be used to block the response in $I^B$ and allow a response in $I^A$.

Once the response in $I^A$ has been blocked, it can be verified that the network solely responds to inputs which contain a response along $I^B$ that is larger than a threshold close to 0.5. Note that, as in Fig.~S5 \textsf{L-M}, different values of $\bar{c}_A$ only minimally affect the exact position of the threshold.

To conclude, we remark that this procedure leaves the freedom of fixing the network parameters in many different configurations.
The complex rank-two architecture leads to larger finite-size effects than the respective unit-rank setup which acts as a single detector of correlations. In particular, the error at the level of the readout is larger but it decays with the system size, as expected for deviations induced by finite-size effects (Fig.~S5 \textsf{N}). 
Finally, note that when the noise in the input stimuli becomes extremely large, the network loses its ability to respond in a totally context-dependent fashion, as strong fluctuations in the non-relevant stimulus become likely to elicit a response.


\subsection{Method details for Main Figures}

\subsubsection{Figure 1}
In this figure, $\Sigma_m=\Sigma_n=1.0$. Note that the precise position of the instability to chaos depends on the value of $\Sigma_m$.  
The connectivity vectors $m$ and $n$ were generated from bivariate Gaussian distributions (means $M_m$ and $M_n$, variances $\Sigma_m$ and $\Sigma_n$, correlation $\rho$). Here we display the case where $m$ and $n$ overlap only along the unitary direction  ($M_m>0$, $M_n>0$, $\rho=0$, see \emph{Methods}). As shown in Fig.~S2, qualitatively similar regimes are obtained when the overlap is defined on an arbitrary direction. 
\textsf{C-D}: Network simulations were performed starting from initial conditions centered around $m$ and $-m$. Activity is integrated up to $T=800$. In simulations, $N=5000$, and statistics are averaged over 15 different connectivity realizations. The error bars, when visible, correspond to the standard deviation of the mean (as in every other figure, if not differently specified).

\subsubsection{Figure 2}
In this figure,  $g=0.8$. Other parameters are set as in Fig.~\ref{fig:p1}.	
\textsf{B}: The asymptotic input parameters are indicated by gray dots in \textsf{D} (middle). The simulation results (dark gray traces) correspond to $20$ trajectories for different network realizations (different trajectories strongly overlap). We simulated $N_{tr}=20$ different networks, each consisting of $N=3500$ units. In every network realization, the random part of the connectivity $\chi_{ij}$ is varied, while the low-rank part $m_in_j$ is kept fixed. $I$ (resp. $m$) scale: 0.7 (resp. 0.25) units.
\textsf{D}: The external input is increased along $n_{\perp}$, the component of $n$ that is perpendicular to the overlap direction.

\subsubsection{Figure 3}
The input and the readout vectors are Gaussian patterns of standard deviation $\Sigma=2$.
\textsf{C} (right): Colored traces: 20 trajectories from different network realizations  (different trajectories strongly overlap). 
		We simulated $N_{tr}=20$ different realizations of the network, each consisting of $N=2500$ units. In every network realization, the random part of the connectivity $\chi_{ij}$ is generated independently, while the low-rank part $m_in_j$ is kept fixed. $I^A$, $I^B$ and $m$ scale: 1.5 units.
\textsf{D}: Here, and in every plot if not differently stated, $\rho$ indicates the Pearson correlation coefficient. 
\textsf{F}: The PC axis are determined by analyzing separately the trials corresponding to the Go (top) and the Nogo (bottom) stimuli. Connectivity is measured as the average reciprocal synaptic strength; it includes both the random and the unit-rank components and it is averaged across network realizations. Note that the value of the correlation coefficient $\rho$ increases with the number of realizations $N_{tr}$ and the structure strength.

\subsubsection{Figure 4}
	The input and the readout vectors are Gaussian patterns of standard deviation $\Sigma=1.2$. The overlap between the connectivity vectors $m$ and $n$ leading to non-linear responses is quantified by $\rho_m=\rho_n=2.0$.
\textsf{B}: The input is generated as white noise of mean $\bar{c} = 0.6$ and standard deviation $\sigma = 0.4$ (the noise trace in the figure is only for illustration purposes). The red dashed line indicates the threshold in the implemented network.
\textsf{C}: The gray bar indicates the time point at which the network output is measured. Here and in the following figures, the readout includes an offset, so that the baseline value is set to zero.
\textsf{D}: 	We simulated many input noise traces for $N_{tr}=4$ different realizations of the network, each consisting of $N=2500$ units.  In every network realization, the random part of the connectivity $\chi_{ij}$ is varied, while the low-rank part $m_in_j$ is kept fixed.  Trajectories are smoothed with a Gaussian filter of standard deviation equal to one normalized time unit. $I$ (resp. $m$) scale: 0.5 (resp. 3.5) units.
\textsf{F}: The structure strength corresponds to the overlap $\rho_m\rho_n$. The effective time scale  is measured as the inverse of the value of the outlier eigenvalue of the stability matrix for $\bar{c}=0$.
\textsf{G}:  The psychometric curve was measured across $N_{tr}=100$ different realizations. The network  produces an output to the stimulus if at the end of the stimulus presentation (vertical gray line in \textsf{B}) the value of the readout $z$ is larger than one half of the largest readout value predicted by the theory. 
\textsf{H}: Details as in Fig.~\ref{fig:p3} \textsf{F}.

\subsubsection{Figure 5}
The stimuli vectors are Gaussian patterns of standard deviation $\Sigma=1.2$. We furthermore set: $g=0.8$, $\beta_m=0.6$, $\beta_n=1$, $\rho_m=3$, $\rho_n=1.6$. The  amplitudes  of  the two context directions  are  fixed to  $[0.08,  -0.14]$ (resp.  $[-0.14,0.08]$) during the Context $A$ (resp.~Context $B$) trials.
\textsf{B}: We consider in this case a unit-rank network as in Fig.~\ref{fig:p2} \textsf{D}, and we show in the two panels the network response for two different values of the input strength along the overlap axis (we set, respectively, $M_I=-0.3$ and 0.6). Details on the effect of contextual modulation on the full rank-two model  are further illustrated in Fig.~S5 \textsf{L-N}.
\textsf{E}: We simulated  $N_{tr}=4$ different realizations of the network, each consisting of $N=3000$ units. In every network realization, the random part of the connectivity $\chi_{ij}$ is varied, while the low-rank part $m_in_j$ is kept fixed. $I^A$ and $I^B$ (resp. $w$) scale: 1.0 (resp. 2.0) units.
\textsf{F}:  The network performance was measured across $N_{tr}=50$ different network realizations of size $N=7500$. The network  produces an output to the stimulus if at the end of the stimulus presentation (vertical gray line in \textsf{D}) the value of the readout $z$ is larger than one half of the largest readout value predicted by the theory.
\textsf{G}: Details as in Fig.~\ref{fig:p3} \textsf{F}.

\subsubsection{Figure 6}
The stimuli vectors are Gaussian patterns of standard deviation $\Sigma=1.2$. We furthermore set: $g=0.8$, $\beta_m=0.6$, $\beta_n=1$, $\rho_m=3$, $\rho_n=1.38$. The  amplitudes  of  the two context directions  are  fixed to  $[0.08,  -0.18]$ (resp.  $[-0.18,0.08]$) during the Context $A$ (resp.~Context $B$) trials.
\textsf{B}: Here $\bar{c}_A=0.6$ and $\bar{c}_B=0.1$, while the standard deviation of the noise in the input is $\sigma=0.3$ (the noise trace in the figure is only for illustration purposes).
\textsf{D}:   We simulated many noisy input traces for $N_{tr}=5$ different realizations of the network, each consisting of $N=4000$ units. In every network realization, the random part of the connectivity $\chi_{ij}$ is varied, while the low-rank part $m_in_j$ is kept fixed. For the sake of clarity, only correct trials have been included. $I^A$ and $I^B$ (resp. $w$) scale: 1 (resp. 1.5) units.
\textsf{E}: Network performance was measured across $N_{tr}=50$ different network realizations of size $N=7500$.

\section{Quantification and Statistical Analysis}

In this section, we briefly describe the analysis techniques that have been applied to the datasets generated from direct simulations of activity in finite-size networks (Figs.~\ref{fig:p2}, \ref{fig:p3}, \ref{fig:p4}, \ref{fig:p5} and \ref{fig:p6}).

\subsubsection{Dimensionality reduction} 
In order to extract from the high-dimensional population activity the low-dimensional subspace which contains most of the relevant dynamics, we performed dimensionality reduction via a standard Principal Component (PC) analysis.

To begin with, we constructed the activation matrix $X$. In $X$, every column corresponds to the time trace of the activation variable $x_i(t)$ for unit $i$, averaged across trials. We indicate as \emph{trials} different network simulations, where different noisy inputs, or different quenched noise in the random connectivity matrix have been generated (details are specified in the figure captions).
The activation matrix $X$ is normalized through Z-scoring: to every column, we subtract its average over time, and we divide by its standard deviation. Note that Z-scoring distorts the shape of the population trajectory in the phase space. For this reason, in order to facilitate the comparison with the trajectory predicted by the mean-field theory, in Fig.~S3 we more simply consider the mean-subtracted matrix $X$. Applying the PCA analysis to one of the two data formats impacts the results from a quantitative point of view, but does not change their general validity.

The principal components (PC) are computed as the normalized eigenvectors $\{e_l\}_{l=1,...,N}$ of the correlation matrix $C=X^TX$. The PC are sorted in decreasing order according to the corresponding real eigenvalue $\lambda_l$. The activation matrix $X$ can be projected on the orthonormal basis generated by the PC vectors by computing: $X' = XE$, where $E$ is the $N\times N$ matrix containing the PC eigenvectors ordered as columns.
The variance explained by the $l$-th PC mode $e_l$ can be computed as the $l$-th entry on the diagonal of the rotated correlation matrix $C'=X'^TX'$.

While in our network models the low-rank part of the connectivity determines a purely low-dimensional dynamics (Fig.~S3 \textsf{A}), the random part of the connectivity generates a continuum of components whose amplitude is determined by strength of the random connectivity $g$ with respect to the connectivity and input vectors. In Fig.~\ref{fig:p2}, where $g=0.8$, the low-dimensional nature of the dynamics is revealed by considering averages across several ($N_{tr}=20$) realizations of the random connectivity. In Fig.~S3 \textsf{B}, we illustrate the result of performing PCA on the activity generated by a single network. In this case, even if more PC components contribute to the total variance, the two first axis bear a strong resemblance with the directions predicted with the theory. In Fig.~S3 \textsf{C} we show that, in the same spirit, a PCA analysis can be used to extract the relevant geometry of the network model also when activity is strongly chaotic.

In order to more easily connect with the theoretical predictions, we systematically applied dimensionality reduction on datasets constructed from the activation variable $x_i$. We verified that our results still hold, from a qualitative point of view, when the analysis is performed on the non-linearly transformed variables $\phi(x_i)$.
In the network models we considered, the activation variables $\phi(x_i)$ indeed form a non-linear but dominantly low-dimensional manifold in the phase space. The axes predicted by the mean-field theory determine the dominant linear geometry of this manifold, and can be still captured (although less precisely) by looking at the first PC components.

\subsubsection{Linear regression} 
In order to estimate how single units in the network are tuned to different task variables (such as input stimuli or decision variables), we used a multi-variate linear regression analysis.

To this end, we considered the full population response $x_i^k(t)$, where $k=1, ..., N_{tr}$ indicates the trial number. Following \citep{Mante2013}, our aim was to describe the network activation variables as linear combinations of the $M$ relevant task variables. In Fig.~\ref{fig:p3}, the two variables we considered were the strength of the Go and of the Nogo inputs, that we indicate here with $c_{\mathrm{Go}}$ and $c_{\mathrm{Nogo}}$:
\begin{equation}
	x_i^k(t) = \beta^{\mathrm{Go}}_{i,t} c_{\mathrm{Go}} (k) + \beta^{\mathrm{Nogo}}_{i,t} c_{\mathrm{Nogo}} (k).
\end{equation}
In a Go, or in a Nogo trial, only one of the two strength coefficients is non-zero.
In Fig.~\ref{fig:p4}, the two relevant task variables are assumed to be the input strength along $I$, quantified by $c$, and the network output, quantified as the value of the readout $z$ at the end of the stimulus presentation:
\begin{equation}
	x_i^k(t) = \beta^\mathrm{{input}}_{i,t} c (k) + \beta^{\mathrm{choice}}_{i,t} z (k).
\end{equation}
In Figs.~\ref{fig:p5} and \ref{fig:p6},  the relevant variables are four: the strength of stimuli $A$ and $B$, the trial context and the network output. We thus have:
\begin{equation}
	x_i^k(t) = \beta^{{A}}_{i,t} c_{{A}} (k) + \beta^{{B}}_{i,t} c_{{B}} (k) + + \beta^{\mathrm{ctx}}_{i,t} y (k) + \beta^{\mathrm{choice}}_{i,t} z (k).
\end{equation}
where the context variable is represented by a unique symbolic variable $y$, which takes value $y=1$ in Context $A$ and $y=-1$ in Context $B$.

More generally, we indicate with $\beta_{i,t}^{\nu}$ the regression coefficient of unit $i$ with respect to the task feature $\nu$ at time $t$. The vector $\beta_{i,t} = \{\beta_{i,t}^{\nu}\}_{\nu=1,.., M}$ indicates the collection of the $M$ variables regressors for a given unit at the time point $t$.
We compute the regression coefficients by defining a matrix $F$ of size  $M \times N_{tr}$, where every row contains the value of the $M$ relevant task variables across trials. The regression coefficient vectors are then estimated by least-square inversion:
\begin{equation}
	\beta_{i,t} = (FF^T)^{-1} F x_{i,t}
\end{equation}
where the vector $x_{i,t}$ is constructed by collecting across trials the value the activation variable of unit $i$ at time $t$.

In order to get rid of the time dependence of our result, we simply consider the coefficients $\beta_{i,t}$ at the time point where the two-dimensional array $\beta_{i,t}$ for every $i$ has maximal norm \citep{Mante2013}.
The resulting set of $M$-dimensional vectors $\beta_i$ contains the regression coefficients of unit $i$ with respect to the $M$ relevant task variables. The $N$-dimensional regression axis for a given task variable $\nu$ is finally constructed by collecting the $\nu$-th components of  $\beta_i$ across different population units: $\{\beta^{\nu}_i\}_{i=1,..,N}$.

\section{Data and Software Availability}

Software was written in the  Python (http://python.org) programming languages. Implementations of algorithms used to compute quantities presented in this study are available at: https://github.com/fmastrogiuseppe/lowrank/.

\clearpage

\renewcommand{\thefigure}{S\arabic{figure}}
\setcounter{figure}{0}


\begin{figure}
	\captionsetup{labelformat=empty}
	\centering
	\includegraphics{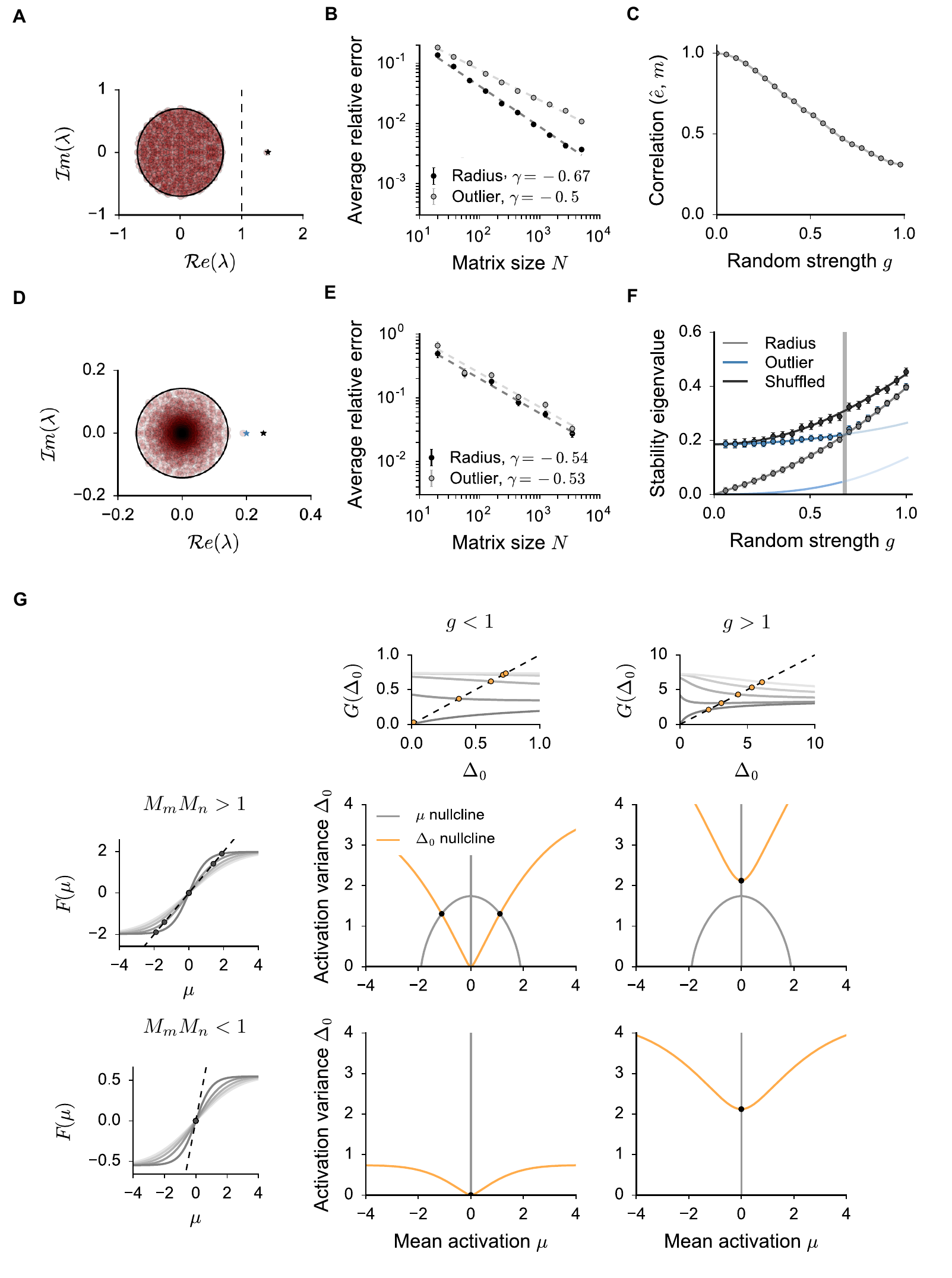}
	\caption{}
\end{figure}

\begin{figure}
	
	\captionsetup{labelformat=adja-page}
	\ContinuedFloat
	\caption{  {\bf \textsf{ Dynamical Mean-Field description of rank-one networks whose right- and left-connectivity vectors overlap solely on the unitary direction}} ($\rho=0$, see \emph{Methods}). Related to Figure 1.} 
	
	{\bf \textsf{(A-B-C)}} Eigenspectrum of the partially structured connectivity matrix $J_{ij}$, related to the stability matrix $S_{ij}$ of the homogeneous fixed points through: $S_{ij}=\phi'(\bar{x})J_{ij}$. 
	\textsf{\bfseries A.} Eigenspectrum of $J_{ij}$ in the complex plane. Red dots: eigenspectrum of a single realization $J_{ij}$ of size $N=1000$.  In black: theoretical prediction. Every matrix $J_{ij}$ consists of a sum of a random and of a fixed unit-rank structure. In the large $N$ limit, the spectrum of the full matrix is given by the sum of the eigenspectra of the two parts. The black circle has radius equal to the total random strength $g$, and the black star indicates the position of the non-zero eigenvalue of the rank-one structure $P_{ij}$. 
	\textsf{\bfseries B.} Mismatch between the statistics measured in finite-size networks ($x_{sim}$) and the theoretical prediction ($x_{th}$) as the network size $N$ is increased. The error is normalized: $|x_{sim}-x_{th}|/x_{th}$. Averages over 100 realizations. The error bars (as in every other figure, if not differently specified) correspond to the standard deviation of the mean. Dashed lines: power-law best fit ($y\propto N^{\gamma}$). The values of $\gamma$ are indicated in the legend.
	\textsf{\bfseries C.} Pearson correlation coefficient between the connectivity eigenvector $m$ and the eigenvector $\hat{e}$ which corresponds to the outlier eigenvalue.
	Choice of the parameters: $\rho=0$, $M_mM_n=1.43$, $\Sigma_m=0.33$, $\Sigma_n=1$. In \textsf{A} and  \textsf{B}, $g=0.7$.
	
	\vspace{0.3cm}
	
	{\bf \textsf{(D-E-F)}} Analysis of the eigenspectrum of the linear stability matrix $S_{ij}=\phi'(\bar{x})J_{ij}$ for heterogeneous stationary solutions.
	\textsf{\bfseries D.} Eigenspectrum of $S_{ij}$ in the complex plane. Red dots: eigenspectrum of a single, finite-size realization of $S_{ij}$, $N=2500$. The radius of the black circle corresponds to the theoretical prediction $r=g\sqrt{\langle[\phi_i'^2]\rangle}$.
	The black star indicates the position of the non-zero eigenvalue of the rank-one structure ${m_i\phi'(x_j^0)n_j}/{N}$, which deviates significantly from the position of the outlier eigenvalue. We thus address the problem of evaluating the position of the outlier eigenvalue through a mean-field stability analysis (Eq.~71), the prediction of which is indicated by the blue star.
	\textsf{\bfseries E.} Mismatch between the  results from simulations and mean-field predictions for the radius and the outlier position.
	The error is measured as an average over $N_{tr}=30$ finite size matrices, and decays as the system size is increased. Details as in \textsf{B}.
	\textsf{\bfseries F.} Radius and outlier of the stability eigenspectrum for increasing random strength values. The dots indicate the results of numerical simulations of networks with $N=2500$ units, averaged over $N_{tr}=30$ realizations of the random and structured connectivities. In grey: radius of the compact bulk (continuous line: mean-field prediction $r$). In blue: position of the outlier eigenvalue (continuous dark and light lines: first and second eigenvalue of matrix $\mathcal{M}$ given in Eq.~71). In black: position of the outlier when $\chi_{ij}$ is shuffled (continuous line: mean-field prediction for the outlier of the structured part ${m_i\phi'(x_j^0)n_j}/{N}$). 
	Choice of the parameters: $\rho=0$, $M_mM_n=2.2$, $\Sigma_m=0.4$, $\Sigma_n=1$.  In \textsf{D} and  \textsf{E}, $g=0.5$.
	
	\vspace{0.3cm}
	
	{\bf \textsf{(G)}} Graphical analysis of stationary solutions. Large figures: nullcline plots for the population-averaged DMF equations in Eq.~83. Black dots indicate the solutions that are stable with respect to the outlier eigenvalue. Four set of parameters (two values for $M_mM_n$, two for $g$) have been selected. Note that the shapes of the $\mu$ and the $\Delta_0$ nullcline depend only on the structure strength $M_mM_n$  and the disorder strength  $g$. For the figures in the first (resp.~second) row, the structure strength $M_mM_n=0.55$ (resp.~$M_mM_n=2.0$) is weak (resp.~strong). For the figures in the first (resp.~second) column: the random strength $g=0.7$ (resp~ $g=2.0$) is weak (resp.~strong).	Note that the stationary states at large $g$ values (right column) are always unstable with respect to the continuous component of their stability eigenspectra (Fig.~1 \textsf{C-D}).
	The small side figures  associated to every row and column show how the $\mu$ (for the rows) and $\Delta_0$ (for the columns) nullclines have been built. We solve $\mu=F(\mu)$ (resp. $\Delta_0=G(\Delta_0)$) for different initial values of $\Delta_0$ (resp.~$\mu$). Different initial conditions are displayed in gray scale. Dark grey refers to $\Delta_0=0$ (resp. $\mu=0$). The dots indicate the solutions for different initial values, which together generate the nullcline curves. Choice of the parameters: $\Sigma_m =1$.
	
	\label{fig:s1}
\end{figure}

\clearpage


\begin{figure}
	\captionsetup{labelformat=empty}
	\centering
	\includegraphics{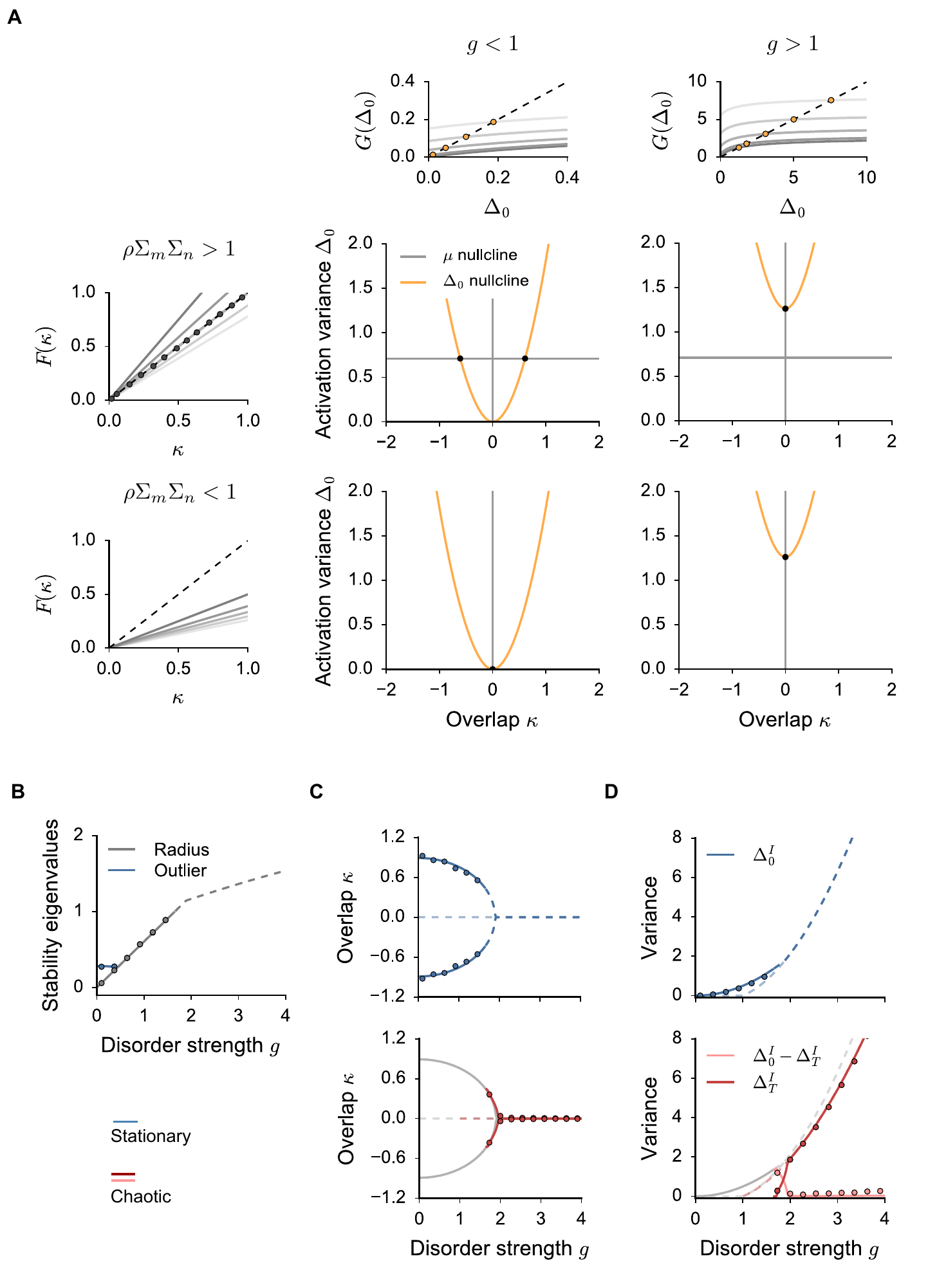}
	\caption{}
\end{figure}

\begin{figure}
	\captionsetup{labelformat=adja-page}
	\ContinuedFloat
	\caption{  {\bf \textsf{ Dynamical Mean-Field description of rank-one networks whose right- and left-connectivity vectors overlap onto an arbitrary direction $y$}} ($M_m=M_n=0$, $\rho \neq 0$, see \emph{Methods}). Related to Figure 1.}
	
	{\bf \textsf{(A)}} Graphical analysis of stationary solutions. Large figures: nullcline plots for the population-averaged DMF equations in Eq.~89. Black dots indicate the solutions that are stable with respect to the outlier eigenvalue. Four set of parameters (two values for $\rho\Sigma_m\Sigma_n$, two for $g$) have been selected. Note that the shapes of the $\kappa$ and the $\Delta_0$ nullcline depend only on the  structure strength $\rho\Sigma_m\Sigma_n$ and the disorder strength $g$.  For the figures in the first (resp.~second) row, the structure strength $\rho\Sigma_m\Sigma_n$ (resp.~$\rho\Sigma_m\Sigma_n$) is weak (resp.~strong).  For the figures in the first (resp.~second) column: the random strength $g=0.5$ (resp.~$g=1.7$) is weak (resp.~strong). Note that the stationary states at large $g$ values (right column) are always unstable with respect to the continuous circular component of their stability eigenspectra (see \textsf{B-C-D}).	The small figures associated to every row and column show how the $\kappa$ (for the rows) and $\Delta_0$ (for the columns) nullclines have been built. We solve $\kappa=F(\kappa)$ (resp. $\Delta_0=G(\Delta_0)$) for different initial values of $\Delta_0$ (resp. $\kappa$). Different initial conditions are displayed in gray scale. Dark grey refers to $\Delta_0=0$ (resp. $\kappa=0$). The dots indicate the solutions for different initial values, which together generate the nullcline curves.
	
	\vspace{0.3cm}
	
	{\bf \textsf{(B-C-D)}} Bifurcation diagram of the activity statistics as the random strength $g$ is increased. Details as in Fig.~1 \textsf{C-D}.
	\textsf{\bfseries B.} Stability eigenspectrum of stationary solutions, mean-field prediction for the radius of the compact part and the outlier position. 
	\textsf{\bfseries C.} Overlap $\kappa = \langle n_i [\phi_i] \rangle$. 
	\textsf{\bfseries D.} Individual second order statistics. The DMF solutions are displayed as continuous (resp.~dashed) lines if they correspond to a stable (resp.~unstable) state. In \textsf{C-D}, top panels display statistics for stationary solutions and bottom panels display statistics for chaotic solutions.
	Dots: we measured activity statistics in finite-size networks, starting from globally positive and negative initial conditions. Activity is integrated up to $T=400$. $N=3500$, average over 8 different network realizations. 
	Choice of the parameters: $\Sigma_m=\Sigma_n=1.5$, $\rho = 2.0 / \Sigma_m\Sigma_n$.
	
	\label{fig:s2}
\end{figure}

\clearpage


\begin{figure}
	\captionsetup{labelformat=empty}
	\centering
	\includegraphics{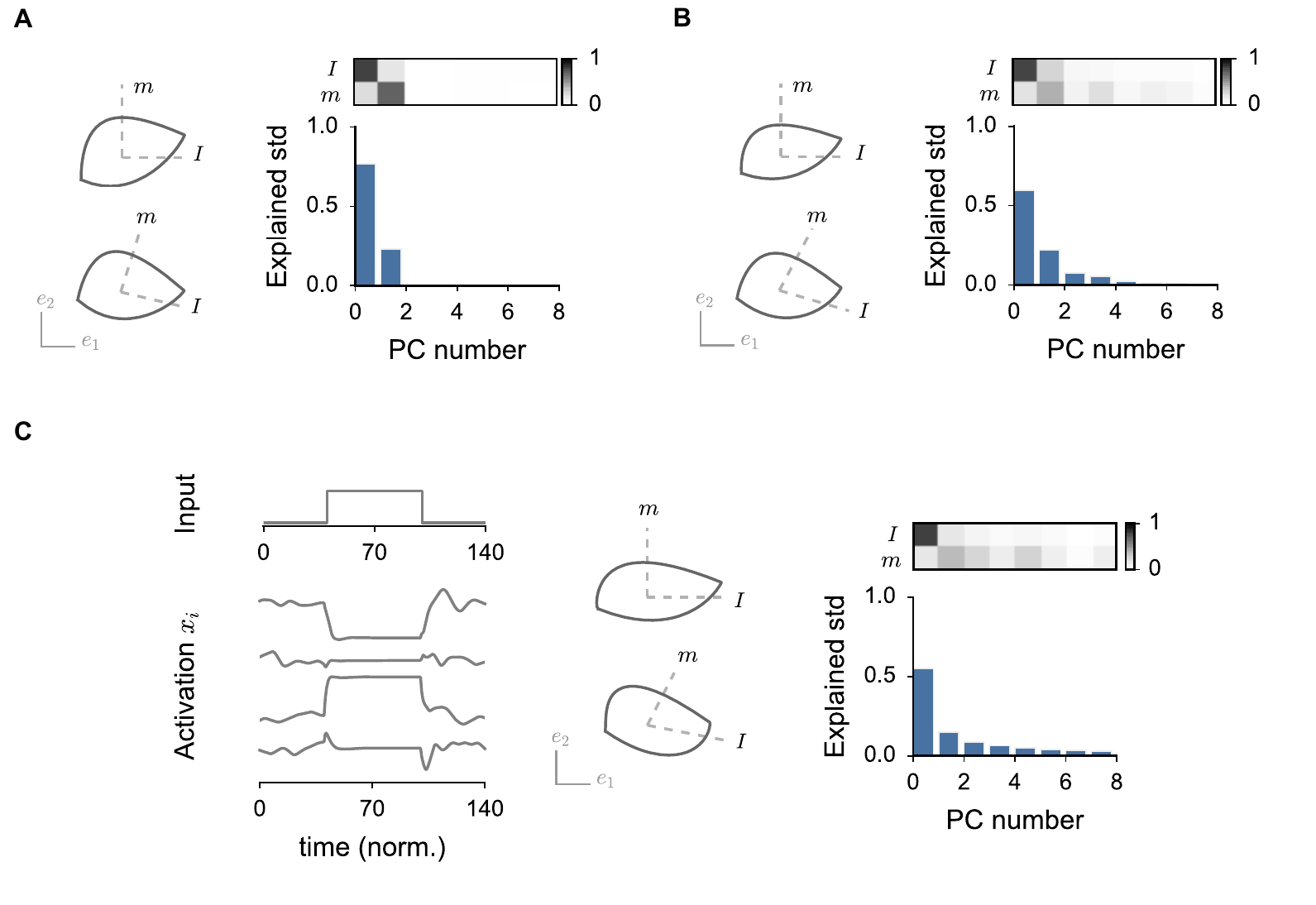}
	\caption{}
\end{figure}

\begin{figure}
	\captionsetup{labelformat=adja-page2}
	\ContinuedFloat
	
	\caption{ {\bf \textsf{ Two-dimensional dynamics in networks with unit-rank structure and external inputs. }} Related to Figure 2.}
	
	We consider a unit-rank network as in Fig.~2 \textsf{B-C}. The connectivity vectors $m$ and $n$  are orthogonal, but the external input vector contains a component along $n$, whose strength (quantified by $\Sigma_{nI}$, see \emph{Methods}) undergoes a step increase from $0.2$ to $2.0$.  We simulate data from networks of size $N=3500$.
	We analyze the dimensionality of the dynamics by comparing the relevant low-dimensional trajectory predicted by the mean-field theory with the strongest modes extracted through dimensionality reduction (Principal Component analysis, see \emph{Methods}). 
	\textsf{\bfseries A.} Analysis for a purely structured network ($g=0$). Left top: the mean-field theory predicts that the low-dimensional network dynamics $x=\{x_i\}$ lies in the plane defined by the right-connectivity vector $m$ and the external input $I$. We thus projected the high-dimensional population activity (dark grey trajectory) on this plane. Left bottom: we projected the network dynamics (continuous), along with the two vectors $m$ and $I$ (dashed), on the plane defined by the first two PC axis $e_1$ and $e_2$. Right top: Pearson correlation coefficient between vectors $m$ and $I$ and the first eight PC. Right bottom: strength of the first eight PC, measured as the fraction of the standard deviation of activity that they explain (see \emph{Methods}). Note that when the network connectivity is fully structured ($g=0$) as in this case, activity is exactly two-dimensional. The first two PC axis span the $m-I$ plane, but they define a rotated set of basis vectors. 
	\textsf{\bfseries B.} Analysis for a network which includes a random term in the connectivity matrix ($g=0.8$). While in Fig.~2 we performed the PC decomposition on trial-averaged data ($N_{tr}=20$), here we considered a single trial, defined as a single realization of the random connectivity matrix. Details as in \textsf{A}. Note that the random component of the connectivity adds noisy contributions in a continuum of PC directions, whose strength depends on the value of $g$ with respect to the amplitude of input and connectivity vectors, and becomes weaker and weaker when averaging with respect to different realizations of $\chi_{ij}$. When $g>0$, vectors $m$ and $I$ are not fully contained in the $e_1-e_2$ plane, so their projections on the PC plane are not orthogonal.
	\textsf{\bfseries C.} Analysis for a network which includes a strong random term in the connectivity matrix ($g=1.8$), such that spontaneous activity is chaotic. In the left-most column, similarly to Fig.~2, we plot the time trajectories of four randomly selected units. The center and the right columns are as in \textsf{A} and \textsf{B}, with PCA  performed on trial-averaged activity ($N_{tr}=20$). The scale of the projections panels is here set arbitrarily.
	
	\label{fig:s3}
\end{figure}

\clearpage


\begin{figure}
	\captionsetup{labelformat=empty}
	\centering
	\includegraphics{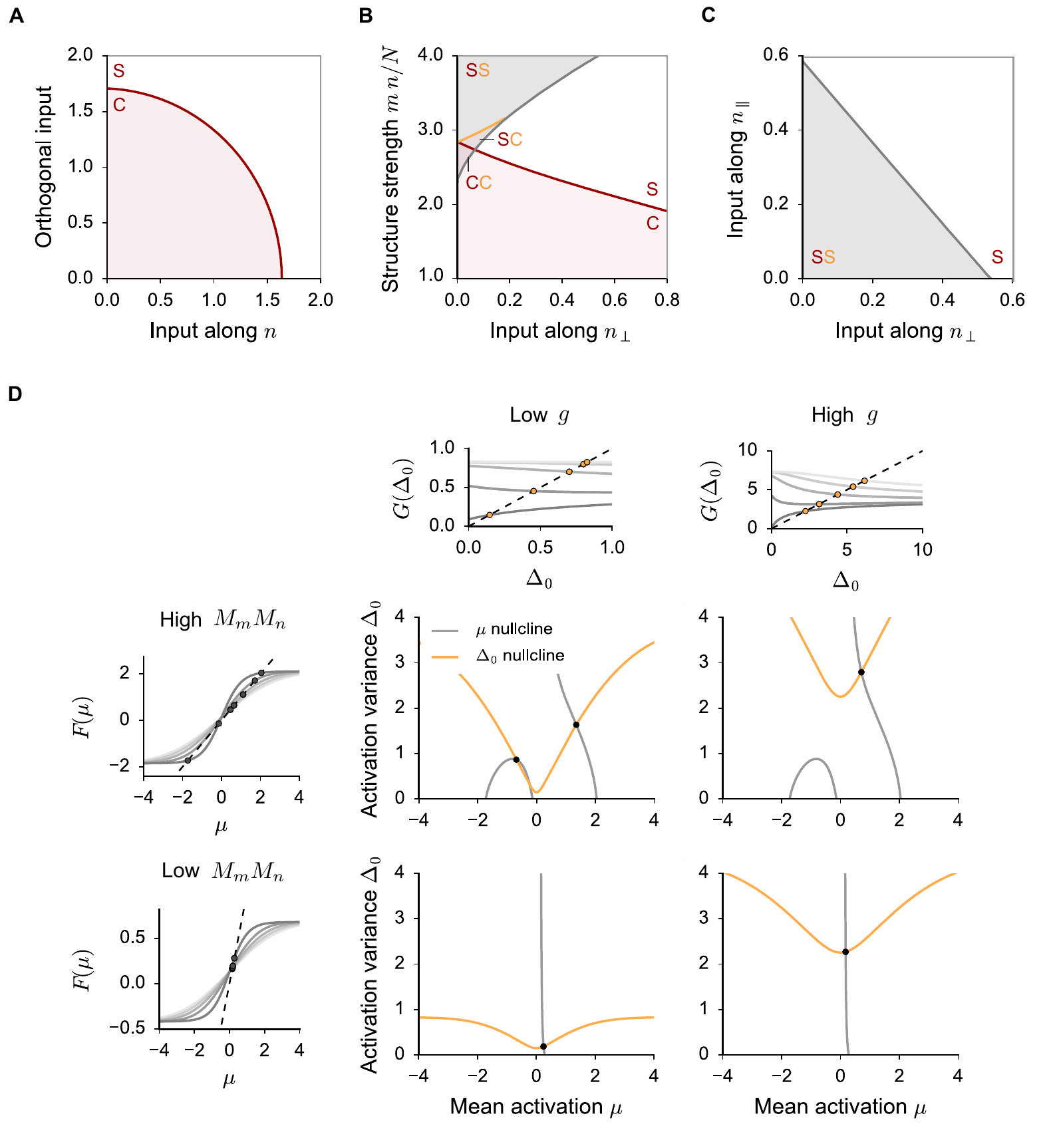}
	\caption{}
\end{figure}

\begin{figure}
	\captionsetup{labelformat=adja-page}
	\ContinuedFloat
	\caption{ {\bf \textsf{ Dynamical Mean-Field description of input-driven dynamics for rank-one networks whose right- and left-connectivity vectors overlap solely on the unitary direction}} ($\rho=0$, see \emph{Methods}). Related to Figure 2.}
	
	\textsf{\bfseries (A-B-C)} Dynamical regimes of the network activity as function of the structure connectivity strength $m^Tn/N$, the random strength $g$ and the input strength.
	Grey shaded areas indicate the parameter regions where the network activity is bistable.  Red shaded areas indicate the phase space regions where network dynamics are chaotic. When two stable solutions exist, the yellow and the red letter indicate whether each of them is stationary (S) or chaotic (C). Note that stationary and chaotic dynamics can coexist (SC region). In \textsf{\bfseries A}, as in Fig.~2 \textsf{D} center, the two connectivity vectors $m$ and $n$ are orthogonal. We varied the external input strength by increasing the amplitude of the component along $n$ (quantified by $\Sigma_{nI}$, see \emph{Methods}) and of the orthogonal one (quantified by $\Sigma_{\perp}$). Note that inputs along both directions contribute to suppressing the amplitude of chaotic fluctuations. In \textsf{\bfseries B}, as in Fig.~2 \textsf{D} right, the two connectivity vectors $m$ and $n$ are not orthogonal, but they share an overlap component along the unitary direction. We varied the structure strength (quantified by $M_mM_n$) and the strength of the input along the direction of $n$ that is perpendicular to the structure overlap, $n_{\perp}$ (again quantified by $\Sigma_{nI}$). Similarly to Fig.~1, strong structure overlaps can lead to the appearance of two bistable solutions. In presence of non trivial external inputs, however, such solutions are not symmetric, and can loose stability on different parameter boundaries. In particular, we observe that external inputs tend to suppress bistable regimes, by favouring one solution over the other. In \textsf{\bfseries C}, finally, the network configuration is similar to \textsf{B}, but we consider external inputs which include a second component along the direction of $n$ that is shared with $m$, $n_{\parallel}$  (quantified by $M_I$). We observe that both input directions play similar roles in reducing the extent of the bistable regime.
	Choice of parameters: $g=2.2$, $\Sigma_m=\Sigma_n=1.0$, $\Sigma_{mI}=0$.
	
	\vspace{0.3cm}
	
	\textsf{\bfseries (D)} Graphical analysis of stationary solutions. In this example, the external input vector overlaps with $n$ on the unitary overlap direction $n_{\parallel}$ ($M_I=0.13$), and includes orthogonal components quantified by $\Sigma_I=0.3$.
	Large figures: nullcline plots for the stationary form of the population-averaged DMF equations in Eq.~98. Black dots indicate the solutions that are stable with respect to the outlier eigenvalue. Four set of parameters (two values for $M_mM_n$, two for $g$) have been selected. Note that the shape of the $\mu$ and the $\Delta_0$ nullcline depends only, respectively, on the structure strength $M_mM_n$ and the disorder $g$ together with the input statistics. For the figures in the first (resp.~second) row, the structure strength $M_mM_n=0.55$ (resp.~$M_mM_n=2.0$) is weak (resp.~strong). For the figures in the first (resp.~second) column: the random strength $g=0.7$ (resp.~$g=2.0$) is weak (resp.~strong).	The small figures associated to every row and column show how the $\mu$ (for the rows) and $\Delta_0$ (for the columns) nullclines have been built. We solve $\mu=F(\mu)$ (resp. $\Delta_0=G(\Delta_0)$) for different initial values of $\Delta_0$ (resp. $\mu$). Different initial conditions are displayed in gray scale. Dark grey refers to $\Delta_0=0$ (resp. $\mu=0$). The dots indicate the solutions for different initial values, which together generate the nullcline curves. Choice of the parameters: $\Sigma_m =1$.
	\label{fig:s4}
\end{figure}

\clearpage


\begin{figure}
	\captionsetup{labelformat=empty}
	\centering
	\begin{adjustwidth}{-0.2in}{0in} 
		\includegraphics{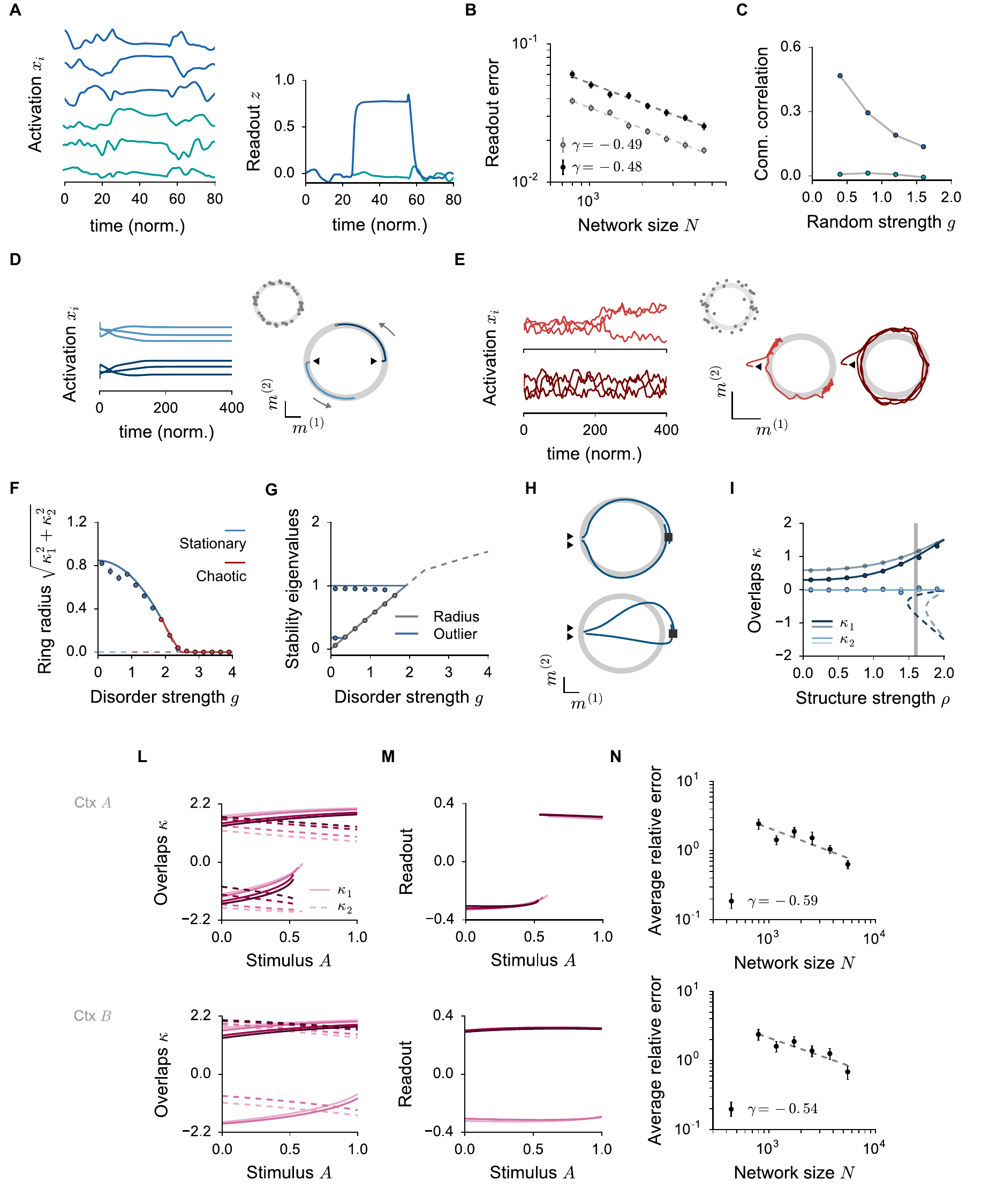}
	\end{adjustwidth}
	\caption{}
\end{figure}

\begin{figure}
	\captionsetup{labelformat=adja-page}
	\ContinuedFloat
	\caption{ {\bf \textsf{ Dynamical Mean-Field description of low-rank networks designed for solving computational tasks. }}Related to Figures 3, 5  and 6.}
	
	\textsf{\bfseries (A-B-C)} Rank-one networks can robustly perform computations also when their dynamics is chaotic due to large random connectivities. Here, we show an example from the Go-Nogo task (Fig.~3). We focus on large random strength values ($g=2.5$), so that spontaneous network dynamics is chaotic. \textsf{\bfseries A.} Left: response of three randomly selected units to the Go pattern $I^A$ (top, blue) and to the Nogo pattern $I^B$ (bottom, green). Right: time trace of the readout  $z(t)$ for the Go (blue) and the Nogo (green)  stimulus. \textsf{\bfseries B.} Absolute, normalized distance between the theoretical prediction and the value of the readout $z$ obtained from finite-size realizations. As expected, the magnitude of the average normalized error decays with the network size as $\sim 1/\sqrt{N}$. In grey: $g=0.8$, in black: $g=2.5$. Averages over 200 network realizations. Details as in Fig.~S1 \textsf{B}. 
	\textsf{\bfseries C.} As in Fig.~3 \textsf{F}, we consider pairs of units and we compute the correlation coefficient between their weights onto the first PC axis and their average reciprocal connectivity strength. The PC axis is computed separately for data corresponding to Go (blue) or the Nogo (green) trials. The correlation coefficient for the Go trials decreases with the amplitude of the random connectivity, although the error in the readout is only weakly affected (panel {\bfseries B.}). For every entry of the connectivity matrix $J_{ij}$, indeed, the random part $g\chi_{ij}$ has larger amplitude then the structured one $P_{ij}$. As a consequence, the random noise can hide a fraction of the strong correlations existing between the PC weights and the rank-one connectivity $P_{ij}$. Note that the absolute value of the correlation coefficient depends on the variance of the rank-one connectivity. Finally, the correlation coefficient increases as the connectivity gets averaged on more and more realizations of the random part.
	Choice of the parameters as in Fig.~3.
	
	\vspace{0.3cm}
	
	\textsf{\bfseries (D-E-F-G-H-I)}  Ring attractor from rank-two connectivity structures with connectivity vectors  characterized by strong internal overlaps (see \emph{Methods}). 
	\textsf{\bfseries D.} Sample of activity from a finite-size realization ($N=4000$) of the rank-two network. Activity is initialized in two different initial conditions (light and dark blue), indicated by the small arrows.  Left: time traces of the activation variables for three randomly selected network units. Note the long time range  on the $x$ axis. Right: population activation $x=\{x_i\}$ projected on the plane spanned by the right vectors $m^{(1)}$ and $m^{(2)}$.  The ring attractor predicted by the mean-field theory is displayed in light gray. The strength of the disorder is $g=0.5$, so that the network is in a stationary regime. In the small inset, we reproduce the theoretical prediction together with the final state of additional $N_{tr}=20$ networks realizations, that are displayed as grey dots. 
	\textsf{\bfseries E.} Sample of activity for two finite-size realizations ($N=4000$) of the structured connectivity matrix (dark and light red). Details as in \textsf{D}. The strength of random connections is  $g=2.1$, so that the network is in a chaotic regime. Chaotic fluctuations can occur together with a slow exploration of the ring (dark red). If two specific states on the ring appear to be more stable, chaotic fluctuations can induce jumps between the two of them (light red). 
	\textsf{\bfseries F-G.} Mean-field characterization of the ring structure: radius of the ring attractor and stability eigenvalues. Details as in Fig.~1. Dots: numerical results from finite-size ($N=4000$) networks, averaged over 6 realizations of the connectivity matrix.
	\textsf{\bfseries H-I.} Input patterns which correlate with the left vector $n^{(1)}$ reduce the ring attractor to a single stable state. Activity is thus projected in the direction spanned by the right vector $m^{(1)}$. 
	In \textsf{\bfseries H}, we show the input response for two finite-size networks. The grey ring displays the mean-field solution in absence of external inputs ($g=0.5$, as in \textsf{D}). In the top panel, the input is weak ($\Sigma_I = 0.2$, see \emph{Methods}). The transient dynamics as well as the equilibrium state lie close to the ring structure. In the bottom panel, the input is strong ($\Sigma_I = 0.6$), and the ring structure is not anymore clearly apparent. 
	In \textsf{\bfseries I}, we plot the values of the overlaps $\kappa_1$ (blue) and $\kappa_2$ (azure) as a function of the  structure strength parameter $\rho$, for fixed input strength. Stable solutions are plotted as continuous lines, unstable ones as dashed. Solid (resp.~transparent) lines refer to weak (resp.~strong) external inputs: $\Sigma_I=0.2$ (resp.~0.6). The vertical gray line indicate the value of $\rho$ that has been used in \textsf{H}. Dots: numerical results as in \textsf{F-G}.
	Choice of the parameters (see \emph{Methods}): $\Sigma = 2.0$, $\rho_1 = \rho_2 =1.6$.
	
	\vspace{0.3cm}
	
	\textsf{\bfseries (L-M-N)} Rank-two structures for implementing non-linear stimuli detection in a context-dependent fashion (Fig.~6): theoretical mean-field predictions.
	\textsf{\bfseries L.} Values of the first-order statistics $\kappa_1$ (continuous) and $\kappa_2$ (dashed) as a function of the overlap strength along the stimulus $I^A$. Results are shown for four increasing values of the overlap strength along the second stimulus $I_B$. Top (resp.~bottom): the contextual gating inputs are such that a response to $I^A$ (resp.~$I^B$) is selected.
	\textsf{\bfseries M.} Readout value, built by summing the values of $\kappa_1$ and $\kappa_2$ (Eq.~158). Note that although $\kappa_1$ and $\kappa_2$ vary with input strength, on each branch their sum is approximately constant. Details as in \textsf{L}.
	\textsf{\bfseries N.} Average normalized error between the DMF predictions and the simulated readout, in the two gating conditions as a function of the network size $N$. Average over 60 network realizations, details as in Fig.~S1 \textsf{B}.
	Parameters as in Fig.~6.
	
	\label{fig:s5}
\end{figure}

\clearpage

\begin{figure}
	\captionsetup{labelformat=empty}
	\centering
	\includegraphics{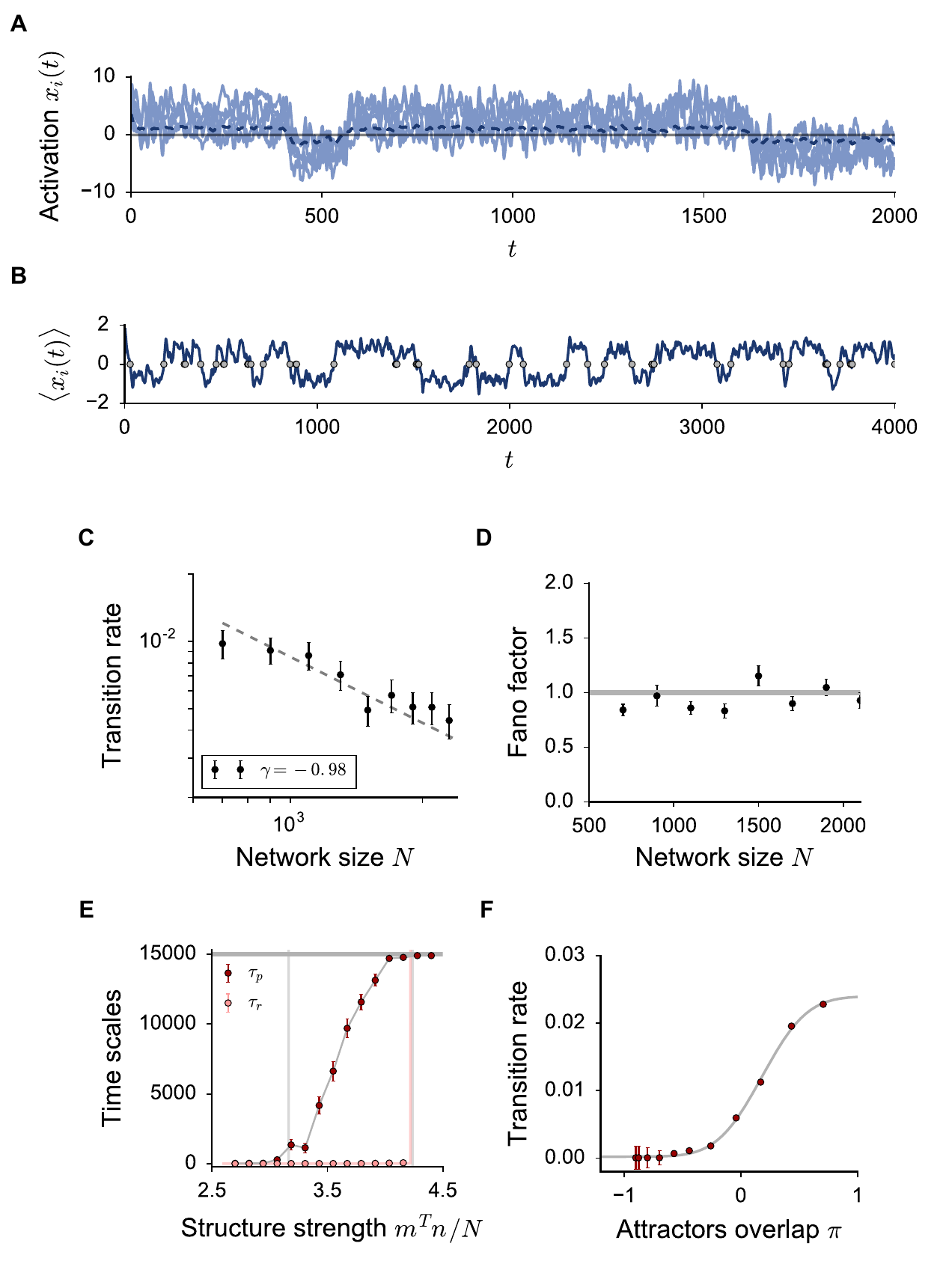}
	\caption{}
\end{figure}

\begin{figure}
	\captionsetup{labelformat=adja-page}
	\ContinuedFloat
	\caption{ {\bf \textsf{ Dynamics of unit-rank networks of finite-size are characterized by two distinct time-scales.}} Related to Figure 1.}
	
	In Fig.~1 we have shown that, when the structure strength is large, the DMF theory predicts the existence of two bistable states, which can display chaotic activity. For those states, the population-averaged statistics of the activation variable $x_i$ are stationary. In the chaotic regime, indeed, irregular temporal fluctuations are decorrelated from one unit to the other, so that the central limit theorem applies at every time step, and the network statistics are constant in time.
	In finite-size networks, however, the network statistics are not stationary: their dynamics display instead two different time-scales.
	The instantaneous population-averaged activity undergoes small fluctuations of amplitude $\mathcal{O}(1/\sqrt{N})$, whose time-scale is given by the relaxation decay of chaotic activity.
	Because of bistability, furthermore, the first-order statistics displays also sharp transitions from positive to negative values and viceversa, which are made possible by the self-sustained temporal fluctuations. In the following, we focus on rank-one structures where the overlap direction is defined along the unitary vector. As a consequence, the relevant first-order statistics is simply the population-average of the activation vector $\mu$ (see \emph{Methods}).
	\textsf{\bfseries A.} Sample from a finite-size network: activation time traces for randomly chosen units displaying attractors jumps. Dashed blue line: time-dependent population average. 
	\textsf{\bfseries B.} Time-dependent population average in a longer trial. 
	\textsf{\bfseries C-D.} We consider transition events as point processes, and we measure the average transition rate. 
	We arbitrarily define a transition point as the time step at which the population-averaged activation crosses zero (grey points in \textsf{B}).
	In \textsf{\bfseries C}, we show that the transition rate decays to zero as the network size $N$ is increased. Details as in Fig.~\ref{fig:s1} \textsf{B}.
	Note that the transition rate depends on the amplitude of finite-size fluctuations measured with respect to the average phase space distance between the two attractors. As a consequence, the transition rate depends on the architecture parameters and on the network size, but also varies strongly from one realization of the connectivity matrix to the other.
	\textsf{\bfseries D.}  Fano factor of the point process for different values of the network size $N$, which noisily oscillates around 1. For every realization of the network, the jumps count is measured in different windows of the total integration time $T=15.000$. The Fano factor is measured for every realization and then averaged over $N_{tr}=30$ different networks.  
	\textsf{\bfseries E-F.} Analysis of the two time-scales displayed by the network dynamics. The first time-scale is measured as the relaxation time constant $\tau_r$, which can be derived within the DMF framework by computing the time decay of the full auto-correlation function $\Delta(\tau)$. The persistence time scale, indicated by $\tau_p$, coincides instead with the average time interval which separates two attractors transitions.
	In \textsf{\bfseries E}, we show that both time scales depend on the network architecture parameters. Here, we fix the random strength $g=3$ and we increase the structure strength. 
	When the structure is weak (left), the network is in the classical homogeneous chaotic state. The persistence time scale coincides here with the relaxation time constant of chaotic fluctuations.
	When the structured and the random components have comparable strengths, instead, two heterogeneous chaotic phases co-exist (center). In this regime, the average persistence time increases monotonically with the structure strength, and reaches arbitrarily large values. Note that the relaxation time undergoes a very slow increase before sharply diverging at the boundary with stationary states, but the increase takes place on a much smaller scale.
	Finally, if the structure is too strong (right), the two bistable states become stationary. In this region, $\tau_r$ is formally infinite, while $\tau_p$ coincides with the total duration of our simulations.
	Pink continuous line: DMF prediction, measured as the full width half maximum of the auto-correlation function $\Delta(\tau)$. Pink dots: a rough estimate of $\tau_r$ from finite size networks is obtained by rectifying the population average signal and we computing the full width half maximum of its auto-correlation function. 
	\textsf{\bfseries F.} We compare the average transition rate with the average overlap between the two attractors in the phase space. 	For every unit, the typical overlap between its positive and its negative trajectories is given by $\pi_i=2(-\mu-\sqrt{\Delta_{\infty}}z+\sqrt{\Delta_0-\Delta_{\infty}})$. We average across the population, yielding: $\pi = 2(-\mu+\sqrt{\Delta_0-\Delta_{\infty}})$. We then normalize $\pi$ through dividing by its value in the unstructured chaotic regimes ($2\Delta_0$). When positive, $\pi$ returns an overlap; when negative, it measures a distance between the two orbits. For every set of the architecture parameters, the theoretical expected value of the overlap can be computed within the DMF framework.
	We show that, in finite-size networks, the transition probability between the two chaotic attractors monotonically increases with the attractors overlap in the phase space. In the figure, the points returned by simulations are fitted with an error function of which we evaluate numerically the amplitude, the offsets and the gain: $f(x) = p_0 + p_1\:\mathrm{erf}(p_2(x-p_3))$.
	Choice of the parameters: $\rho=0$, $g=3.0$, $\Sigma_m=0$.
	
	\label{fig:s6}
\end{figure}


\begin{figure}
	\captionsetup{labelformat=empty}
	\centering
	\includegraphics{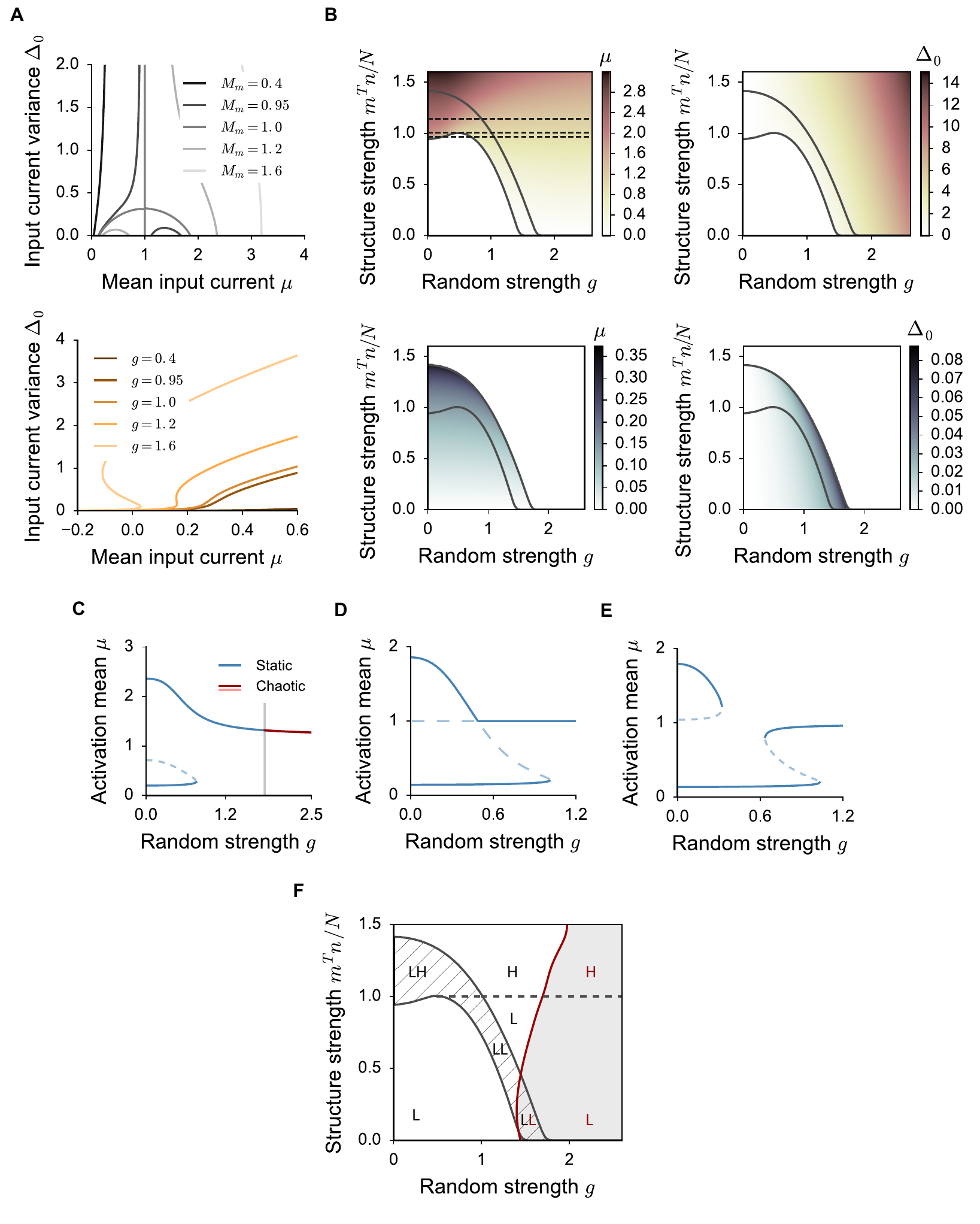}
	\caption{}
	\label{fig:s7}
\end{figure}

\begin{figure}
	\captionsetup{labelformat=adja-page}
	\ContinuedFloat
	\caption{ {\bf \textsf{ Dynamics of unit-rank networks with positively-defined activation functions.}} Related to Figure 1.} 
	
	In the main text, we performed our analysis of low-rank networks by adopting a completely symmetric network model, whose input-free solutions are invariant under the sign transformation $x_i(t) \rightarrow - x_i(t)$. Such symmetry is broken when a more biologically-plausible, positively-defined activation function $\phi(x)$ is adopted.
	Here, we investigate the effect of changing the transfer function to: $\phi(x) = 1 + \tanh(c(x-\gamma))$. Note that adding a shift $\gamma$ is equivalent to including an external and constant negative input. The parameter $c$, instead, rescales the slope of $\phi(x)$ at the inflection point. For simplicity, we fix $\gamma=1$ and $c=1.5$.
	We furthermore restrict the analysis to the case of unit-rank structures whose right- and left-connectivity vectors solely overlap on the unitary direction ($\rho=0$, see \emph{Methods}). The Dynamical Mean Field (DMF) sets of equations were derived for an arbitrary activation function, so they can directly be adapted to the present scenario.
	\textsf{\bfseries A.} We start by graphically analysing the stationary solutions (Eq.~83), and we plot the two nullclines of the system for different values of the architecture parameters.
	The top panel displays the $\mu$  nullclines for different $M_mM_n$ values. At $M_mM_n=1$, the unstable branch coincides with $\mu=1$, and the stable ones are symmetric. Around $M_mM_n=1$, the perfect pitchfork is broken in one or the other direction, generating a first stable continuous branch and a second one, where one unstable and one stable solution merge at low or high firing rate. For extremely low (resp. high) $M_mM_n$ values, finally, there's just one nullcline at low (resp. high) $\mu$ values. 
	The $\Delta_0$  nullcline (bottom panel) displays a more complex behaviour compared to the symmetric $\phi(x)=\tanh(x)$ case. When $g$ is sufficiently large, indeed, it can become a non-monotonic function of the mean input $\mu$, transforming into a \emph{S}-shaped nullcline. As it is shown in the following, this more complex shape is able to induce bistable activity even when the $\mu$ nullcline is reduced to a single continuous branch. This situation is reminiscent of the \emph{fluctuations driven} bistable regime in [Renart et al, 2007]. 
	\textsf{\bfseries B.} Stationary stable solutions plotted as color maps on the parameter space defined by the random and the structure strengths. The mean-field system admits two classes of stable solutions. The first one, illustrated in the top row, takes large mean and variance values. It suddenly disappears on the leftmost grey boundary of the plot, in a parameter region which co-exists with the second solution. The second solution, plotted in the bottom row, takes typically small values of $\mu$ and $\Delta_0$, and disappears on the right-most boundary with a first-order transition as well. 
	\textsf{\bfseries C-D-E.} In order to dissect more systematically the nature of those solutions, we fix the value of the structure strength (dashed lines in in \textsf{B}), and we gradually increase the random strength $g$.
	In \textsf{\bfseries C}, we fix the structure strength to high values: $M_mM_n=1.2$. The bifurcation pattern occurring in this case resembles what we observed in the original case with $\phi(x)=\tanh(x)$. At low values of $g$, two stable fixed points are built, respectively, on the high and on the low branches of the $\mu$ nullcline. For that reason, we call this state LH (cfr with \textsf{F}). When the random connectivity is too strong, the low firing rate fixed point annihilates, and only one high firing solution survives (H state).  
	In \textsf{\bfseries D},  $M_mM_n$ is exactly equal to unity.
	At small $g$ values, similarly to the previous case, network activity is bistable and admits one L and one H stationary state. As $g$ increases, the $\Delta_0$ intersect the high firing rate branch at smaller and smaller values of $\mu$. Finally, the H state is lost, and the second stable fixed point is realized on the intermediate branch at $\mu=1$. This bistable state is thus formally a LI state. Finally, at large $g$ values, the two intersections on the low rate branch collapse together and disappear. Bistability is lost and only one intermediate (I) state exists.
	In \textsf{\bfseries E}, we consider slightly smaller values of $M_mM_n$. A classical LH state exists at small $g$ values, the bistable state at large random strengths involves two stable solutions which originate both a low firing rates (LL state). The two states strongly differ in the value of their variance. When $g$ is sufficiently large, one unique low firing rate, high variance state (L) survives. 
	\textsf{\bfseries F.} The different activity states are finally sketched in the phase diagram. Note that I states separate the phase diagram in \textsf{F} in two macro areas: below the dashed line, every stationary and chaotic solution is built on the same low firing rate branch of the $\mu$ nullcline, and is thus formally a L state. Finally, the exact shape of the phase diagram depends on the value of the parameters $c$ and $\gamma$. 
	Choice of the parameters: $\rho=0$, $\Sigma_m=0$.
	
\end{figure}


\begin{figure}
	\captionsetup{labelformat=empty}
	\centering
	\includegraphics{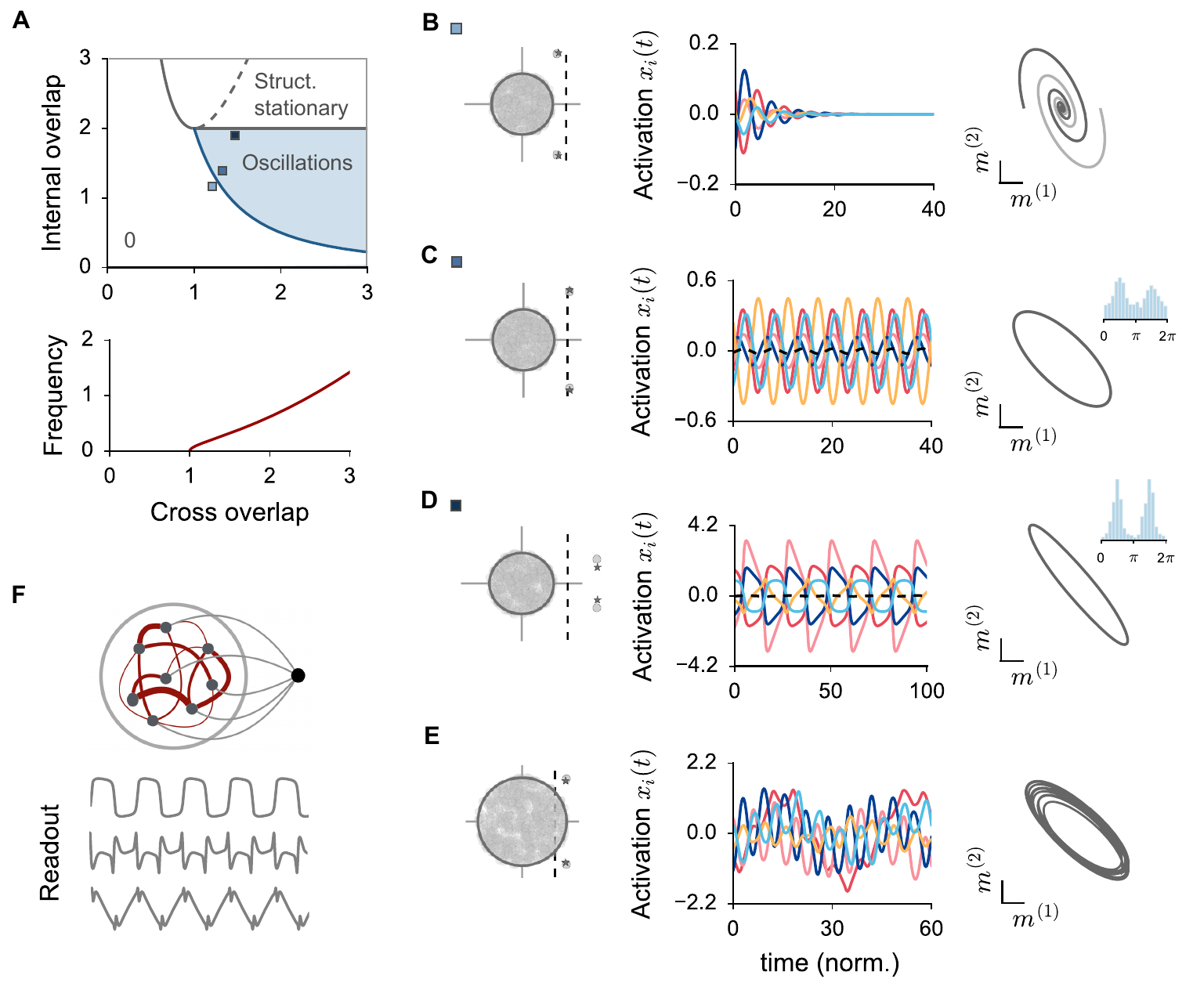}
	\caption{}
	\label{fig:s7}
\end{figure}

\clearpage

\begin{figure}
	\captionsetup{labelformat=adja-page}
	\ContinuedFloat
	\caption{ {\bf \textsf {Oscillatory activity from rank-two structures that include a cross overlap between  left- and right-connectivity vectors. }}}
	
	\textsf{\bfseries A.} Top: phase diagram for the rank-two structure with negative cross-overlap (see \emph{Methods}). For different values of the internal and the cross overlaps, the trivial fixed point can lose stability and give rise to oscillatory or stationary structured activity. The Hopf bifurcation is indicated in blue, the instability to stationary activity in grey. The light-blue parameter region corresponds to sustained non-linear oscillations. Bottom: frequency of oscillations along the Hopf bifurcation boundary, in units defined by the implicit time scale of the network dynamics.
	\textsf{\bfseries B-C-D-E.} Samples of activity for different connectivity parameters. From left to right: stability eigenspectrum of the trivial fixed point (theory and simulations), sample of activation trajectories (the population average is indicated in dashed black), and population dynamics obtained by projecting the population activation $x$ on the right-connectivity vectors $m^{(1)}$ and $m^{(2)}$. The parameters that have been used for every sample are indicated in \textsf{A}.   \textsf{\bfseries B}: Oscillatory transients in the fixed point regime. \textsf{\bfseries C}: Stable oscillations above the Hopf instability. The elongated shape of the closed trajectory on the $m^{(1)}-m^{(2)}$ plane is inherited by the phase distribution across the population, and can be tuned by slightly modifying the parameters of the rank-two structure (see \emph{Methods}).  \textsf{\bfseries D}: Highly non-linear oscillations close to the boundary with bistable activity. \textsf{\bfseries E}: Oscillatory activity at high $g$ values ($g=1.35$), where dynamics include a chaotic component.
	\textsf{\bfseries F.} When oscillations are strongly non-linear, their spectrum includes a large variety of frequencies that can be used to reproduce highly non-linear periodic patterns. We designed three random readout vectors and we linearly decoded activity from the dynamical regime in \textsf{D} to generate periodic non-linear outputs, which are displayed in grey.
	
\end{figure}

\end{document}